%% file: main.tex
\documentclass[twocolumn,twocolappendix]{aastex631}

\usepackage{graphicx}
\usepackage{float}
\usepackage{amsmath}
\usepackage{amssymb}
\usepackage{longtable}
\usepackage{array}
\newcommand{\PreserveBackslash}[1]{\let\temp=\\#1\let\\=\temp}
\newcolumntype{C}[1]{>{\PreserveBackslash\centering}p{#1}}
\usepackage{hyperref}
\hypersetup{linkcolor=blue,citecolor=blue,filecolor=cyan,urlcolor=black}
\defcitealias{Oey2005}{OWK2005}
\defcitealias{Navarete2011}{NFD2011}
\defcitealias{Bik2012}{Bik2012}
\defcitealias{Kiminki2015}{KKB2015}
\defcitealias{Roman-Lopes2019}{RL2019}

\input{commands.tex}

\revised{\today}
\submitjournal{ApJ}

\shorttitle{Red Stellar Populations and Dust Extinction Toward W3}
\shortauthors{Campbell et al.}


\newcommand\T{\rule{0pt}{2.6ex}} % Top strut
\newcommand\B{\rule[-1.2ex]{0pt}{0pt}} % Bottom strut

\providecommand{\sorthelp}[1]{}

\begin{document}

%%%%%%%%%%%%%%%%%%%%%%%%
%                                                                      %
%                TITLE AND AUTHOR                    %
%                                                                      %
%%%%%%%%%%%%%%%%%%%%%%%%

\title{Red Stellar Populations and Dust Extinction toward W3}

\correspondingauthor{Campbell, J. L.}
\email{campbell@astro.utoronto.ca}

\author[0000-0002-2511-5256]{J. L. Campbell}
\affil{David A. Dunlap Department of Astronomy \& Astrophysics, University of Toronto, 50 St. George Street, Toronto, ON M5S 3H4, Canada}
\affil{Canadian Institute for Theoretical Astrophysics, University of Toronto, 60 St. George Street, Toronto, ON M5S 3H8, Canada}

\author[0000-0002-5236-3896]{P. G. Martin}
\affiliation{Canadian Institute for Theoretical Astrophysics, University of Toronto, 60 St. George Street, Toronto, ON M5S 3H8, Canada}

\author[0009-0001-0559-4392]{S. Song}
\affiliation{Department of Information Technology and Electrical Engineering, ETH Zurich, Gloriastrasse 35,
8092 Zurich, Schweiz}

\author[0000-0003-1842-6096]{M. Rahman}
\affiliation{Sidrat Research, 124 Merton Street, Suite 507, Toronto, ON M4S 2Z2, Canada }

\author[0009-0005-6205-5810]{L. Einstein}
\affiliation{Stanford Intelligent Systems Laboratory, Durand Building, 496 Lomita Mall, Stanford, CA 94305, USA}

%%%%%%%%%%%%%%%%%%%%%%
%                                                                %
%                     ABSTRACT                        %
%                                                                %
%%%%%%%%%%%%%%%%%%%%%%

\begin{abstract}

We explore red stellar populations toward the W3 giant molecular cloud through the use of optical-to-infrared (IR) photometry and Gaia DR 3 data, simultaneously characterizing stellar content and properties of dust in the molecular medium. 
We use a Rayleigh-Jeans Color Excess (RJCE) method modified to de-redden stellar observations of both red giants (RGs) and OB stars, and construct an IR Hertzsprung-Russell diagram validated against the Besan\c{c}on Galactic model. 
Taking advantage of the near-universal IR interstellar extinction law and precise Gaia measurements, we develop a method for obtaining the spectral classification, foreground extinction, and distance moduli of stars, validated by spectroscopically-confirmed OB stars. 
We constrain the observed parallax and proper motion of OB stars in W3, demonstrating the importance of considering systematic effects in the parallax bias, and assign parallax- and proper motion-based cloud membership to our stellar samples. 
While it has been assumed that all spectroscopic OB stars are inside the W3 cloud, we find evidence of seven background B stars and three potential runaway OB stars. 
The methods developed here based on known stellar populations enable us to identify 82 new OB candidates that are confidently within the cloud.
We quantify several dust-to-dust empirical correlations, in particular the IR color excess $E(H-[4.5])$ and the optical depth $\tau_1$ of submillimeter dust emission at 1\,THz using RGs behind W3, measuring a best fit of 
$E\HG = (1.07 \pm 0.04) \times 10^3\, \tauu_\mathrm{,\,HOTT} + (0.00 \pm 0.02)\,\mags$. 
\end{abstract}


%\tableofcontents

%%%%%%%%%%%%%%%%%%%%%%%
%                                                                    %
%               INTRODUCTION                         %
%                                                                    %
%%%%%%%%%%%%%%%%%%%%%%%

\section{Introduction}\label{sec:intro}

Interstellar dust plays a crucial role in many astrophysical processes of the interstellar medium (ISM) and continues to pose a significant challenge for astronomical observations. Despite the increasing attention to interstellar dust, many details such as chemical composition \citep{Mathis1977, Draine2003, Hensley2021}, size distribution \citep{Kim1994, Weingartner2001, Clayton2003}, 3D distribution \citep{Green2019, Lallement2019, Leike2019}, grain alignment \citep{Draine1996, Draine1997, Lazarian2003, Weingartner2003, Lazarian2007, Hoang2009, Reissl2020}, and environmental effects \citep{Zasowski2009, planck2011-7.13, Kiminki2015} remain unclear. 

It is exceptionally difficult to characterize the dust properties of any individual line of sight unambiguously, such as the line-of-sight distribution of dust or the shape of the extinction curve. These observational challenges are exacerbated at low Galactic latitudes where most massive molecular clouds reside and dust extinction is both high and non-uniform. While advances have been made in understanding the dust properties of the diffuse ISM, interstellar dust associated with the molecular medium is more evolved and less well understood. For instance, the slope of the optical portion of the extinction curve is known to decrease toward molecular clouds with high levels of extinction, possibly attributed to grain growth and coagulation \citep{Draine2003}. There is also evidence of increased dust emissivity in the molecular ISM compared to the atomic medium that is again attributed to dust coagulation in dense environments \citep[e.g.,][]{planck2011-7.13}.

The observational challenges imposed by interstellar dust have an impact on the characterization of Galactic stellar populations. While spectroscopic information provides a confident assessment of stellar spectral classification, often photometric methods are necessary because spectroscopy is time-consuming and expensive. The use of near-infrared (NIR) colors is widely used for measuring interstellar reddening because dust extinction is less and the variability of its wavelength dependence less severe \citep{Cardelli1989, Martin1990, Indebetouw2005} and the variation in stellar intrinsic color is smaller \citep{Majewski2011}. While in the optical the reddening vector and stellar locus are often either parallel or have multiple intersection points in color-color diagrams, this degeneracy between dust reddening and intrinsic stellar color can be broken by combining NIR with mid-infrared (MIR) photometry \citep{Majewski2011}. Nevertheless, photometric classification in the infrared (IR) is still often hampered by unknown intrinsic colors, impacting color excesses, and highly uncertain or unknown distances. 

The advancement of high-precision stellar parallaxes with Gaia \citep{Gaia2018, Gaia2019, Gaia2021a} provides essential distance information for billions of nearby stars. However, Gaia relies on optical measurements that are severely impacted by regions of high extinction. A multi-pronged approach that combines the advantages of IR photometry and Gaia astrometry is essential for studying stars that are embedded in or behind molecular clouds and thus highly obscured by dust. 

Westerhout 3 (W3; \citealp{Westerhout1958}) is a nearby ($d\,{\sim}\,2\,\kpc$; \citealp{Hachisuka2006, Xu2006, Navarete2011}) giant molecular cloud (GMC) among the most massive (${\sim}4{\times}10^5\,\Msun$; \citealp{Moore2007, Polychroni2012}) star-forming regions in the outer Galaxy \citep{RI2011}. This GMC contains numerous compact {\HII} regions \citep{Megeath2008}, massive stars at varying degrees of evolution \citep{Tieftrunk1997}, and both triggered \citep{Oey2005} and rare \citep{Kerton2008} events of spontaneous star formation. This had made W3 a rich environment for targeted observational studies \citep[e.g.,][]{Elmegreen1980, Moore2007, Ruch2007, Megeath2008, Motte2010, RI2011, RI2015, Navarete2019}. 

In particular, the above challenging issues for spectral classification were encountered in a seminal photographic study of W3 by \citet{Elmegreen1980} in which 137 bright infrared stars (`the BIRS') were identified, whose photographic magnitudes at $I$($\lambdaeff \approx$ 0.82\,\micron) were much brighter than at $R$($\lambdaeff \approx$ 0.64\,\micron). The two-band photographic photometry at the time was insufficient for unambiguous spectral classification, the possible classifications ranging from giants and supergiants to massive early or pre-main-sequence stars. More study was recommended given the possibility that some of the BIRS were born and embedded in W3. Of course, Gaia parallax measurements were not yet available for distance determination.

In this paper, we use a wide variety of optical through IR photometry and Gaia DR3 astrometry to assess the spectral classification of the BIRS and other reddened stars, emphasizing the value of using spectroscopically-confirmed OB stars (OBs) and red giants (RGs) as fiducial points of comparison. Simultaneously, we investigate the properties of dust extinction in W3.

This paper is organized as follows. 
% 2
The data are described in Section \ref{sec:data}, with supplementary details in Appendix \ref{app:astrometry}. 
% 3
We present a modern assessment of the BIRS in Section \ref{sec:results}, spanning positions, cross matches to spectroscopic OBs and RGs, optical $r$ and $i$ photometry, association with thermal dust emission, and parallaxes. 
% 4
Section~\ref{sec:spectral-class} 
takes advantage of modern IR broadband photometric data and Gaia parallaxes to provide an initial classification of the BIRS based on their placement in various familiar photometric-based diagrams. 
% 5
Section~\ref{sec:BIRS-membership} uses Gaia data to establish parallax- and proper-motion-based cloud membership criteria to apply to our stellar samples. 
% 6
Section~\ref{sec:migrate} investigates potential runaway OB stars, field OB stars, and field RGs. 
% 7
We present our multiband-photometry spectral energy distributions (SEDs) from optical to IR in Section~\ref{sec:SEDs}. 
% 8
Then in Section~\ref{sec:SEDclass} we refine our spectral classifications by fitting these SEDs, incorporating models of the frequency dependence of dust extinction. 
%9
We search for new W3 OB star candidates based on IR photometry and Gaia astrometry in Section~\ref{sec:newOBcandidates}.
% 10
In Section~\ref{sec:distantprobes} we quantify the empirical correlation between IR color excess and the optical depth of submillimeter dust emission, using individual distant RGs as probes. 
% 11
Lastly, our conclusions are presented in Section~\ref{sec:conclusions}.

%%%%%%%%%%%%%%%%%%%%
%                                                         %
%                     DATA                           %
%                                                         %
%%%%%%%%%%%%%%%%%%%%

\section{Data}
\label{sec:data}

We use a variety of observational data for stars and thermal dust emission toward the W3 GMC in the outer Galaxy. Our stellar data also includes the western edge of the adjacent W4 GMC on which several of the BIRS are projected \citep{Elmegreen1980}. We also used an extinction curve for the dust spanning the range of the optical through IR photometry.

%%%%%%%%%%%%%%%%%%%%%%%
%                                                                   %
%                     ASTROMETRY                     %
%                                                                   %
%%%%%%%%%%%%%%%%%%%%%%%

\subsection{Photographic Positions}
\label{sec:opp}

Locations of the BIRS are tabulated in \citet{Elmegreen1980} and annotated on a print of the $I$ band photographic plate, her plate 9 or figure 1. We obtained digital images of many portions of this annotated print\footnote{Chris Sasaki, private communication.} and uploaded them to {\tt Astrometry.net}\footnote{\url{http://astrometry.net/}; Dustin Lang, private communication.} to produce digital images with World Coordinate System (WCS) coordinates. We used these and the stellar asterisms near the BIRS in combination with digital images from modern surveys to confirm the identifications and refine the coordinates, as described in Appendix \ref{app:astrometry}.

%%%%%%%%%%%%%%%%%%%%%%%
%                                                                   %
%                     PHOTOMETRY                     %
%                                                                   %
%%%%%%%%%%%%%%%%%%%%%%%

\subsection{Photometry}\label{subsec:photometry}

Our multi-band stellar photometry is based largely on data downloaded from publicly-available catalogs from optical and IR photometric surveys.  In somewhat historical order of our use, these are the Two Micron All Sky Survey (2MASS; \citealp{Skrutskie2006}), INT Photometric H-Alpha Survey (IPHAS) DR2 \citep{Drew2005}, American Association of Variable Star Observers (AAVSO) Photometric All-Sky Survey (APASS) DR9 \citep{Henden2014, Henden2016}), Panoramic Survey Telescope and Rapid Response System (Pan-STARRS1; \citealp{Chambers2016}), Asteroid Terrestrial-impact Last Alert System (ATLAS) DR1 \citep{Tonry2018}, GLIMPSE-360 \citep{Whitney2008}, All Wide-field Infrared Survey Explorer (AllWISE; \citealp{Cutri2014}), and unWISE (unWISE; \citealp{Lang2014, Schlafly2019}). Details and quality control filtering are described in Appendix \ref{app:qualityfilter}, along with some unpublished data from the Canada France Hawaii Telescope (CFHT), the Dunlap Institute Telescope (DIT), and Spitzer/IRAC. 

Finally we used lists of $J$, $H$, \Ks\ photometry from the Large Binocular Telescope \citep{Bik2012} and of $V$, $R$, $I$ photometry from the 2.3 m Bok Telescope \citep{Kiminki2015} that were provided for spectroscopically-confirmed OB stars in W3.

%%%%%%%%%%%%%%%%%%%%%%%%
%                                                                      %
%                     SPECTROSCOPY                   %
%                                                                      %
%%%%%%%%%%%%%%%%%%%%%%%%

\subsection{Spectroscopy}\label{subsec:spectroscopy}

We compiled lists from targeted IR (\citealp{Navarete2011} (\citetalias{Navarete2011}); \citealp{Bik2012} (\citetalias{Bik2012}); \citealp{Roman-Lopes2019} (\citetalias{Roman-Lopes2019})) and optical (\citealp{Oey2005} (\citetalias{Oey2005}); \citealp{Kiminki2015} (\citetalias{Kiminki2015}))\footnote{Although we tend to refer to them as OB stars, 89 of the 91 \citetalias{Kiminki2015} stars are B stars, not the more massive O stars capable of producing a significant \HII region.} observations of 116 spectroscopically-confirmed OB stars toward W3.

We use data from the SDSS-IV/Apache Point Observatory Galactic Evolution Experiment (APOGEE, \citealp{Majewski2012, Majewski2017}) DR14 \citep{Zasowski2013} to identify 215 RGs in a pointing with some coincidental overlap with W3. We obtained their effective temperature ({\Teff}), surface gravity (\logg), and metallicity ({\FeH}). 

We complement these data with unpublished spectra of a few targeted BIRS near KR 140 using the TripleSpec NIR spectrograph on the 5 m Hale telescope at Palomar Observatory. 
Using synthetic spectra in the 1.5--1.75\,\micron\ range, we found the following tentative classifications:
BIRS 128: M2 I,
BIRS 129: K4 I,
BIRS 130: K5 III,
BIRS 131: K5 III,
and
BIRS 132: OB.
These are refined by alternative approaches below. 

%%%%%%%%%%%%%%%%%%%%%%%%
%                                                                       %
%                   GAIA ASTROMETRY                  %
%                                                                       %
%%%%%%%%%%%%%%%%%%%%%%%%

\subsection{Gaia Astrometry}\label{subsec:astrometry}

We make use of the Gaia Early DR3 (EDR3; \citealp{Gaia2021a}) catalog for its improved astrometric positions, stellar parallaxes, and proper motions compared to DR2 \citep{Gaia2018}, noting that these particular quantities did not change between EDR3 and DR3. Despite the high-quality DR3 astrometry \citep{Lindegren2021astr}, many sources have spurious astrometric solutions that must be excised \citep{Fabricius2021}. We required that ${\tt (parallax+parallax\_error)}>0$ after the zero-point correction. We filtered according to the Renormalized Unit Weight Error ({\ruwe}) such that ${\ruwe}<1.4$ \citep{Fabricius2021} and the astrometric solution quality indicator \citep{Rybizki2022}, based on the Gaia Catalog of Nearby Stars \citep{Gaia2021b}, such that ${\tt astrometric\_fidelity}>0.5$.

%%%%%%%%%%%%%%%%%%%%%%
%                                                               %
%                     VARIABLES                      %
%                                                               %
%%%%%%%%%%%%%%%%%%%%%%

\subsection{Variable Stars}\label{subsec:variables}

We use both the first catalog of variable stars measured by ATLAS  \citep{atlasvar,Heinze2018} and that from Gaia\footnote{\url{https://gea.esac.esa.int/archive/documentation/GDR2/Catalogue_consolidation/chap_cu9val_cu9val/sec_cu9val_946_timeseries/}} \citep{Gaia2019, Eyer2022}.
%variable star catalogs.

%%%%%%%%%%%%%%%%%%%%%%%%
%                                                                       %
%                  EXTINCTION CURVE                  %
%                                                                       %
%%%%%%%%%%%%%%%%%%%%%%%%

\subsection{Extinction Curve}
\label{sec:extinction-curve}

The interstellar extinction curve in the NIR ($\sim$ 0.9--2\,\micron) can be described by a simple power-law of the form $A_\lambda\propto\lambda^{-\powerlaw}$. A power-law exponent in the range of 1.61 \citep{Rieke1985,Cardelli1989} to $\sim$ 1.8 \citep{Draine1989,Martin1990,Mathis1990} is often employed. This shape likely flattens out in the MIR ($\sim$ 2--8\,\micron, including for molecular clouds \citep{Flaherty2007}. We adopted the `Astrodust' extinction curve \citep{Hensley2020} for $\lambda\,\geq$\,2\,\micron\ to accommodate this flattening. 

We opted not to use the entire Astrodust extinction curve because it is aimed at the high Galactic latitude sky, whereas dust is likely more evolved in the molecular medium and exhibits different properties compared to the diffuse ISM \citep{Kim1994}. For example, the ratio of total to selective extinction {\RV} was found to be higher for W3 (3.6; \citealp{Kiminki2015}) compared to the Galactic average value for the diffuse ISM (3.1; \citealp{Draine2003}). The CCM extinction curve \citep{Cardelli1989} parameterized by {\RV} covers the optical and UV; we adopted this for our application in the more limited range $\sim$ 0.45--0.9\,\micron, along with its power law extension to 2\,\micron. Because of field coverage, the extinction curve might be different for most of the APOGEE RGs compared to stars that are within or behind W3. Therefore, we explored four extinction curves with the following parameterizations: an {\RV} of 3.1 or 3.6, and an {\powerlaw} of 1.61 or 1.8, using a normalization of $\AKs=1\,\mags$ at $\Ks$. Results shown below are for a fiducial extintion curve with $\RV = 3.6$ and $\powerlaw = 1.8.$ plus the Astrodust extension in the MIR.

From the normalized extinction curve, one can compute the normalized extinction at individual passbands, or differences between passbands (color excesses). Reddening vectors, including the slope and magnitude, can be computed as needed.  All of these computed values are related to the corresponding actual values by a common scale factor that accounts for the actual dust column in front of a star; this is a fitting parameter in SED fits (Section \ref{sec:SEDclass}). 

\begin{figure*}[t!]
\centering
\includegraphics[width=18cm]{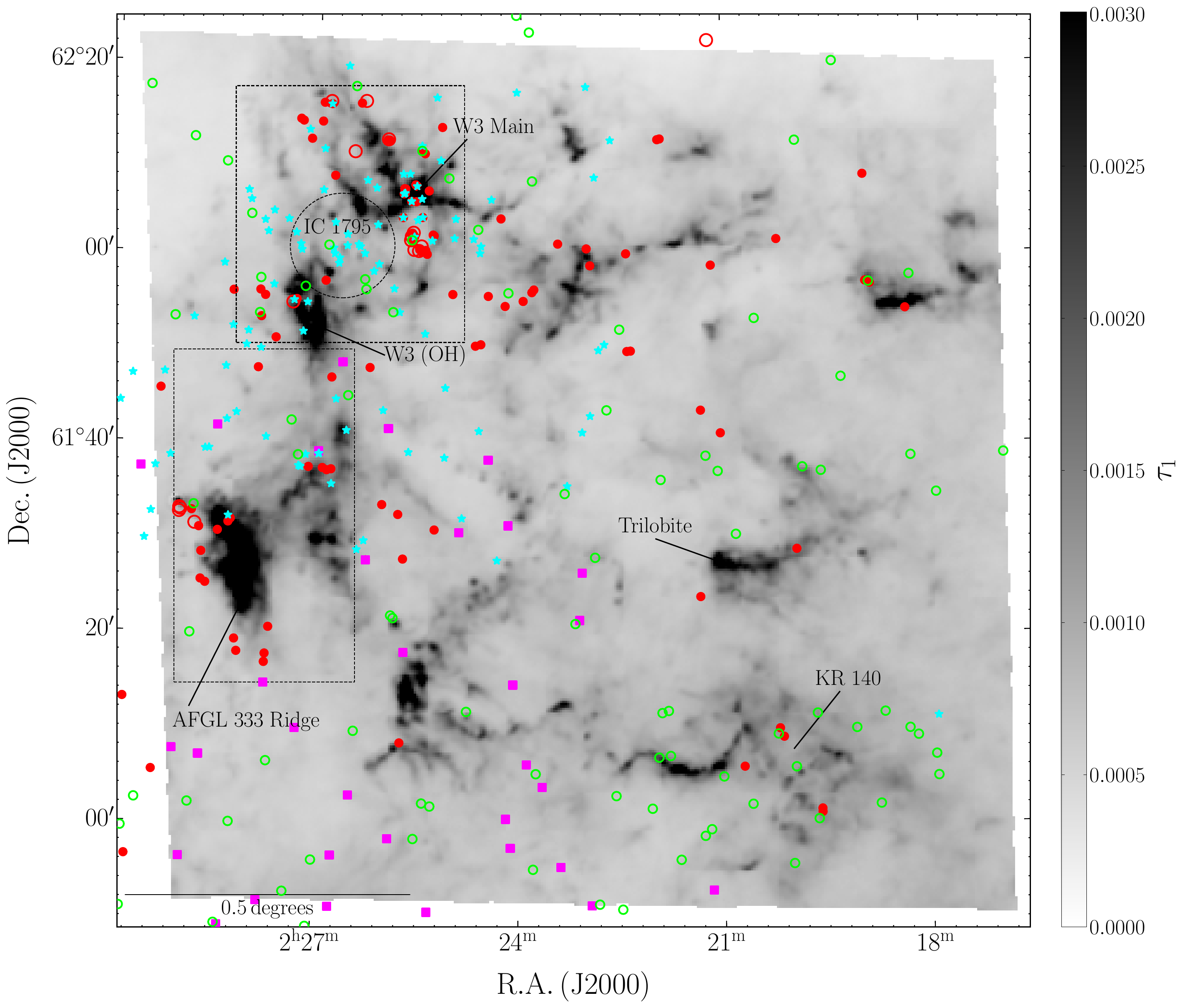}
\caption{HOTT $\tauu$ showing our corrected BIRS positions (red circles) along with positions of KKB2015 spectroscopically-confirmed OB stars (cyan stars) and APOGEE RGs (magenta squares) overlaid. The BIRS `imposters' are indicated with open circles. New OB candidates are shown with open lime circles (see Section \ref{sec:newOBcandidates}). Some of the most prominent star formation sites are identified, including W3 Main, W3 (OH), IC 1795, AFGL 333 Ridge, KR 140, and Trilobite. The upper dashed rectangle designates the $24\arcmin{\times}28\arcmin$ W3 Main Complex sub-region centered on $(\alpha_{\rm J2000},\delta_{\rm J2000})=(2^{\rm h}26^{\rm m}32^{\rm s}.4, +62{\deg}04{\arcmin}00{\arcsec})$, while the lower dashed rectangle designates the $19\arcmin{\times}35\arcmin$ AFGL 333 Ridge sub-region centered on $(\alpha_{\rm J2000},\delta_{\rm J2000})=(2^{\rm h}27^{\rm m}45^{\rm s}.00, +61{\deg}32{\arcmin}10{\arcsec})$ (see Section \ref{subsubsec:subr}). Within the W3 Main Complex is the IC 1795 OB cluster, identified with a $330\arcsec$ radius circle centered on $(\alpha_{\rm J2000},\delta_{\rm J2000})=(2^{\rm h}26^{\rm m}39^{\rm s}.0, +62{\deg}00{\arcmin}41{\arcsec}$; \citealp{Roccatagliata2011}), just a few arc seconds from BD+61 411 (IC 1795 89; \citealp{Ogura1976}), a main ionizing star \citep{Maiz2016}. The $0\fdg5$ angular scale shown in the lower left corner corresponds to a physical scale of $17\,\pc$ at a cloud distance of 2\,kpc or parallax of 0.5\,mas (Section \ref{subsec:ucalc}). This and subsequent figures can be zoomed to good effect to distinguish the fine detail.}
\label{fig:tau-map}
\end{figure*}

\subsection{Herschel Thermal Dust Emission}
\label{sec:hotttau}

For spatial context within the W3 GMC, in Figure \ref{fig:tau-map} we show the HOTT $\tauu$ map of the submillimeter optical depth $\tau$ at 1 THz (300\,\micron) from the Herschel Optimized Tau and Temperature (HOTT) analysis by \citet{Singh2022}.  Unlike the $\tau$ map used in the investigations reported by \citet{RI2013}, there is no correction for dust along the line of sight in front of or behind the W3 molecular cloud. 

The positions of the BIRS (red), spectroscopic OB (cyan), and APOGEE RG (magenta) are overlaid. Many of the BIRS are found toward regions of intermediate $\tauu$, rather than peak dust columns where optical magnitudes could be highly attenuated by dust; i.e., there are selection effects (see Section \ref{subsec:dust}).

%%%%%%%%%%%%%%%%%%%%%%%
%                                                                  %
%                   SUB-REGIONS                      %
%                                                                  %
%%%%%%%%%%%%%%%%%%%%%%%

\subsubsection{Sub-regions} \label{subsubsec:subr}

The BIRS appear in projection against various sub-regions of W3.
To avoid possible confusion with the terminology used in the literature, we define ours here and show these sub-regions on the $\tauu$ map in Figure \ref{fig:tau-map}. We refer to the entire radio structure and molecular cloud as `W3' \citep[e.g.,][]{RI2011}. The high density layer (HDL) adjoining W4 comprises the AFGL 333 Ridge, W3 Main, and W3(OH) substructures. While some targeted studies use the term `W3 Complex' to refer to the combined W3 Main and W3(OH) substructures \citep[e.g.,][]{Roman-Zuniga2015, Kiminki2015}, we use the term `W3 Main Complex' for clarity. We refer to the `AFGL 333 Ridge' as its own separate sub-region within the HDL. Some have used the terminology `W3 Cluster' to refer to the IC 1795 cluster \citep[e.g.,][]{Navarete2019}, but we opt to use the original `IC 1795' terminology.

%%%%%%%%%%%%%%%%%%%%%
%                                                             %
%                     RESULTS                        %
%                                                             %
%%%%%%%%%%%%%%%%%%%%%

\section{A modern assessment of the BIRS}\label{sec:results}

%%%%%%%%%%%%%%%%%%%%%%%%
%                                                                       %
%                     BIRS ASTROMETRY                %
%                                                                       %
%%%%%%%%%%%%%%%%%%%%%%%%

\subsection{Gaia Astrometric Positions of the BIRS}\label{subsec:corrected-astrometry}

\begin{figure}[t!]
\centering
\includegraphics[width=8.5cm]{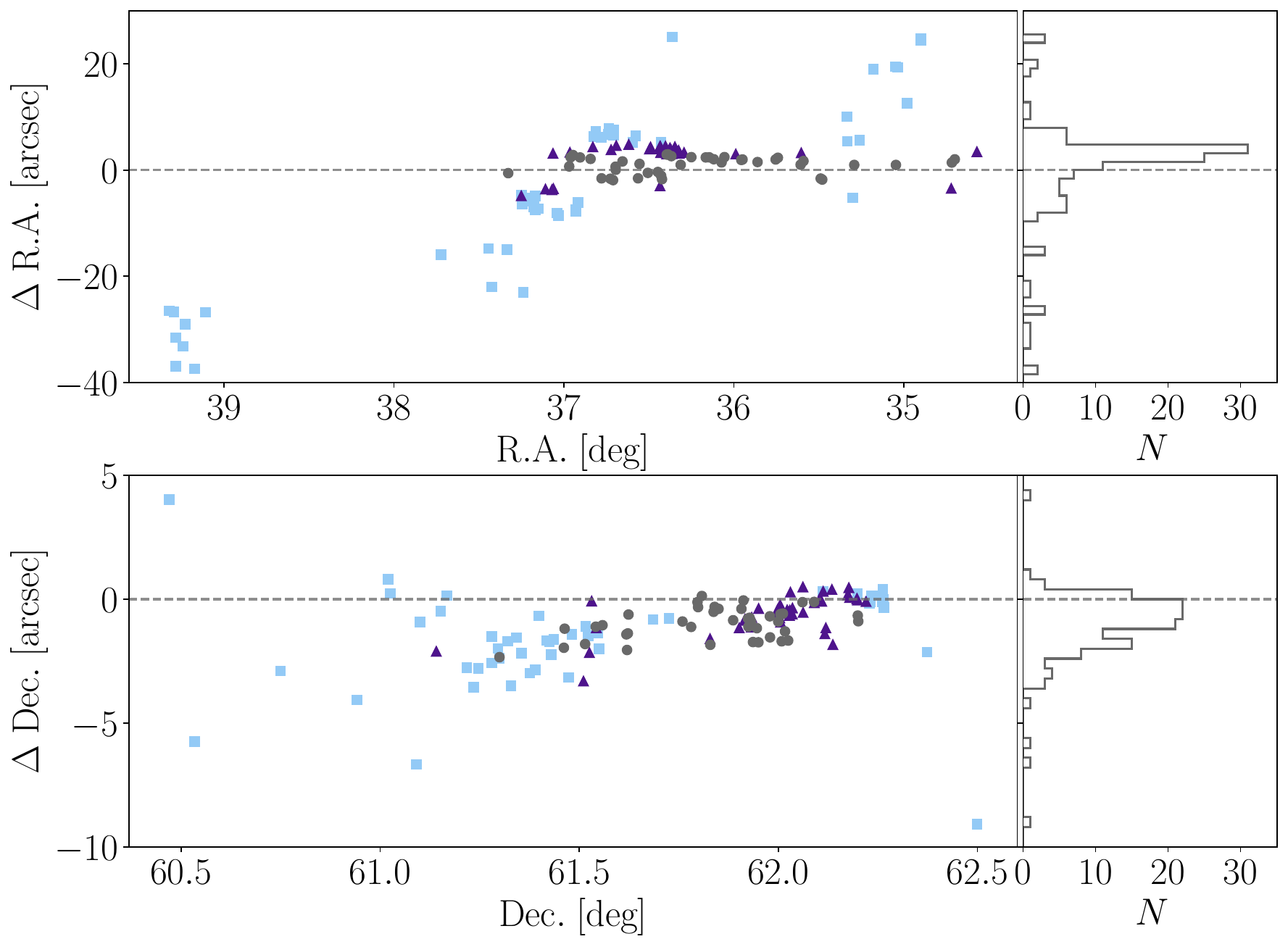}
\caption{Offsets to the BIRS J2000 positions in R.A. (top panel) and Dec.\ (bottom panel) as a function of position (left panel), and their overall distributions (right panel). BIRS with total offsets $<3\arcsec$ are shown in grey (circles), between $3\arcsec$ and $5\arcsec$ in purple (triangles), and $\geq5\arcsec$ in blue (squares).}
\label{fig:pos-offsets}
\end{figure}

Upon first inspection of 2MASS images, we found that the tabulated positions of the BIRS in \citet{Elmegreen1980} were inaccurate and in need of updating. This can be seen by searching the BIRS by identifier in SIMBAD\footnote{\url{http://simbad.cds.unistra.fr/simbad/sim-fid}} and inspecting the 2MASS and DSS previewers. For example, BIRS 128 to 132 near KR 140 show a clear systematic positional offset indicative of an inaccurate photographic plate solution in this region. Details of our work leading to Gaia DR3 positions for the BIRS are described in Appendix \ref{app:astrometry}, with updated positions tabulated in Table \ref{table:birs-positions}.

Figure \ref{fig:pos-offsets} shows the offsets to the BIRS J2000 positions\footnote{The original BIRS positions were tabulated as B1950, and so we first transformed them to J2000 before computing offsets.} in R.A. (top panel) and Dec.\ (bottom panel) as a function of position (left panel), and their overall distributions (right panel). The BIRS required positional offsets on the order of several to tens of arcseconds, which is larger than the estimated 0{\farcs}9 positional accuracies \citep{Elmegreen1980}. The systematic nature of the positional offsets suggests a degrading quality of the plate solution away from the central pointing of the photographic plates. The magnitudes of the positional offsets, particularly in R.A., are largest beyond the eastern edge of the HOTT dust map of W3 (Figure \ref{fig:tau-map}) that we use for our comparisons with thermal dust emission.

One star in particular (BIRS 20) was found to have significant positional offsets between survey epochs. We investigated the proper motion of this candidate foreground star \citep{Elmegreen1980} in detail in Appendix \ref{subsec:BIRS20}. 

We also found three BIRS (4, 5, and 41) to be confused nebulosity structures rather than stellar objects. BIRS 4 (in W3 North along with BIRS 5 -- see below, and BIRS 6 -- see Section \ref{subsec:birs6}) is completely obscured by nebulosity in Pan-STARRS $g_{P1}$ and $r_{P1}$ as well as Gaia. BIRS 5 (in the W3 North nebulosity) and BIRS 41 (between BIRS 40 and BIRS 42 below IRS N3 in W3 J \citealp{Bik2012}) are diffuse ridges with no indication of stellar objects in either Pan-STARRS or Gaia data.  In addition, BIRS 106 appears to be a blend of two unresolved stars. These four objects were therefore discarded from our analysis and lack updated positional measurements (or photometry).

%%%%%%%%%%%%%%%%%%%%%%%%%%
%                                                                             %
%                   BIRS CROSS MATCHES                  %
%                                                                             %
%%%%%%%%%%%%%%%%%%%%%%%%%%

\subsection{BIRS Cross Matches to Spectroscopic OB stars and RGs}\label{subsec:BIRS-OB stars-RGs}

We cross matched the Gaia positions of the BIRS in Table \ref{table:birs-positions} to positions in lists of 116 spectroscopically-confirmed OBs and 215 APOGEE RGs (Section \ref{subsec:spectroscopy}). To determine the cross-match radius for a list, we measured the mean ($\mu$) and standard deviation ($\sigma$) of the peak positional offset using the nearest Gaia matches to the positions in the list and used $\mu+5\sigma$ as the cross-match radius with the BIRS. All but two BIRS matched to \citetalias{Bik2012} OB stars also have \citetalias{Kiminki2015} matches. For the one OB star with a match to \citetalias{Navarete2011}, we used the minimum cross-match radius for this star. For stars with matches to both \citetalias{Bik2012} and \citetalias{Kiminki2015} lists, we opted to use their listed optical positions. For the two OB stars with only \citetalias{Bik2012} positions, we used a cross-match radius that captures both matches. Our cross-match radii for the spectroscopic lists are $0\farcs9$, $0\farcs4$, $0\farcs7$, $0\farcs5$, and $0\farcs4$ for the \citetalias{Navarete2011} OBs, \citetalias{Bik2012} OBs, \citetalias{Kiminki2015} OBs, \citetalias{Roman-Lopes2019} OBs, and APOGEE RGs, respectively. 

Of the 133 BIRS (not including the three that we identified as being diffuse structures, and 106), we found matches for seven as OBs (BIRS 6, 24, 26, 28, 30, 31, and 35) and three as RGs (BIRS 85, 94, and 126), noting that the APOGEE list has very limited coverage of W3. These matches are noted in column 2 of the summary Table \ref{table:birs-results} and further details on the spectral type from the literature can be found in the cited lists (table footnote b).

\input{tables/table1.tex}

%%%%%%%%%%%%%%%%%%%%%%%%%
%                                                                         %
%               R AND I PHOTOMETRY                   %
%                                                                         %
%%%%%%%%%%%%%%%%%%%%%%%%%

\begin{figure}[t!]
\centering
\includegraphics[width=8.5cm]{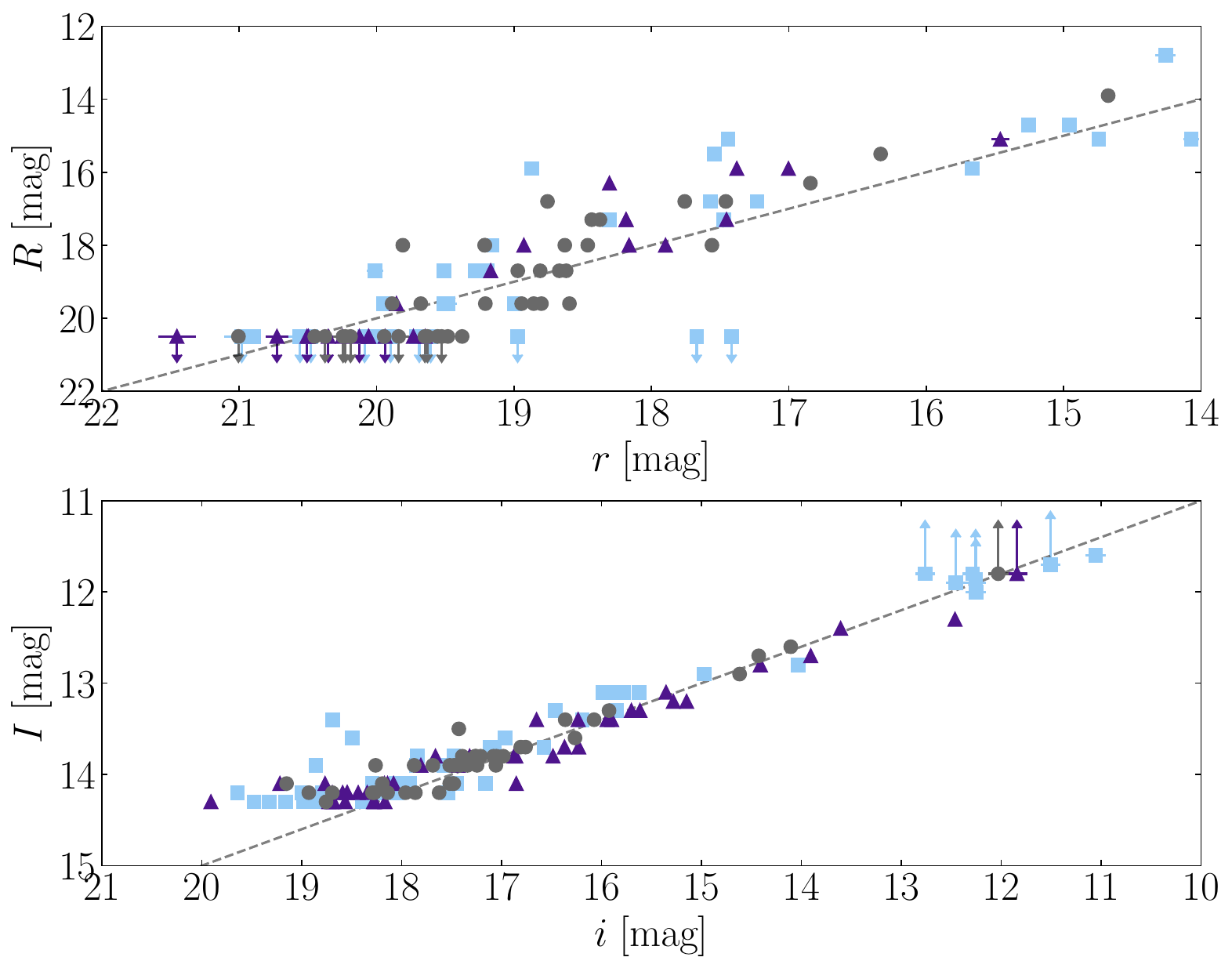}
\caption{Photographic magnitudes vs.\ ATLAS photometric magnitudes of the BIRS using the same color scheme as that in Figure \ref{fig:pos-offsets}. Top panel: $R$ vs.\ $r$ with a line of unit slope shown as the dashed line. Bottom panel:  $I$ vs.\ $i$ with, by contrast, a line of slope 0.4.}
\label{fig:rvsr_ivsi}
\end{figure}

\subsection{Photometry in the R and I and $r$ and $i$ bands}
\label{subsec:RI-photometry}

We compared the original tabulated $R$ and $I$ photographic magnitudes of the BIRS \citep{Elmegreen1980} to modern ATLAS $r$ and $i$ AB photometric magnitudes (see Table \ref{table:birs-positions}). Figure \ref{fig:rvsr_ivsi} shows $R$ vs.\ $r$ (top panel) and $I$ vs.\ $i$ (bottom panel) using the same color scheme as that in Figure \ref{fig:pos-offsets}. While detected at the $I$ band, many of the BIRS have only upper-limit photographic measurements in the $R$ band where dust extinction and/or intrinsically red objects cause fainter measurements. The $I$ and $i$ magnitudes are well correlated and there is no observable trend in the scatter of $I$ vs.\ $i$ or $R$ vs.\ $r$ with the magnitude of the positional offsets, validating our BIRS identifications and corrected positions. 

While there is significantly less scatter in $I$ vs.\ $i$, the $R$ vs.\ $r$ magnitudes more closely follow a slope of one while $I$ vs.\ $i$ magnitudes follow a slope of 0.4. Further, the $I$ magnitudes are $3.5\pm0.1\,\mags$ brighter than the $i$ magnitudes, resulting in photographic $(R-I)$ colors that are significantly redder than their photometric ${\ri}$ counterparts. The effective wavelength of the photographic $I$ magnitudes (0.82\,\micron) is closer to that of ATLAS $z$ (0.866\,\micron) than $i$ (0.752\,\micron), biasing photographic $(R-I)$ to redder colors. The $I$ magnitudes are, however, equally as poorly correlated with $z$ magnitudes.

The photographic magnitudes of the BIRS were measured using the magnitude-diameter relation, which was intended for determining approximate magnitude ranges and not for quantitative photometry \citep{King1977}. The BIRS photographic magnitudes are therefore imprecise and lack tabulated uncertainties. A global $\sim1$ mag uncertainty was estimated, which might be higher for the seven brightest BIRS because of a change in slope of the magnitude-diameter relation \citep{Elmegreen1980}.

\begin{figure}[t!]
\centering
\includegraphics[width=8.5cm]{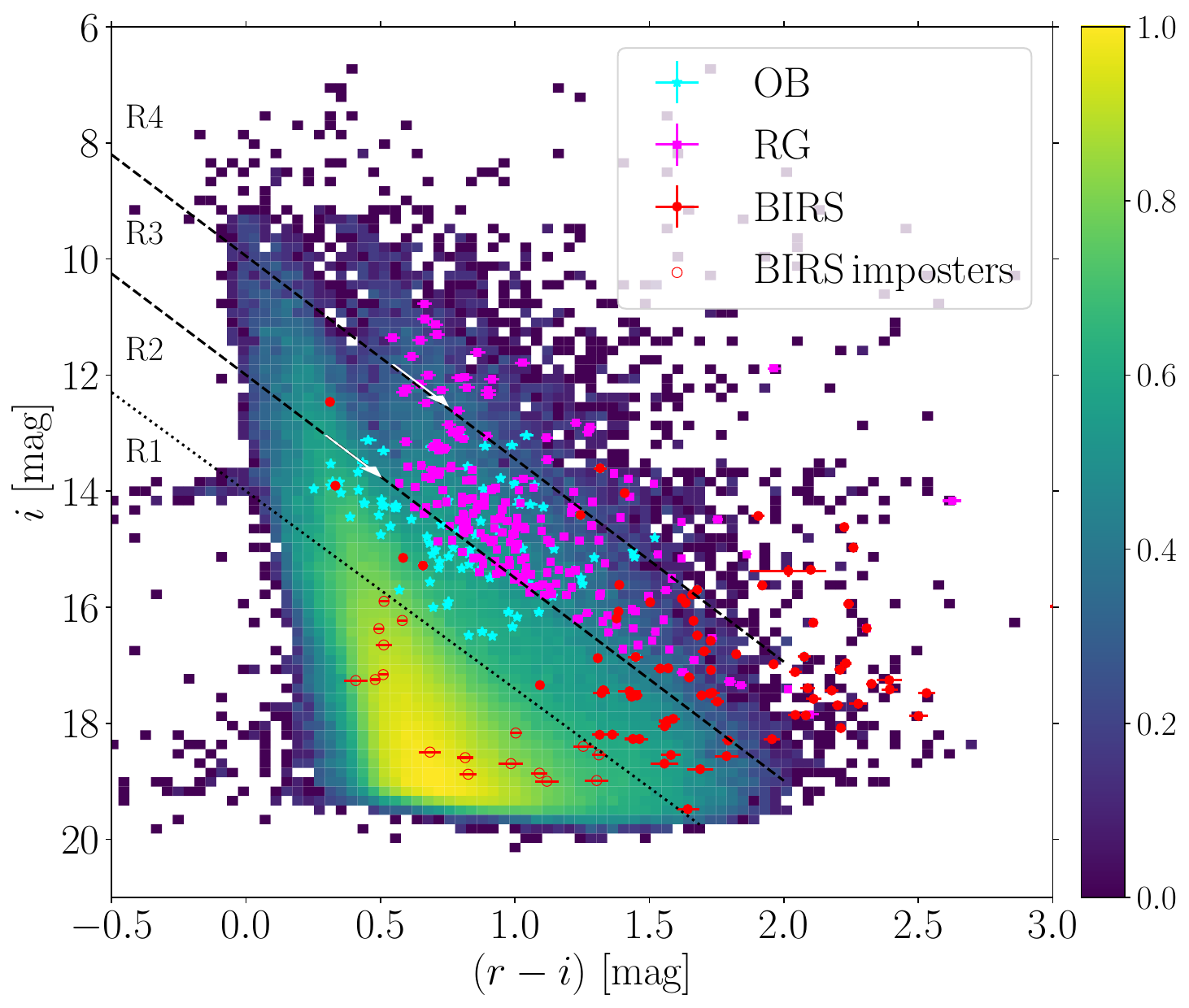}
\caption{ATLAS CMD, as a 2D histogram of $i$ vs.\ {\ri}, of stars toward W3 in AB magnitudes. The color bar for the 2D histogram shows the number of stars relative to the maximum for the chosen pixelization. Overlaid are the BIRS (red circles), OB stars (cyan stars), and APOGEE RGs (magenta squares). The lower dotted black diagonal line indicates a color-magnitude cut of $i<3.41(r-i)+14\,\mags$ to identify faint and unreddened BIRS `imposters' (open red circles). The arrows (white) designate reddening vectors of typical OB stars (lower) and RGs (upper) for $\AV=1$ while the dashed diagonal lines show the extension of the reddening vector for greater \AV. Regions 1 to 4 (R1-4) designate different stellar populations in the CMD as discussed in the text.
}
\label{fig:ivsri}
\end{figure}

\subsection{BIRS designation}\label{subsec:desig}

As the BIRS acronym encodes, the original BIRS designation was based on both $I$ photographic magnitudes being fairly bright and a red $R-I$ color, which should be evident in a color magnitude diagram (CMD). We show an ATLAS $i$ vs.\ {\ri} CMD in Figure \ref{fig:ivsri} from our catalogs of the field, overlaid with the BIRS (red circles), OB stars (cyan stars), and APOGEE RGs (magenta squares). The two approximately vertical concentrations near the top of the underlying distribution are unreddened bands of main sequence (MS) stars (left) and RGs (right), with mixing of stars of the same spectral classification vertically, because $i$ is an apparent magnitude dependent on distance modulus. The white arrows designate reddening vectors of typical OB stars (lower) and RGs (upper) for ${\AV}=1\,\mags$, while the dashed diagonal lines toward the lower right show the extensions for larger values of {\AV}.\footnote{Reddening vectors are calculated approximately from monochromatic effective wavelengths and the adopted extinction curve with $\RV = 3.6$ and $\powerlaw = 1.8$ (Section \ref{sec:extinction-curve}).} 

A subset of the BIRSs are found near the unreddened low-mass MS band and have no cross matches with spectroscopic OB stars or RGs. These stars, which with modern photometry are neither particularly bright nor red, we dub `imposters' because they contradict their original `BIRS' designation. The lower dotted black diagonal line, $i<3.41\, {\ri}+14\,\mags$, separates 17 imposters shown as open circles (BIRS 7, 9, 13, 16, 33, 34, 36, 40, 44, 45, 51, 78, 96, 97, 98, 100, 117). All of the BIRS imposters were found toward the HDL and W3 Main Complex regions, except BIRS 117 toward W4. We make note of the BIRS imposters in the last column of Table \ref{table:birs-results}.

We identified four distinct regions in the ATLAS $i$ vs.\ {\ri} CMD containing different stellar populations. While Region 1 (R1) contains mostly unreddened low-mass MS stars (including BIRS imposters), R2 contains early-type stars with varying degrees of extinction. Spectral classification becomes increasingly degenerate in R3 and R4, particularly to the right of the unreddened RG band, where we see mixing between OB stars and RGs over a wide range in extinction. This highlights the challenge of separating OB stars and RGs with such limited data. The degeneracy between dust reddening and intrinsically red stellar photospheres must be lifted using other data, as we investigate further in Section \ref{sec:spectral-class}.

Using modern photometry, four BIRS in R2 and R3 are not particularly red. BIRS 32 and 42 are found near the unreddened MS band 
in R3 with 
[$i, \ri$] at [12.46, 0.32]\,\mags\ and R2 at [13.91, 0.33]\,\mags, respectively. While they plot among particularly bright and minimally-reddened OB stars, their parallaxes indicate that they are lower mass MS foreground stars. 
On the other hand, BIRS 35 (spectroscopic OB star \citetalias{Kiminki2015} 26) and 37, both in R2 at [15.28, 0.66]\,\mags\ and [15.15, 0.58]\,\mags, respectively, are moderately reddened late B stars at the parallax of W3 or beyond. 

\citet{Elmegreen1980} noted BIRS 20 as a candidate foreground star from its proper motion. It is among the reddest of the BIRS, in R4 at [15.4, 2.0]\,\mags. It is a late M dwarf (see Appendix \ref{subsec:BIRS20}).

\subsubsection{BIRS-like stars}
\label{subsubsec:birslike}

With modern photometry, we can define a wider population of `BIRS-like' stars using a CMD like Figure \ref{fig:ivsri}. Depending on the goals of such a designation, one might include stars brighter than $i<3.4\, {\ri}+12\,\mags$ (the R2 -- R3 boundary), i.e., in R3 and R4, and redder than $\ri>1$ (avoiding the MS). Depending on the survey, one might also adopt a cutoff at low brightness, because significant reddening and extinction can challenge optical sensitivity limits.

\begin{figure}[t!]
\centering
\includegraphics[width=8.5cm]{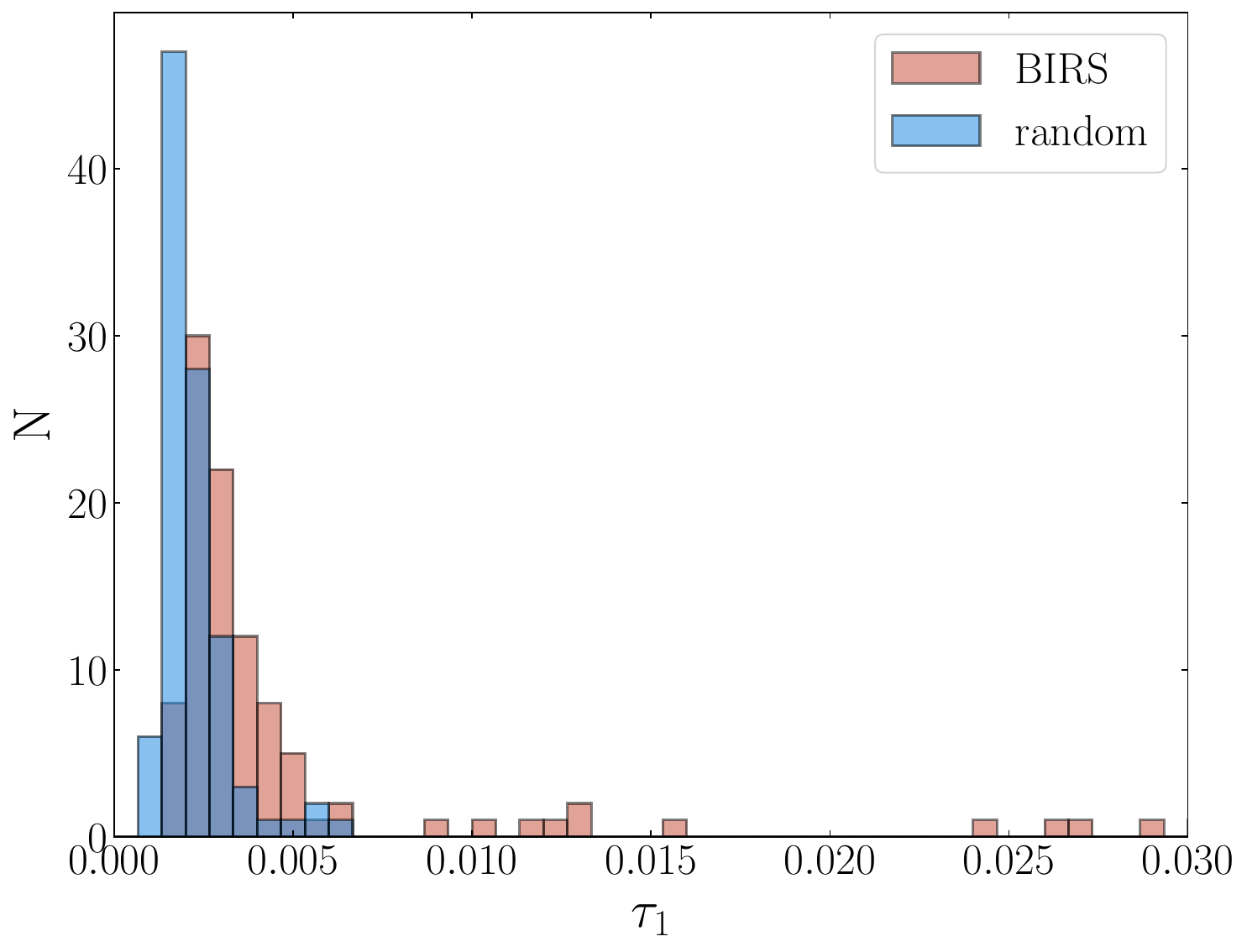}
\caption{Distribution of HOTT $\tau_1$ for the BIRS sightlines (red) compared to random sightlines toward W3 (blue).}
\label{fig:columndenhist}
\end{figure}

%%%%%%%%%%%%%%%%%%%%%%%
%                                                                    %
%                     THERMAL DUST                   %
%                                                                    %
%%%%%%%%%%%%%%%%%%%%%%%

\subsection{Relationship of the BIRS and HOTT $\tauu$}\label{subsec:dust}

\subsubsection{Histogram of $\tauu$ toward the BIRS}
\label{subsec:histtau1}

\citet{Elmegreen1980} found that the BIRS tend to be located in areas of low star counts so that their $(R-I)$ colors could be attributed to foreground dust reddening (although this is somewhat undermined by the significant positional errors). We investigated this idea further by searching for evidence that the BIRS are associated with increased HOTT $\tauu$. We interpolated the $\tauu$ map (resolution 36\farcs7, pixel size 14\arcsec) for the BIRS positions using a bicubic spline interpolation.

\begin{figure}[t!]
\centering
\includegraphics[width=8.5cm]{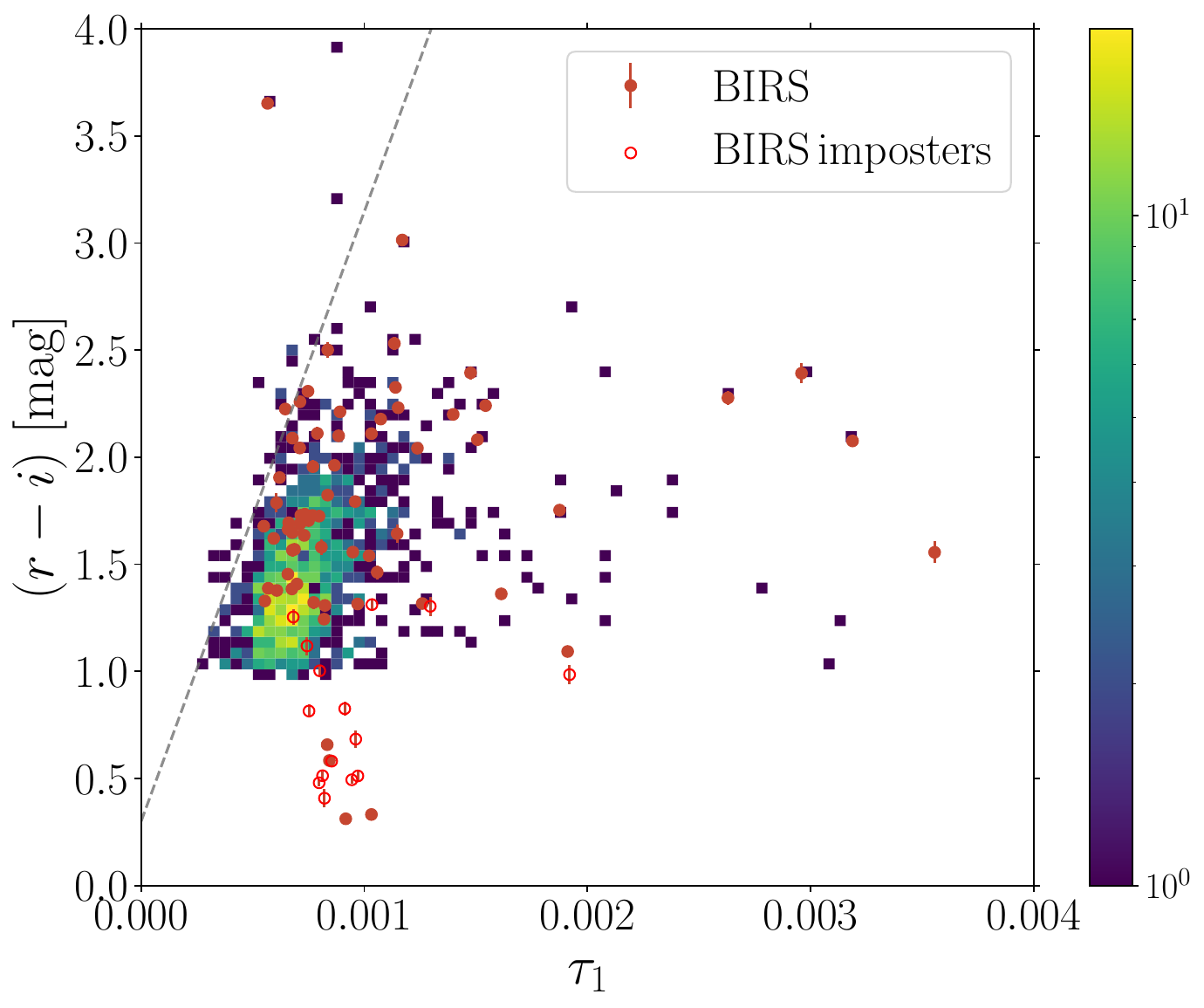}
\caption{2D histogram of ATLAS {\ri} color vs.\ HOTT ${\tauu}$ for BIRS-like stars toward W3 with $i<3.4\, {\ri}+12\,\mags$ and redder than ${\ri}>1$. Relative color bar as in Figure \ref{fig:ivsri}. Plotted in red for comparison are the original BIRS and BIRS imposters.  The diagonal dashed line shows an approximate upper envelope consistent with distant stars sampling the full column density measured by $\tauu$.  The intercept assumes an intrinsic $(r-i)$ color of $\sim 0.3$\,mag for background RGs.
}
\label{fig:ri-vs-tau}
\end{figure}

In Figure \ref{fig:columndenhist}, we compare the distribution of HOTT $\tauu$ toward the BIRS (red) to the distribution for an equal number of random positions in the map (blue).  The distribution of $\tauu$ toward the BIRS sightlines is skewed to slightly higher optical depths, with a median value ($\tau_1=0.003$) that is 1.6 times higher than that of the random cloud positions. The BIRS sightlines trace a much broader range in \tauu ($0.001<\tauu<0.03$) than the random cloud positions ($0.0.001<\tauu<0.006$) with a standard deviation ($\sigma=0.006$) that is 6.9 times that of the random cloud positions. This is consistent with the BIRS having an $(R-I)$ color selection threshold \citep{Elmegreen1980}, which might imply higher dust columns.

\begin{figure*}[t!]
\centering
\includegraphics[width=18cm]{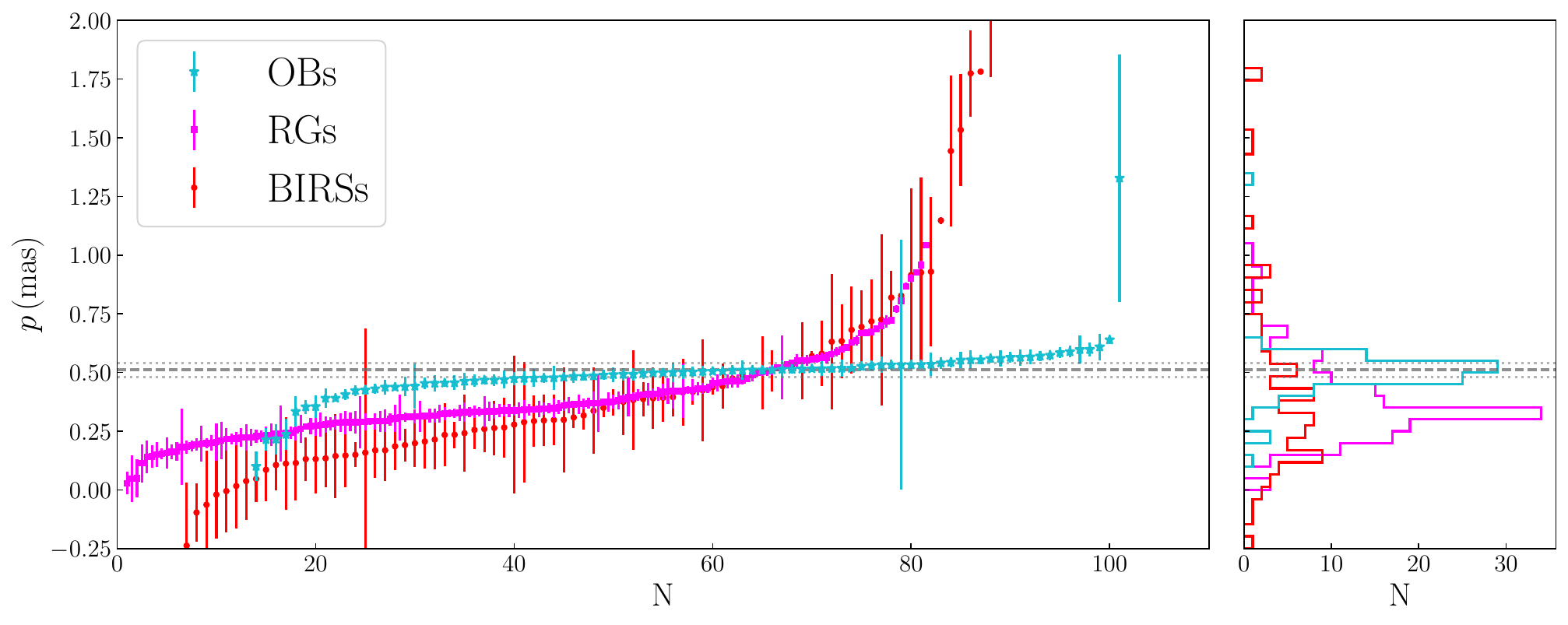}
\caption{Gaia parallaxes of the BIRS (red circles), spectroscopic OB stars (cyan stars), and APOGEE RGs (magenta squares) toward W3 sorted by increasing parallax (`S-plot'; left) along with their respective distributions (right). The error bars represent the $1\sigma$ parallax uncertainty. The dashed horizontal line denotes the W3 maser parallax of $\varpi=0.512\pm0.010\,{\rm mas}$ \citep{Xu2006} while the dotted lines denote the $\pm0.03\,\mas$ parallax distribution width around the median OB parallax used for our proper motion analysis (see Section \ref{subsec:propermotion}). The OB and RG S-plots were shifted horizontally such that they cross the maser parallax at approximately the same location along the x-axis. We note that the BIRS with the four largest parallaxes (BIRS 20, 90, 104, 105, and 136) are not shown here for increased readability.}
\label{fig:Splot}
\end{figure*}

\subsubsection{Color \ri\ and $\tauu$}
\label{subsec:rmitau1}

If dust reddening is a significant factor in determining how red the colors of the BIRS and BIRS-like samples are, then their {\ri} colors should be correlated with other tracers of dust like $\tauu$. Figure \ref{fig:ri-vs-tau} shows the ATLAS {\ri} color vs.\ HOTT ${\tauu}$ for a possible population of BIRS-like stars selected from R3 and R4 in the ATLAS $i$ vs.\ {\ri} CMD (see Section \ref{subsubsec:birslike}), with the BIRS (solid red circles) and BIRS imposters (open red circles) overlaid. 

The BIRS and BIRS-like samples are seen in projection. Their {\ri} colors are reddened only by dust foreground to each star, which is not necessarily the full line-of-sight dust column measured by $\tauu$. The {\ri} colors and $\tauu$ should be more strongly correlated for distant stars that trace more of the dust column compared to foreground stars, forming an `upper envelope' on the distribution, notionally given by the dashed diagonal line (slope $\sim 2800$), along which many of the BIRS are also found. But not surprisingly given the selection effects, many of the BIRS do not trace the entire dust column.

Because {\ri} is a color and not a color excess, the intercept would be the intrinsic color of the dominant class of background stars. For example, if these were red clump stars, thus early K giants (RGs), the intrinsic {\ri} color would be about 0.3 \citep{Covey2007,Ruiz2018}. We explore the upper-envelope relation in Section \ref{subsec:stellarprobes} to establish the color excess per unit optical depth for the high column density molecular medium in W3.

There are a number of BIRS that are found well below this envelope, most of which are BIRS imposters. The four BIRS near $\ri \sim 0.5$ and $\tauu \sim 0.006$ (BIRS 32, 35, 37, and 42) are the four low reddening MS stars discussed in Section \ref{subsec:desig}, two of which (BIRS 32 and 42) are distinctly lower mass foreground stars based on their particularly large Gaia parallaxes (Section \ref{subsec:parallax}). 

%%%%%%%%%%%%%%%%%%%%%%%
%                                                                   %
%                   W3 PARALLAX                       %
%                                                                   %
%%%%%%%%%%%%%%%%%%%%%%%

\subsection{Gaia Parallaxes and Zero Point Correction} \label{subsec:parallax}

We took advantage of Gaia astrometric data to investigate whether the parallaxes of the BIRS are consistent with the conclusion by \citet{Elmegreen1980} that most the BIRS are embedded within the molecular cloud. 

The absolute parallax of a star is $\varpi - Z$, where $\varpi$ is the tabulated value in the EDR3 release and $Z$ is a parallax bias (zero point) to adjust for systematic errors \citep{Gaia2021a}. The global zero point from quasars is $-17\,\uas$. 

However, $Z$ for stars contains systematic errors on the order of a few tens of microarcseconds as a function of magnitude, color, and ecliptic latitude \citep{Lindegren2021bias}. We used the {\tt gaiadr3\_zeropoint}\footnote{\url{https://gitlab.com/icc-ub/public/gaiadr3\_zeropoint}} Python package to compute $Z$ for sources with 5p (${\tt astrometric\_params\_solved}=31$) and 6p (${\tt astrometric\_params\_solved}=95$) solutions. In cases where {\tt nu\_eff\_used\_in\_astrometry} is missing, we supplemented this with the astrometrically-estimated effective wavenumber, {\tt pseudocolour}. We found median values of $Z$ that are significantly more negative in the W3 region compared to the $-17\,\uas$ global average: $-30\pm9\,\uas$ for the entire population of stars in our W3 field, $-38\pm9\,\uas$ for the OB stars, $-44\pm12\,\uas$ for the RGs, and $-53\pm15\,\uas$ for the BIRS, where the quoted $\pm$ values are the respective standard deviations of $Z$ for the sample, not the typical uncertainty $\sigma_Z$ for a given star.

The angular power spectrum of the parallax bias in Gaia EDR3 data indicates that $\sigma_Z$ is $\sim23\,\uas$ on $0\fdg5$ angular scales and saturates at $26\,\uas$ on $\sim0\fdg1$ angular scales \citep{Lindegren2021astr}. We adopt $\sigma_Z = 0.026\,\mas$ in the usual units of Gaia DR3 reported parallax uncertainties, $\sigma_\varpi$. We note that it is not taken into account in the {\tt gaiadr3\_zeropoint} Python package. For the best measured Gaia parallaxes reported, like for the OB stars, $\sigma_Z$ is comparable to $\sigma_\varpi$.  For precision analysis of the W3 parallax and membership (Sections \ref{subsec:parallax-systematics} and \ref{subsec:parallax-membership}), we add $\sigma_\varpi$ and $\sigma_Z$ in quadrature.

The BIRS absolute (zero point corrected) parallaxes and $Z$ are summarized in the third and eighth columns, respectively, of Table \ref{table:birs-results}. The tabulated parallax uncertainties are the reported Gaia EDR3 $\sigma_\varpi$ and do not include $\sigma_Z$.

Figure \ref{fig:Splot} shows the parallaxes of the BIRS (red), spectroscopic OB stars (blue), and APOGEE RGs (magenta) in order of increasing parallax (`S-plot'; left) and their corresponding distributions (right). The 1-$\sigma$ uncertainties do not include $\sigma_Z$. The dashed horizontal line denotes the W3(OH) methanol maser parallax of $0.512 \pm 0.010\,\mas$ \citep{Xu2006}. The dotted horizontal lines specify the parallax range $\pm 0.03\,\mas$ (Section \ref{subsec:parallax-systematics} used in our proper motion analysis (see Section \ref{subsec:propermotion}). We shifted the BIRS and OB S-plots horizontally such that they cross the W3(OH) maser parallax at approximately the same location along the x-axis.  

To avoid duplicated stars in this plot, we discarded BIRS that have spectroscopic cross matches (Section \ref{subsec:BIRS-OB stars-RGs}): five with spectroscopic OB stars (two are embedded stars with no parallax) and three with RGs (one with no parallax, the others background to W3). Five BIRS with parallaxes greater than $2\,\mas$ (BIRS 20, 90, 136, 105, and 104) are not shown for increased readability. We examine the proper motion of the foreground field star BIRS 20 in detail in Appendix \ref{subsec:BIRS20}. 

Qualitatively, based on the distribution of BIRS parallaxes being strongly inconsistent with that of the OB population, it appears that most of the BIRS are not embedded within the W3 molecular cloud. This is quantified further in Section \ref{sec:BIRS-membership}.

%%%%%%%%%%%%%%%%%%%%%%%%%%
%                                                                             %
%            SPECTRAL CLASSIFICATION                %
%                                                                             %
%%%%%%%%%%%%%%%%%%%%%%%%%%

\section{Initial Spectral Classification using IR Photometry and Gaia Parallaxes}
\label{sec:spectral-class}

We take advantage of modern broadband photometric data, particularly in the IR, and Gaia parallaxes to classify the BIRS using their placement in various familiar photometric-based diagrams.

While CMDs can be useful for spectral classification, the ambiguities introduced by not knowing the distances and the effects of dust extinction can cause significant challenges to their interpretation \citep{Majewski2011}. We complement our optical $i$ vs.\ {\ri} CMD in Figure \ref{fig:ivsri} with color-color diagrams (CCDs) that are neutral to distance, employing two NIR color combinations that separate stellar spectral classes.

Hertzsprung-Russell diagrams (\hrd s) are theoretically stellar loci of luminosity vs.\ effective temperature (\Teff). We make use of an observational NIR \hrd\ of absolute magnitude {\MKs} vs.\ intrinsic (de-reddened) color ${\JK}_0$ as a powerful tool for classifying stars. Although it is more challenging to construct, this color-magnitude combination maximizes the separation between the originally reddened OB and RG populations.

%%%%%%%%%%%%%%%%%%%%%
%                                                             %
%                 (J-H) VS (H-Ks)                   %
%                                                             %
%%%%%%%%%%%%%%%%%%%%%

\subsection{2MASS CCD}\label{subsec:2MASS-CCD}

\begin{figure}[t!]
\centering
\includegraphics[width=8.5cm]{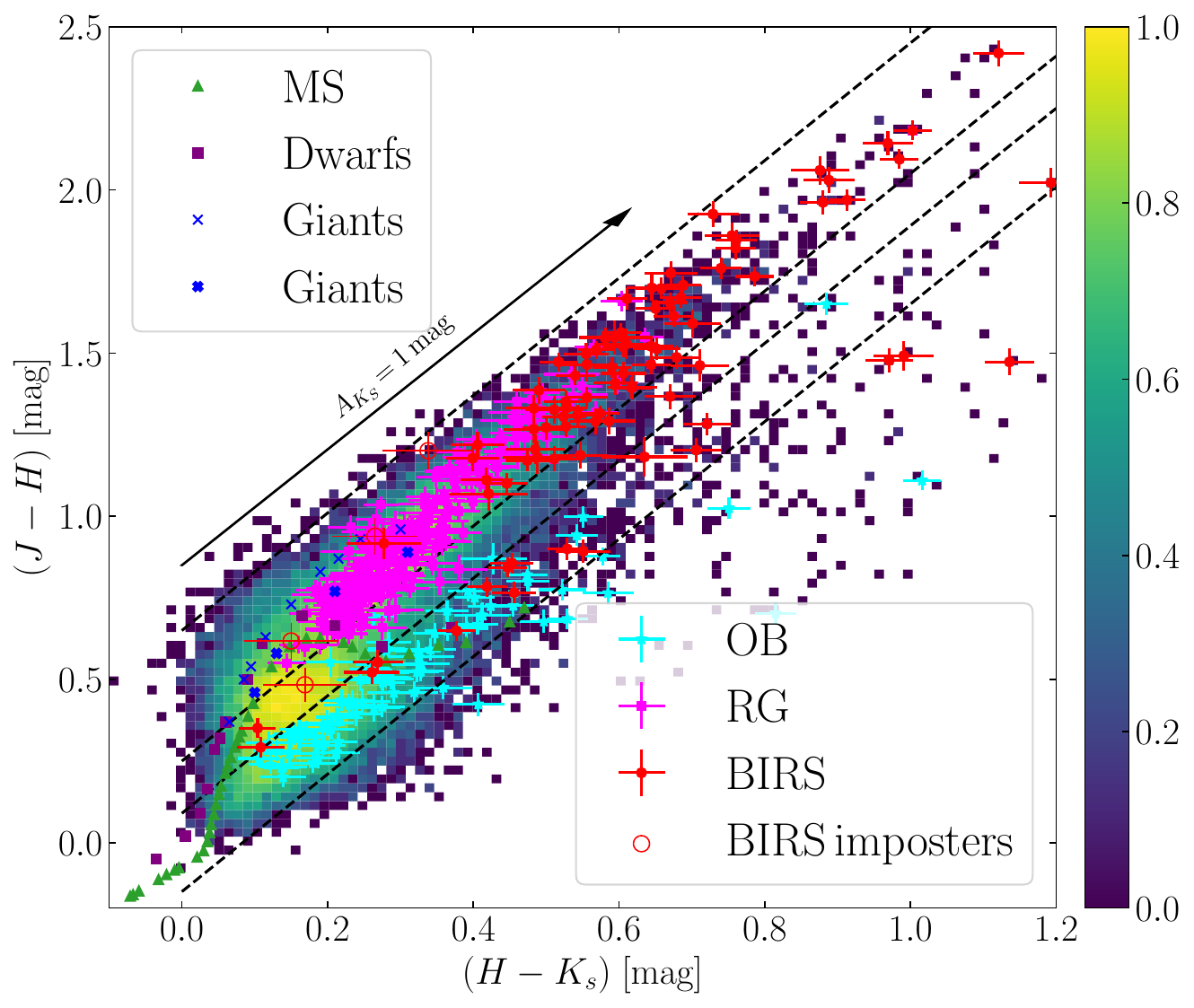}
\caption{2MASS CCD, as a 2D histogram of {\JH} vs.\ {\HK}, of stars toward W3. Relative color bar as in Figure \ref{fig:ivsri}. Pairs of loci of intrinsic colors are plotted for MS dwarfs and evolved giants (see text): dwarfs (09 V -- M9 V; green triangles) and giants (\Teff\ 5000, 4500,  4000, and 3700\,K; filled blue crosses), and in a slightly different filter system, dwarfs (B8 V -- M6 V; purple squares) and giants (G0 III -- M7 III; blue crosses). Overlaid are individual measurements of fiducial samples of reddened RGs (magenta squares) and OB stars (cyan stars), along with the BIRS (red circles) and BIRS imposter (open circles) that are to be classified. Reddened RG and OB \bands\ parallel to the black reddening vector at the top are demarcated by the top two and bottom two dashed lines, respectively (see text). Stars with an IRE are found below the reddened OB \band.}
\label{fig:jhhk}
\end{figure}

Figure \ref{fig:jhhk} shows a 2MASS CCD \citep{Jones1980} of our W3 field as a 2D histogram of {\JH} vs.\ {\HK}.

We indicate the locus of intrinsic colors of MS stars: 09 V to M9 V (green triangles; \citealp{Pecaut2013}).\footnote{By this citation throughout, we are referring to the online updated Version 2022.04.16: \url{https://www.pas.rochester.edu/~emamajek/EEM_dwarf_UBVIJHK_colors_Teff.txt} \label{foot:EEM}} For comparison we show the locus B8 V to M6 V (purple squares; \citealp{Bessell1988}), albeit on a slightly different filter system \citep{bessel2005}. The other important locus is for G0 to M7 giants (blue crosses; \citealp{Bessell1988}). Recent results by \citet{Jian2017} on the 2MASS system are close to this (filled blue crosses for \Teff\ 5000, 4500,  4000, and 3700\,K).

Overlaid on the 2D histogram are individual measurements for three fiducial samples of stars, the RGs from APOGEE (magenta squares), known OB stars (cyan stars), and the BIRS (red circles) that are to be further classified (the BIRS imposters are shown with open circles). The \band\ of reddened RGs is reasonably well demarcated by the color criteria $\JH > 1.8\, \HK + 0.65$ and $+0.25$ (top two dashed lines) and the \band\ of reddened OBs by the color criteria $\JH > 1.8\, \HK + 0.09$ and $-0.15$ (bottom two dashed diagonal lines). The slope (1.8) of these chosen boundaries is close to the slope (1.77) of the reddening line shown, which is for the extinction curves with $\powerlaw=1.8$ (Section \ref{sec:extinction-curve}); the slope (1.69) for $\powerlaw=1.61$ is too shallow.

We find a small number of OBs and BIRS below the reddened OB \band; we verify that this is caused by the presence of an IR excess (IRE) using their SEDs in Section \ref{sec:SEDs}.

%%%%%%%%%%%%%%%%%%%%%%%%%%%%%
%                                                                                       %
%         COMMENTS ON BIRS CLASSIFICATION           %
%                                                                                       %
%%%%%%%%%%%%%%%%%%%%%%%%%%%%%

\subsubsection{Comments on the BIRS classification}
\label{subsec:birs2massccd}

Enforcing photometric quality criteria (Appendix \ref{app:qualityfilter}), there are 92 BIRS (of a possible 133; 69\%) included on our 2MASS CCD. Of these, we find 69 (75\%) on the reddened RG \band\ and only 13 (14\%) on the reddened OB \band. While the intrinsic late-type MS colors overlap with the reddened OB star \band, late-type MS stars are significantly less luminous than their early-type counterparts and are less likely to be observed at high degrees of reddening. Therefore, the BIRS that are found far along the OB reddening vector are more likely to be luminous early-type MS stars. Note, however, the high proper motion star BIRS 20 that has observed 2MASS colors of $\JH=0.65\pm0.03\,\mags$ and $\HK=0.37\pm0.03\,\mags$, consistent with a very cool and unreddened M6 V star. We refine the spectral classification of BIRS 20 in Appendix \ref{subsec:BIRS20}.

Seven BIRS (8\%) are found in an ambiguous region between the reddened MS and RG \bands\ and three (3\%) have significant IREs. Of the four BIRS imposters on this CCD, two (BIRS 34 and 36) have IR colors that are significantly less red than the APOGEE RGs, BIRS 36 being located in the ambiguous region between the OB stars and RGs. While the remaining two BIRS imposters, BIRS 40 and 100, are found approximately one-third and halfway along the reddened APOGEE RG \band, respectively, they are among the BIRS with the smallest 2MASS colors. This is consistent with the BIRS imposters being poor tracers of dust extinction. 

%%%%%%%%%%%%%%%%%%%%
%                                                          %
%                       RJCE jk                      %
%                                                          %
%%%%%%%%%%%%%%%%%%%%

\subsection{Combined 2MASS -- GLIMPSE CCD}\label{subsec:RJCEjk}

\begin{figure}[t!]
\centering
\includegraphics[width=8.5cm]{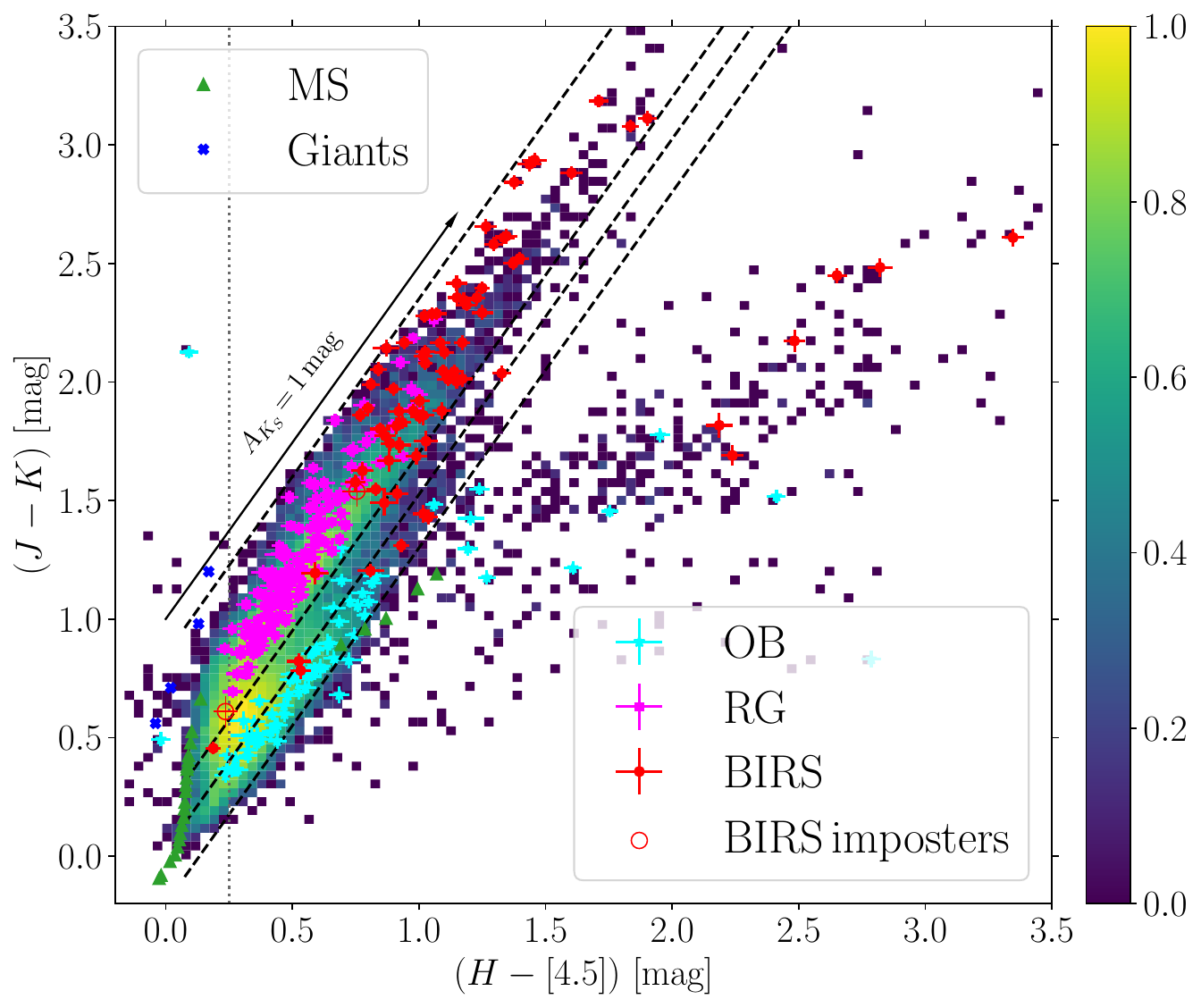}
\caption{Similar to Figure \ref{fig:jhhk}, but for combined 2MASS-GLIMPSE colors {\JK} vs.\ {\HG}. Approximate loci of intrinsic colors are identified (see text): MS (B5 V -- M9 V; green triangles) and giants (\Teff\ of 5000, 4500, 4000, and 3700\,K; filled blue crosses). The color scheme of overlaid measurements of fiducial stars is the same as Figure \ref{fig:jhhk}. The \bands\ with slope 1.5 are close to parallel to the reddening vector shown (see text).}
\label{fig:RJCE_JK-HG2}
\end{figure}

Figure \ref{fig:RJCE_JK-HG2} shows an alternative 2MASS-GLIMPSE CCD of our W3 field as a 2D histogram of \JK\ vs.\ \HG, using the same color scheme as Figure \ref{fig:jhhk}. 

The x-axis is the color \HG\ associated with the Rayleigh-Jeans Color Excess (RJCE) method \citep{Majewski2011}. We used the GLIMPSE-360 survey 4.5\, \micron\ magnitudes \citep{Whitney2008} instead of the WISE magnitudes because of the higher angular resolution at the same frequency and matched them to the 2MASS data using a $0{\farcs}4$ cross-match radius. On the y-axis we used \JK\ rather than \JH\ to bring in the color used in the \hrd\ below in Section \ref{subsec:HRD}.
Compared to Figure \ref{fig:jhhk}, the range of the x-axis is $\sim$\,3 times greater and of the y-axis is $\sim$\,1.5 times greater.

We plot loci of intrinsic colors: B3 V to M9 V MS stars (green triangles; \citealp{Pecaut2013}) and giants (filled crosses for \Teff\ of 5000, 4500, 4000, and 3700\,K; \citealp{Jian2017}). Overlaid on the 2D histogram are individual measurements for the three important samples of stars as in Figure \ref{fig:jhhk}. The \band\ of reddened RGs is reasonably well demarcated by the color criteria $\JK>1.5\,\HG+0.85$ and $+0.2$ (top two dashed diagonal lines) and the \band\ of reddened OBs by the color criteria $\JK>1.5\,\HG+0.025$ and $-0.25$ (bottom two dashed diagonal lines). The slope (1.5) of these chosen boundaries is equal to the slope of the reddening line shown, which is for the extinction curves with $\powerlaw=1.8$ (Section \ref{sec:extinction-curve}); the slope (1.34) for $\powerlaw=1.61$ is again too shallow.

The separation between the reddened OB and RG bands found here and in Figure \ref{fig:jhhk} enables a confident classification of the BIRS and underlying population.

The underlying populations in the OB band on both the 2MASS and 2MASS-GLIMPSE CCDs suggest that there are more OB stars in W3 than have been spectroscopically confirmed to date (Section \ref{subsec:spectroscopy}). While these spectroscopic surveys concentrate coverage in the HDL, our photometric search for new OB star candidates in Section \ref{sec:newOBcandidates} significantly extends the spatial coverage to the entire W3 cloud and the western edge of the adjacent W4 GMC.

%%%%%%%%%%%%%%%%%%%%%%%%%%%%%
%                                                                                       %
%         COMMENTS ON BIRS CLASSIFICATION           %
%                                                                                       %
%%%%%%%%%%%%%%%%%%%%%%%%%%%%%

\subsubsection{Comments on the BIRS classification}
\label{subsec:birs2massglimpseccd}

Of the 80 BIRS (60\%) on our 2MASS-GLIMPSE CCD, we found 63 (79\%) on the reddened RG \band\ and only five (6\%) on the reddened OB \band. Five (6\%) are found to be borderline in the region between the reddened MS and RG \bands. We again found seven BIRS (9\%) below the reddened MS \band\ that, along with the OB stars found there, show evidence of a \mire\ in their SEDs. The BIRS that have `\uns' SEDs (see right panel of Figure \ref{fig:BIRS-SED-examples} below) and those with the most significant IR excesses are found toward the far right of this CCD.

Two spectroscopic OB stars are found to have erroneous IR colors on this CCD: \citetalias{Kiminki2015} 64\footnote{Interestingly, this star is found within the reddened MS band of the 2MASS CCD, having 2MASS colors that are consistent with OB stars.} and \citetalias{Kiminki2015} 86.\footnote{This star is also found within the reddened MS band of the 2MASS CCD, having a sensible NIR-to-optical SED.} \citetalias{Kiminki2015} 64 is well above the reddened RG band with a very low value of $\HG=0.09\pm0.04$ and $\JK=2.13\pm0.03$. This star has a poorly behaved SED that is clearly affected by systematic errors, where overlapping surveys do not have consistent photometry. It is noted that \citetalias{Kiminki2015} 64 has \halpha\ and Paschen emission \citep{Kiminki2015}, but this cannot explain the observed SED.  \citetalias{Kiminki2015} 86 is located at the lower left end of the reddened RG band with $\HG=-0.02\pm0.04$ and $\JK=0.49\pm0.03$. While the GLIMPSE, WISE, and Spitzer/IRAC photometric data for this star are all consistent, they have a dimmer asymptotic behavior than expected from 2MASS, which gives rise to an erroneous \HG\ color. 

%%%%%%%%%%%%%%%%%%%%%%
%                                                                 %
%                     HR DIAGRAM                     %
%                                                                 %
%%%%%%%%%%%%%%%%%%%%%%

\subsection{HR Diagram}\label{subsec:HRD}

%%%%%%%%%%%%%%%%%%%%%%%%%%
%                                                                             %
%                OBSERVED HR DIAGRAM                 %
%                                                                             %
%%%%%%%%%%%%%%%%%%%%%%%%%%

\subsubsection{Observational HR Diagram}\label{subsubsec:HRD-obs}

We constructed an observational \hrd\, \MKs\ vs.\ $\JK_0$, using our combined NIR-MIR photometry and Gaia parallaxes, with a cross-match radius of $0{\farcs}2$ between the Gaia and combined 2MASS-GLIMPSE catalogs. Color criteria $\JK>1.5\,\HG-0.15\,\mags$ and $\HG>0.25\,\mags$ in our 2MASS-GLIMSPSE CCD (Section \ref{subsec:RJCEjk}) removed not-photospheric stars with \ire s. For Gaia we required ${\tt parallax\_error}<0.15\,\mas$. We computed $\JK_0$ and $\AKs$ using our modified RJCE technique (Appendix \ref{subsec:RJCEex}). From the apparent magnitude \Ks\ we obtained the absolute magnitude \MKs\ by 

\begin{equation}
    \MKs = \Ks - \AKs - \dm \, ,
\label{eq:mks}
\end{equation}

{\noindent}where the distance modulus {\dm} expressed in terms of parallax rather than distance is given by

\begin{equation}\label{eq:dmod}
    \dm = 5 \log_{10}\left\{ \frac{100}{(\varpi -Z)\, [{\mathrm {mas}}]} \right\} \, \mags \, .
\end{equation}

\begin{figure}[t!]
\centering
\includegraphics[width=8.5cm]{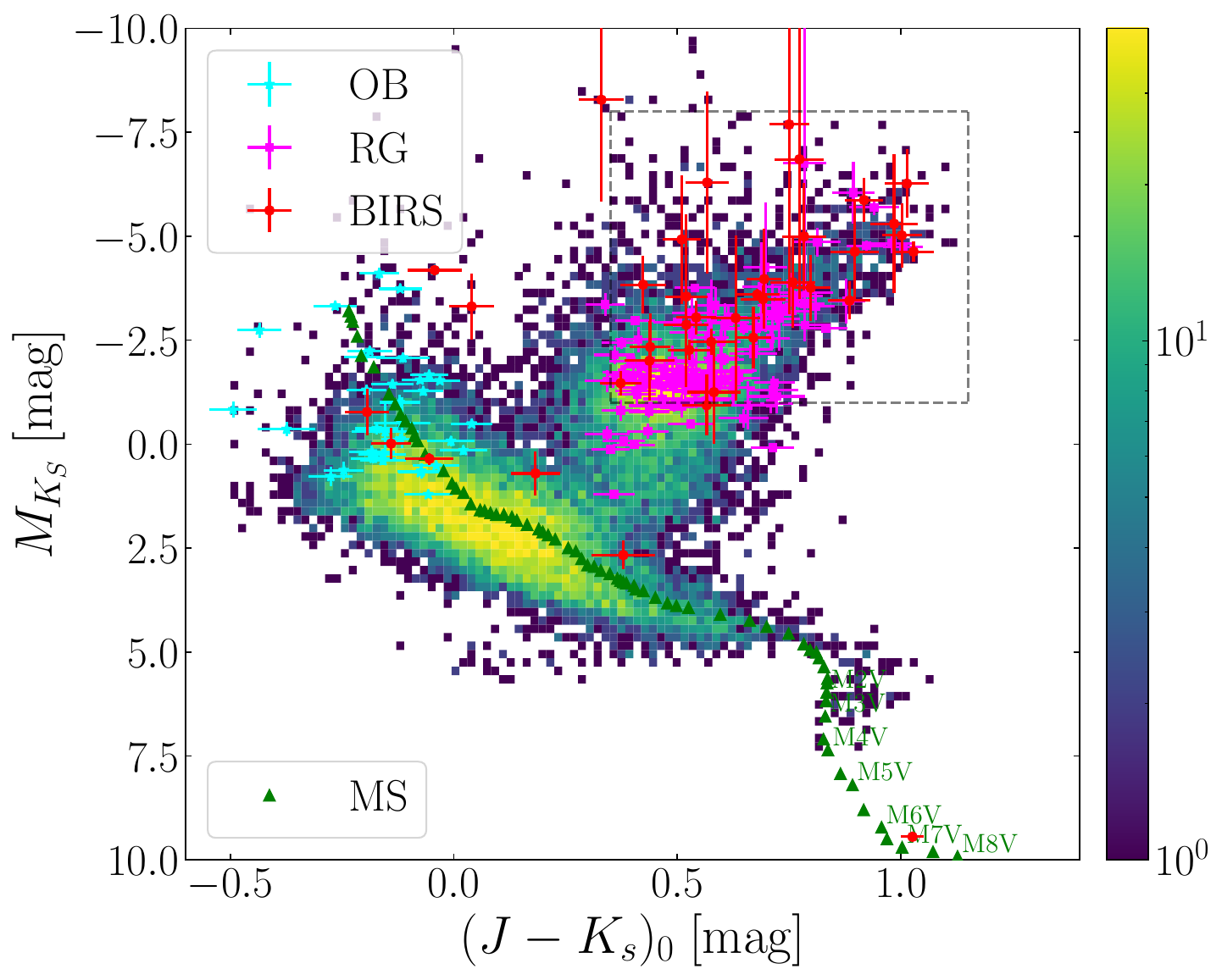}
\caption{2MASS \hrd, as a 2D histogram of \MKs\ vs.\ $\JK_0$, for stars toward W3. Note the logarithmic color bar for the number of stars given the sample and pixelization. The intrinsic MS locus (09 V -- M9 V; green triangles) and the region selecting for RGs (dashed rectangle; see Section \ref{subsec:stellarprobes}) are identified (see text). As in Figure \ref{fig:jhhk}, individual stars from the two fiducial samples, OB stars (cyan stars) and RGs (magenta squares), and the BIRS (red circles), are overlaid.}
\label{fig:hrd-obs}
\end{figure}

\begin{figure}[t!]
\centering
\includegraphics[width=8.75cm]{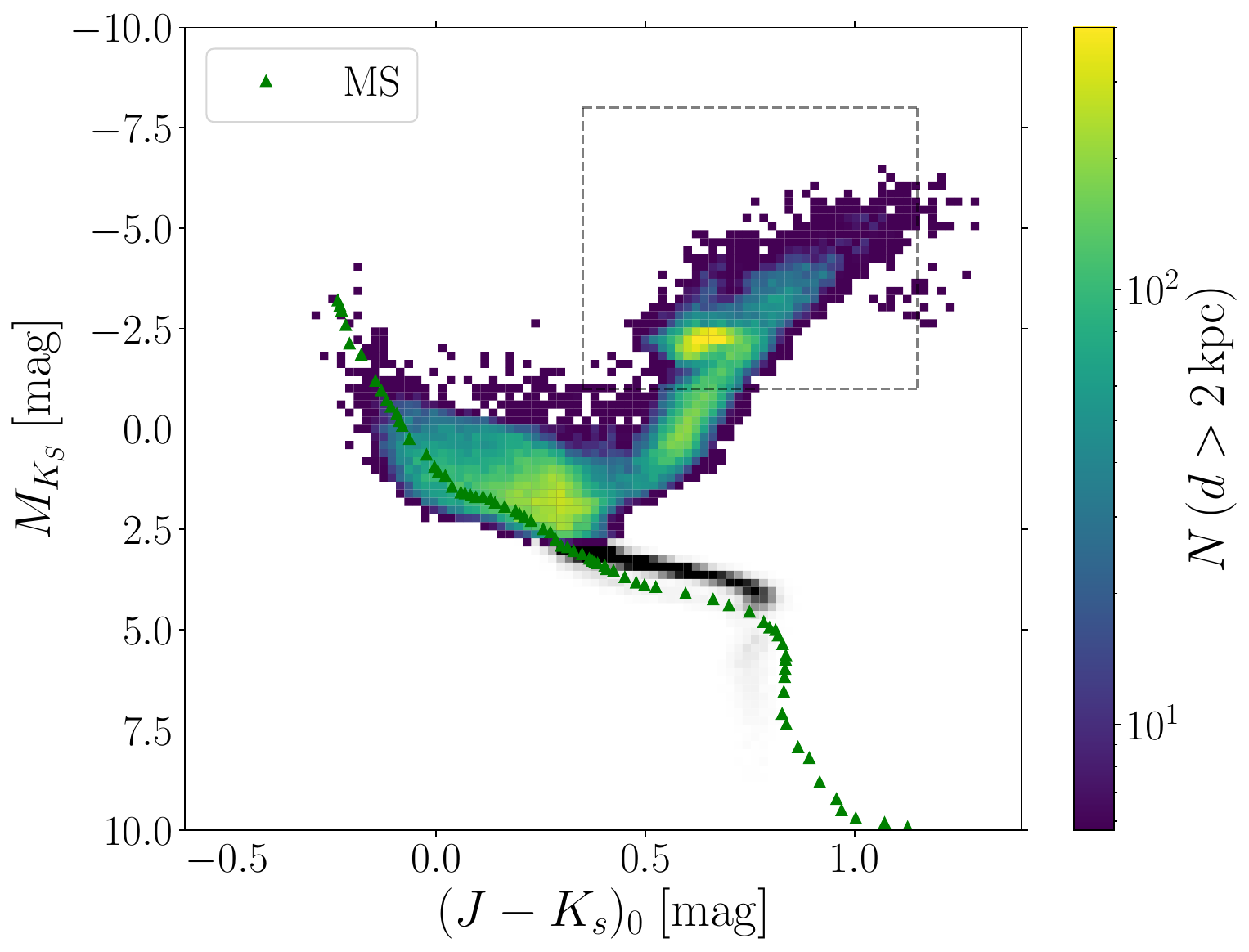}
\includegraphics[width=8.75cm]{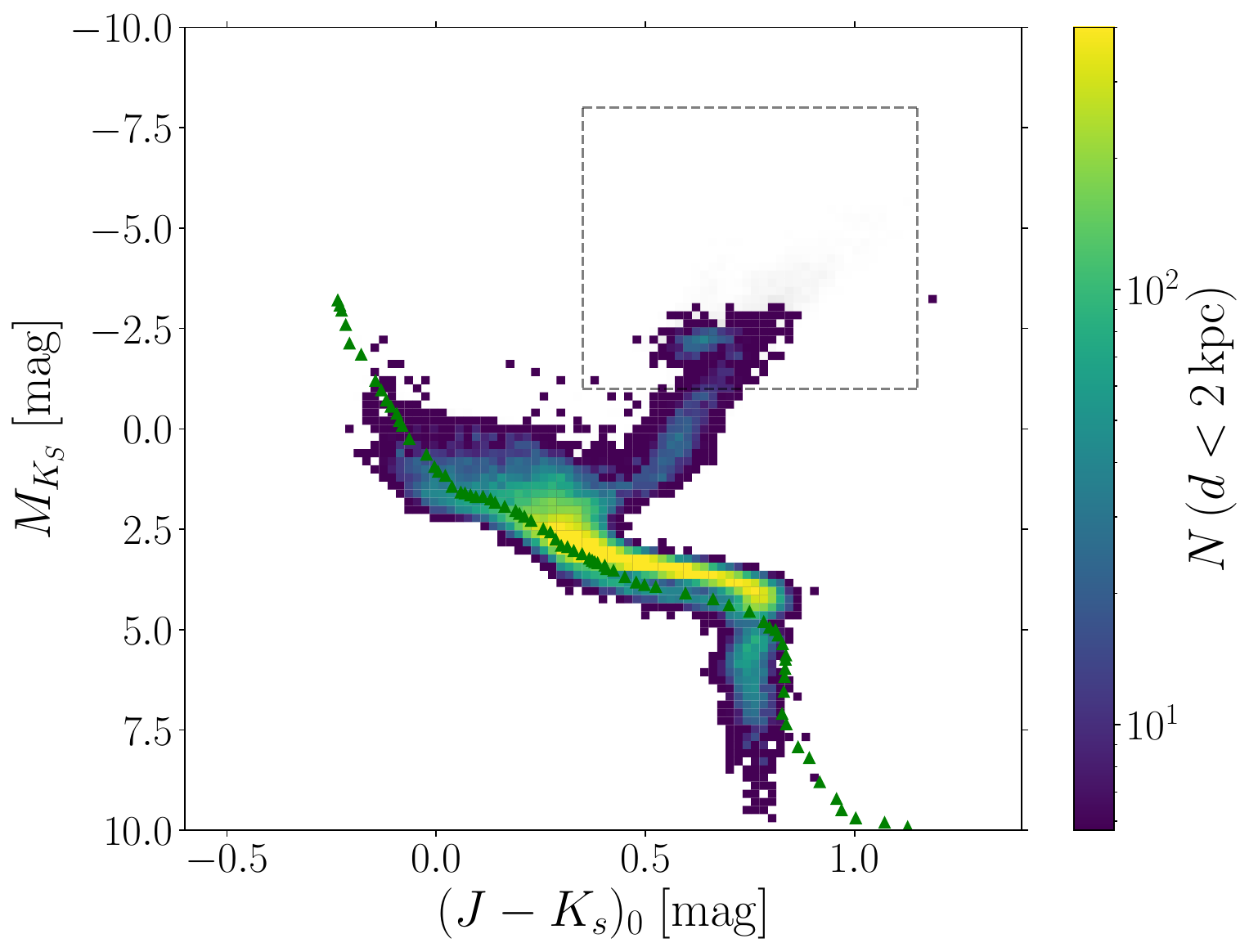}
\caption{HRDs like Figure \ref{fig:hrd-obs}, but for stars toward W3 using unreddened data from the Besan\c{c}on Galactic model (see text). Top: stars at a distance $d>2\,\kpc$. Bottom: $d<2\,\kpc$. The intrinsic MS (09V-M9V) locus (green triangles) and color-magnitude selection for RGs (dashed rectangle) are the same as those in Figure \ref{fig:hrd-obs}. The dynamic ranges of the 2D histograms are the same, top and bottom, and are adjusted to approximate the colors in Figure \ref{fig:hrd-obs}.}
\label{fig:hrd-besancon}
\end{figure}

The \hrd\ is displayed as a 2D histogram in Figure \ref{fig:hrd-obs}. The MS locus (09 V -- M9.5 V) is indicated with green triangles \citep{Pecaut2013}. As in Figure \ref{fig:jhhk}, individual stars from the two fiducial samples and the BIRS are overlaid. Despite the simple approximation made in the extinction correction, the field stars and fiducial groups are in their expected positions (see Section \ref{subsubsec:HRD-BGM}) and are well separated, so that classifying the BIRS is straightforward. 

Only four BIRS are consistent with OB stars (BIRS 6, 15, 37, and 62),  while 31 are consistent with RGs (BIRS 2, 11, 12, 19, 38, 49, 50, 54, 59, 63, 65, 66, 67, 69, 70, 75, 80, 86, 91, 94, 108, 111, 116, 123, 125, 126, 128, 129, 130, 131, 134). This observed \hrd\ corroborates what we concluded using IR CCDs: the vast majority of the BIRS are RGs, populating the APOGEE red clump (RC) and RG branch (RGB) portions of the \hrd. 

Massive protostars with high accretion rates can swell up to such large radii that they appear like red supergiants on an \hrd\ \citep{Morino1998, Linz2009}; however, we found no cross matches of the BIRS with protostars in 2MASS-Spitzer catalogs of the young stellar content in W3 \citep{RI2011}.

At the bottom right of the \hrd\ are cool M dwarfs that are well in the foreground of W3 and of minor interest to this paper. The formulaic extinction correction does not work well for these stars because they are so intrinsically red, reaching $\HG_0 >  0.6\,\mags$. Because the IR extinction is small for nearby stars like these, we simply made no extinction correction for stars with $\varpi > 3\,\mas$ (distance $< 0.333$\,kpc). One such star is the high proper motion star BIRS 20 (Appendix \ref{subsec:BIRS20}) that has 2MASS CCD colors consistent with M6 V (Section \ref{subsec:2MASS-CCD}) and a placement on this \hrd\  
at $[\MKs, \JK_0] = [9.44, 1.03]$\,\mags, consistent with M6--7 V dwarfs.

The dashed rectangle defined by 0.35 $<\JK_0 <$ 1.15\,\mags\ and $-8 < \MKs < - 1\, \mags$ selects RGs used for a correlation analysis of stellar reddening and thermal dust emission optical depth (see Section \ref{subsec:stellarprobes}). 

%%%%%%%%%%%%%%%%%%%%%%%%%%
%                                                                             %
%                BESANCON HR DIAGRAM                 %
%                                                                             %
%%%%%%%%%%%%%%%%%%%%%%%%%%

\subsubsection{Insight from the Besan\c{c}on Galactic Model of Stellar Populations}
\label{subsubsec:HRD-BGM}

We compared our observational \hrd\ to that of the Besan\c{c}on Galactic model (BGM; \citealp{Robin2003}) of stellar populations. We selected stars centered on W3 at $(\ell,b) = (133.9\deg, +1.1\deg)$ including a $6\deg$ range in longitude and $2\deg$ range in latitude. The model does not reproduce individual cloud structures and would likely underestimate dust reddening in sightlines toward W3 and so we used the unreddened BGM data. We applied cuts to the $H$- and {\Ks} apparent magnitudes using the 2MASS sensitivity levels ($H<15\,\mags$, $\Ks < 14.3\,\mags$) and saturation levels ($H > 8.5\,\mags$, $\Ks > 8\,\mags$) such that the modeled data resembles our observational data.  But because of the effect of extinction in our original data, the BGM would have a relative overpopulation of distant stars at the sensitivity limits, e.g., lower MS stars.

Figure \ref{fig:hrd-besancon} shows two BGM {\MKs} vs.\ $\JK_0$ \hrd s split by distance $d$ into `distant' and `nearby' populations: $d>2$\,\kpc\ (top) and $d<2$\,\kpc\ (bottom), respectively. The intrinsic MS colors (09V-M9V; green triangles; \citealp{Pecaut2013}), RG selection dashed rectangle, and dynamic range of the colors in the 2D histograms are the same as those shown in Figure \ref{fig:hrd-obs}, each revealing aspects of consistency with our observational \hrd. Some differences in detail could be due to our extinction correction and others could be due to the underlying absolute magnitudes, colors, and populations included in the BGM model and no filtering for the effects of extinction. This is beyond the scope of this investigation.

In Figure \ref{fig:hrd-besancon}, nearby stars occupy predominantly the MS band, whereas stars that are significantly more distant populate the upper MS, turnoff, and RC and RGB star regions. This provides a consistent basis for distant RG populations being effective tracers of the entire line-of-sight dust column through W3, which we take advantage of in Section \ref{subsec:stellarprobes} where we compare stellar reddening to thermal dust emission optical depth. 

%%%%%%%%%%%%%%%%%%%%%%%%
%                                                                       %
%                BIRS CLASSIFICATION                %
%                                                                       %
%%%%%%%%%%%%%%%%%%%%%%%%

\subsection{Generic BIRS Classification}\label{subsec:generic-BIRS-class}

The superscripts in the last column of Table \ref{table:birs-results} describes information on the method of BIRS spectral classification, much of which is established through the diagrams in this section. We began, however, by identifying the generic spectral classification of the BIRS by trying to find cross matches in catalogs of spectroscopically-confirmed OB stars \citep{Bik2012, Kiminki2015, Navarete2019}, APOGEE RGs \citep{Zasowski2013}, variable objects \citep{Heinze2018, Gaia2019}, and young stellar objects (YSOs; \citealp{RI2011}); there were no cross matches in the latter.  These are noted in column 2.

Next, we used the placement of individual BIRS in various photometric-based diagrams and compared them to reddened \bands\ of OB stars and RGs, in the following order of superscripts in Table \ref{table:birs-results}: (2) observational \hrd\ (Figure \ref{fig:hrd-obs}), (3) 2MASS-GLIMPSE \JK\ vs.\ \HG\ CCD (Figure \ref{fig:RJCE_JK-HG2}), (4) 2MASS \JH\ vs.\ \HK\ CCD (Figure \ref{fig:jhhk}), and (5) ATLAS $i$ vs.\ \ri\ (Figure \ref{fig:ivsri}). 

Of the 133 possible BIRS, we were able to assign a generic photometric-based classification (i.e., MS or RG) to 98 (74\%). Of these, we found that 68 (69\%) are intrinsically-red RGs with only twelve (12\%) OB stars and 18 (18\%) low-mass MS stars (including BIRS 20). Of the 17 that we previously identified as `imposters,' 13 (76\%) are identified as MS stars.

For completeness, we note that (1) we fit BIRS SEDs with dust extinction curves and assign spectral types in Section \ref{subsec:refinedspec-BIRS}.

%%%%%%%%%%%%%%%%%%%%
%                                                          %
%                       KR 140                       %
%                                                          %
%%%%%%%%%%%%%%%%%%%%

\subsubsection{BIRS in the direction of KR 140}
\label{sec:kr140}

Few studies in the literature have discussed the nature of the BIRS since the discussion in \citet{Elmegreen1980}. \citet{Ballantyne2000} mapped the radio emission of the KR 140 \HII\ region at 1420 and 408 MHz using the Dominion Radio Astrophysical Observatory Synthesis Telescope \citep{Roger1973, Veidt1985}, covering BIRS 128-132. They noted a possible overabundance of BIRS toward KR 140 but also that it is difficult to determine if they are physically associated with the \HII\ region. \citet{Kerton2001} mapped the dust continuum emission toward KR 140 using the James Clerk Maxwell Telescope at 450\,\micron\ and 850\,\micron, again covering BIRS 128-132. They found no significant submillimeter or FIR emission associated with these BIRS, placing an estimated upper limit of $A_V \sim 1\,\mags$ on extinction by the associated material in KR 140. They proposed that these stars are likely intrinsically red, such as pre-MS stars or RGs, rather than early MS stars, and that they may be attenuated by background material behind KR 140. TripleSpec spectroscopy of BIRS 128-132 (Section \ref{subsec:spectroscopy}) and our SED fitting (Section \ref{subsec:refinedspec-BIRS}) confirmed that for BIRS 128-131 that is the case.

Using the methods above, we found that BIRS 128-131 are all RGs behind W3. Assuming an intrinsic color of $\HG_0 = 0.08\,\mags$ for RGs and an extinction curve satisfying $\AV/\AKs = 9.41$, we estimate 
$\AV$ values in the range 8.8 -- 14.9\,\mags toward these stars. If the visual extinction foreground to KR 140 is $\AV \sim 6\,\mags$ \citep{Kerton1999}, this corresponds to an extra extinction of $\AV \sim$ 3 -- 9\,\mags\ for these stars projected on the KR 140 region. While this is higher than the previous estimate, for distant stars dust that is background to KR 140 might contribute to this increased estimate of the extinction.

%%%%%%%%%%%%%%%%%%%%%%%%%%%%%%%%
%                                                                                                %
%       W3 CLOUD KINEMATICS AND BIRS MEMBERSHIP     %
%                                                                                                %
%%%%%%%%%%%%%%%%%%%%%%%%%%%%%%%%

\section{W3 Stellar Kinematics and Cloud Membership} \label{sec:BIRS-membership}

%%%%%%%%%%%%%%%%%%%%%%%%%%
%                                                                             %
%                  USEFUL BENCHMARKS                   %
%                                                                             %
%%%%%%%%%%%%%%%%%%%%%%%%%%

\subsection{Useful Benchmarks}
\label{subsec:ucalc}

The following equations quantify geometrical and dynamical quantities relating to the stellar parallax ($\varpi$) and proper motion ($\mu$) data available from Gaia.

The plane-of-sky width $s$ corresponding to a subtended angle $\theta$ is given by

\begin{align}\label{eq:arc}
    s = 17.45\, \left(\frac{0.5\, \mathrm{mas}}{\varpi}\right)\, \left(\frac{\theta}{0\fdg5}\right)\, \mathrm{pc} \, .
\end{align}

{\noindent}If the line-of-sight depth of the cloud is equal to the transverse width $s$, then the fractional parallax change from front to back of the cloud is equal to

\begin{align}\label{eq:fracpar}
    \frac{\varpi^\prime}{\varpi} =  \left(1 - 0.0087\, \frac{\theta}{0\fdg5}\right)\, ,
\end{align}

{\noindent}and the corresponding change in parallax from the front to back of the cloud is 

\begin{align}\label{eq:depthpar}
    \Delta\varpi = - 0.0044\, \left(\frac{\varpi}{0.5\, \mathrm{mas}}\right)\, \left(\frac{\theta}{0\fdg5}\right)\, \mathrm{mas}\, .
\end{align}

{\noindent}Clearly, $\Delta\varpi$ is small compared to the parallax and zero point uncertainties of individual measurements (Section \ref{subsec:parallax}).

A star moving with proper motion $\mu$ at parallax $\varpi$ has a transverse velocity of

\begin{align}\label{eq:vtrans}
    v_{\mathrm{trans}} = 9.5\, 
    \left(\frac{\mu}{1.0\, \mathrm{mas\, yr}^{-1}}\right) \left(\frac{0.5\, \mathrm{mas}}{\varpi}\right) \, \kms \, .
\end{align}

{\noindent}and moves a transverse distance in time $t$ of

\begin{align}\label{eq:strans}
    s_{\mathrm{trans}} = 9.7\, 
    \left(\frac{\mu}{1.0\, \mathrm{mas\, yr}^{-1}}\right) \left(\frac{0.5\, \mathrm{mas}}{\varpi}\right) \left(\frac{t}{1\, \mathrm{Myr}}\right)\, \mathrm{pc} \, ,
\end{align}

{\noindent}or an angle of

\begin{align}\label{eq:ttrans}
    \theta_{\mathrm{trans}} = 0.28\, 
    \left(\frac{\mu}{1.0\, \mathrm{mas\, yr}^{-1}}\right) \left(\frac{t}{1\, \mathrm{Myr}}\right)\, \mathrm{deg} \, .
\end{align}

If the radial motion of the star were comparable to its transverse motion, then the change in parallax would still be very small. 

Often the peculiar motion $\delta \mu$ of a star with respect to the local standard of rest of an ensemble of stars at that parallax would be more relevant, in which case $\delta \mu$ would be used in the above formulae.

%%%%%%%%%%%%%%%%%%%%%%%%%%
%                                                                              %
%                PARALLAX  EXPECTATIONS              %
%                                                                              %
%%%%%%%%%%%%%%%%%%%%%%%%%%

\subsection{W3 Parallax Based on OB Stars}
\label{subsec:parallax-systematics}

\subsubsection{Expectations}
\label{subsubsec:expect}

While the development of Gaia has advanced our ability to measure distances, instrumental effects are a limiting factor in parallax precision. We illustrate this with the following simple geometrical argument, concentrating on the HDL, a prominent cloud structure where the majority of known OB stars reside. Maser parallax measurements of W3(OH) in the HDL suggest a parallax of $\sim 0.5\,\mas$ \citep{Xu2006, Hachisuka2006}. 

If we make the common assumption that the line-of-sight cloud thickness is approximately equal to the plane-of-sky dimension of $\sim 0\fdg5$, this yields a line-of-sight cloud thickness of $\sim 17\,\pc$ or a range in parallax of only $\sim 4\,\uas$ (Section \ref{subsec:ucalc}). This of course might be too simplistic.  The HDL is a feature influenced by the ionizing stars to the east in IC 1805.  If it were a limb-brightened wall of a chimney, then it might be slightly larger along the line of sight; but the scale height of the medium could be as small as 25 pc \citep{basu1999}.

The uncertainty of the zero point correction on this scale, $\sigma_Z= 23\,\uas$, is quite large (Section \ref{subsec:parallax}) and so the corresponding 90 pc distance uncertainty at the parallax of W3 is much larger than the expected W3 depth range just discussed. This sets a practical limit on the precision with which one can measure the parallax (distance) of W3 and differentiate any small changes across the cloud due to cloud substructure.

In a recent study, \citet{Navarete2019} measured the parallaxes of three cloud substructures in W3 using Gaia DR2 data and concluded that W3 Main ($\varpi = 0.458 \pm 0.045\,\mas$), IC 1795 (which the authors refer to as the `W3 Cluster'; $\varpi = 0.471 \pm 0.040\,\mas$), and W3(OH) ($\varpi = 0.506 \pm 0.048\,\mas$) reside at different distances. The number of stellar parallaxes used is small and given the quoted uncertainties, these parallax measurements are not significantly different from one another. See also footnote \ref{foot:contam}. Taken at face value, this parallax range of $\sim 0.048\,\mas$ would correspond to a cloud thickness of 190\,pc (Section \ref{subsec:ucalc}), which is astoundingly large compared to the angular separation between these substructures. However, from our discussion above, $\sigma_Z$ in combination with $\sigma_\varpi$ frustrates such precise differentiation among the parallaxes of these substructures.

\subsubsection{Estimate of the W3 parallax}
\label{subsubsec:estimate}

The parallax distribution of spectroscopically-confirmed OB stars in W3 (Figure \ref{fig:Splot}) shows a considerable range, much of which reflects their discrete and heterogeneous measurement errors. We used a Bayes-informed iterative rectification technique \citep{Lucy1974} to reveal the underlying distribution. Beginning with a flat prior, a strong parallax peak emerged at $0.5\,\mas$ in only a few iterations. 

To refine the peak parallax, we concentrated on the 43 stars with the smallest parallax uncertainties ($<0.025\,\mas$), which we combined in quadrature with $\sigma_Z$. This combination removed minor secondary peaks in the distribution, demonstrating the importance of considering systematic parallax errors. These high-precision parallax stars were found primarily along the HDL with a few elsewhere in lower-density regions.

We fit the parallax peak of these stars with a Gaussian function, finding that the central parallax value was stable at $0.511\,\mas$ with an estimated uncertainty of $0.005\,\mas$ (or as high as $0.01\,\mas$ allowing for the systematics of the Gaussian approximations and using stars with less precise parallaxes). The dispersion of the Gaussian decreased to $\sim 0.03\,\mas$ at the fourth iteration. 

This central parallax is very close to the W3(OH) maser parallax measurements \citep{Xu2006, Hachisuka2006}. There is no evidence that the parallax peak from the 20 or so star subset in the W3 Main Complex (W3 Main, IC 1795, W3(OH); see the annotations in Figure \ref{fig:tau-map}) is any different than that of the remaining 23 stars in the field. It is difficult to be precise about the depth of the cloud because the dispersion of the Gaussian is still reflective of the parallax measurement errors. 

In the following, we adopt a narrow parallax range of $0.48<\varpi<0.54\,\mas$ for the entire W3 cloud. Recalling the discussion in Section \ref{subsubsec:expect}, this is a range in which measured parallaxes would be found because of measurement uncertainties, not a measure of depth.  We assess whether an individual star has a parallax consistent with this, taking into account its parallax error $\sigma_\varpi$ in quadrature with $\sigma_Z$.

Looking at the full list of OB stars (mostly B), there is evidence that some have much lower parallaxes than the central value, which we confirmed in Sections \ref{subsec:fieldOBs} and \ref{subsec:refinedspec-OB}. 

%%%%%%%%%%%%%%%%%%%%%
%                                                             %
%                       PARALLAX                    %
%                                                             %
%%%%%%%%%%%%%%%%%%%%%

\subsubsection{BIRS Parallax-Based Cloud Membership}\label{subsec:parallax-membership}

The BIRS are considerably fainter than the spectroscopically-confirmed OB stars and have significantly larger parallax uncertainties, and so we required that ${\tt parallax\_error} < 0.2\,\mas$ before using their parallaxes to assign cloud membership. 

Of the 77 BIRS with high-enough quality parallax measurements, we found that 30 (39\%) have parallaxes that are consistent with being at the distance of W3 (BIRS 2, 6, 10, 11, 15, 23, 25, 28, 30, 31, 33, 35, 37, 40, 43, 45, 51, 54, 60, 61, 62, 66, 86, 91, 92, 93, 98, 107, 116, 125). These are candidate members of W3, pending further characterization below. 

Another 36 (47\%) were found to be background to the cloud (BIRS 1, 12, 19, 38, 47, 48, 49, 50, 53, 55, 56, 58, 59, 63, 65, 67, 68, 69, 70, 71, 75, 80, 84, 94, 108, 111, 123, 126, 127, 128, 129, 130, 131, 132, 134, 135). The volume sampled in a survey of a finite angular region is greater at larger distances, increasing the probability of finding more distant stars, so long as sensitivty limits are not reached. Furthermore, the $(R-I)$ color selection criterion of the BIRS also imposes an observational bias toward larger distances because of increased extinction. In Section \ref{subsec:stellarprobes}, we verified our identification of background BIRS by comparing their $E{\HG}$ vs.\ $\tauu$ correlation to that of distant RGs. 

Eleven BIRS (14\%) were found to be foreground to the cloud (BIRS 20, 29, 32, 34, 36, 42, 44, 88, 90, 124, 136). 

Of the four BIRS in the KR 140 region, only one has a parallax consistent with W3. We discuss KR 140 further in Section \ref{subsec:VES}. 
Of the eight BIRS in W4 (BIRS 111-118), only one (BIRS 116) has a parallax that is consistent with that of W3, and none have proper motions that are consistent with W3. 

For the BIRS, our parallax-based possible cloud membership by sub-region (defined in Section \ref{subsubsec:subr}) is encoded in a three-digit number in the ninth column of Table \ref{table:birs-results}.
The first digit corresponds to projection on a cloud sub-region: (1) the W3 Main Complex, (2) the AFGL 333 Ridge or (3) the Field, covered by neither (1) nor (2). The second digit specifies our parallax-based possible cloud membership: a star satisfying the W3 parallax criterion is assigned the same value as the first digit and a zero if it does not (missing astrometry or a parallax uncertainty $> 0.2\,\mas$ is assigned a dash). The third digit specifies our proper-motion-based cloud possible membership, as determined in the following section.

\begin{figure*}[t!]
\centering
\includegraphics[height=8cm]{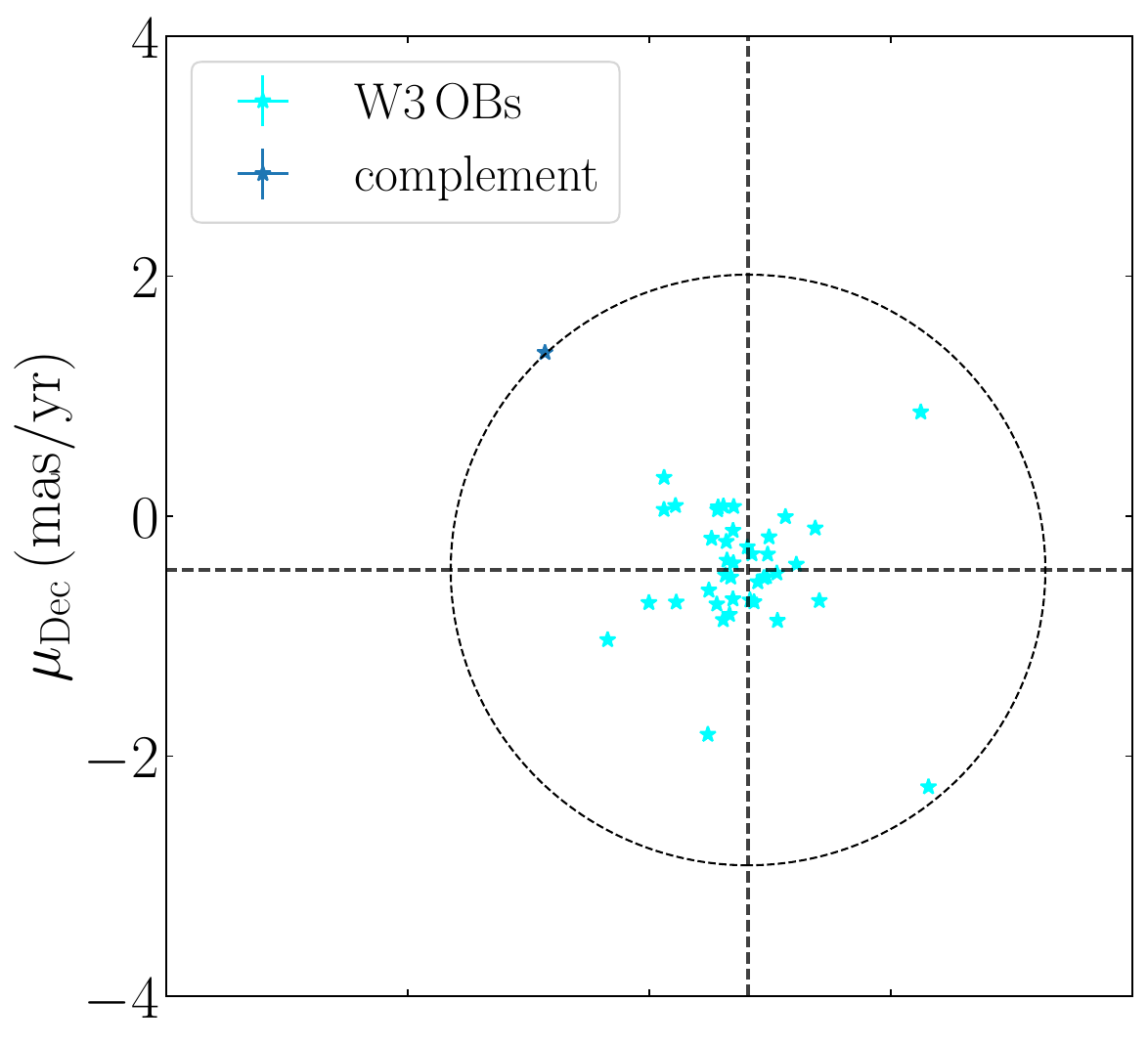}%\hspace*{-3pt}
\includegraphics[height=8cm]{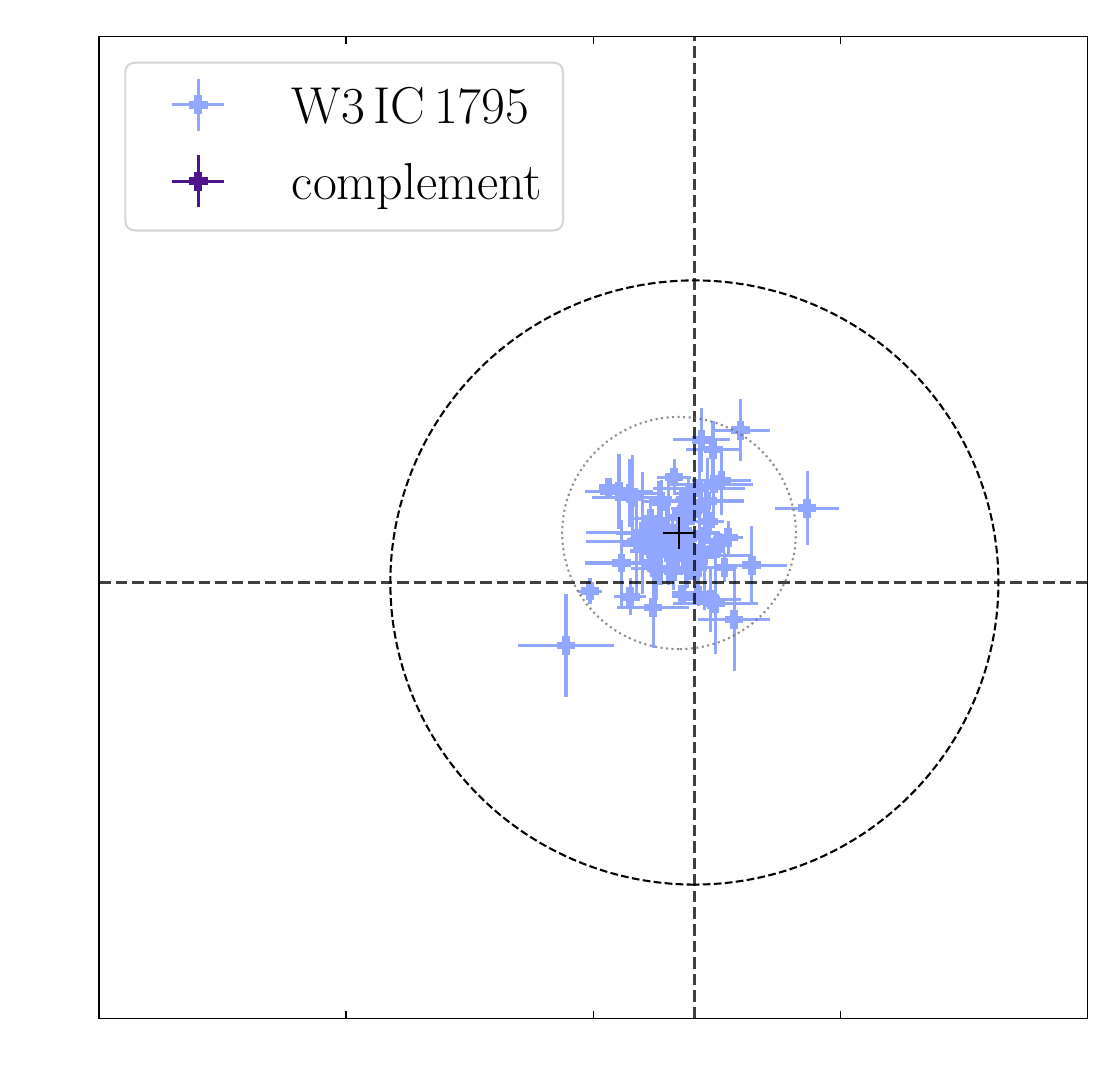}\\
\hspace{15pt}\includegraphics[height=8.7cm]{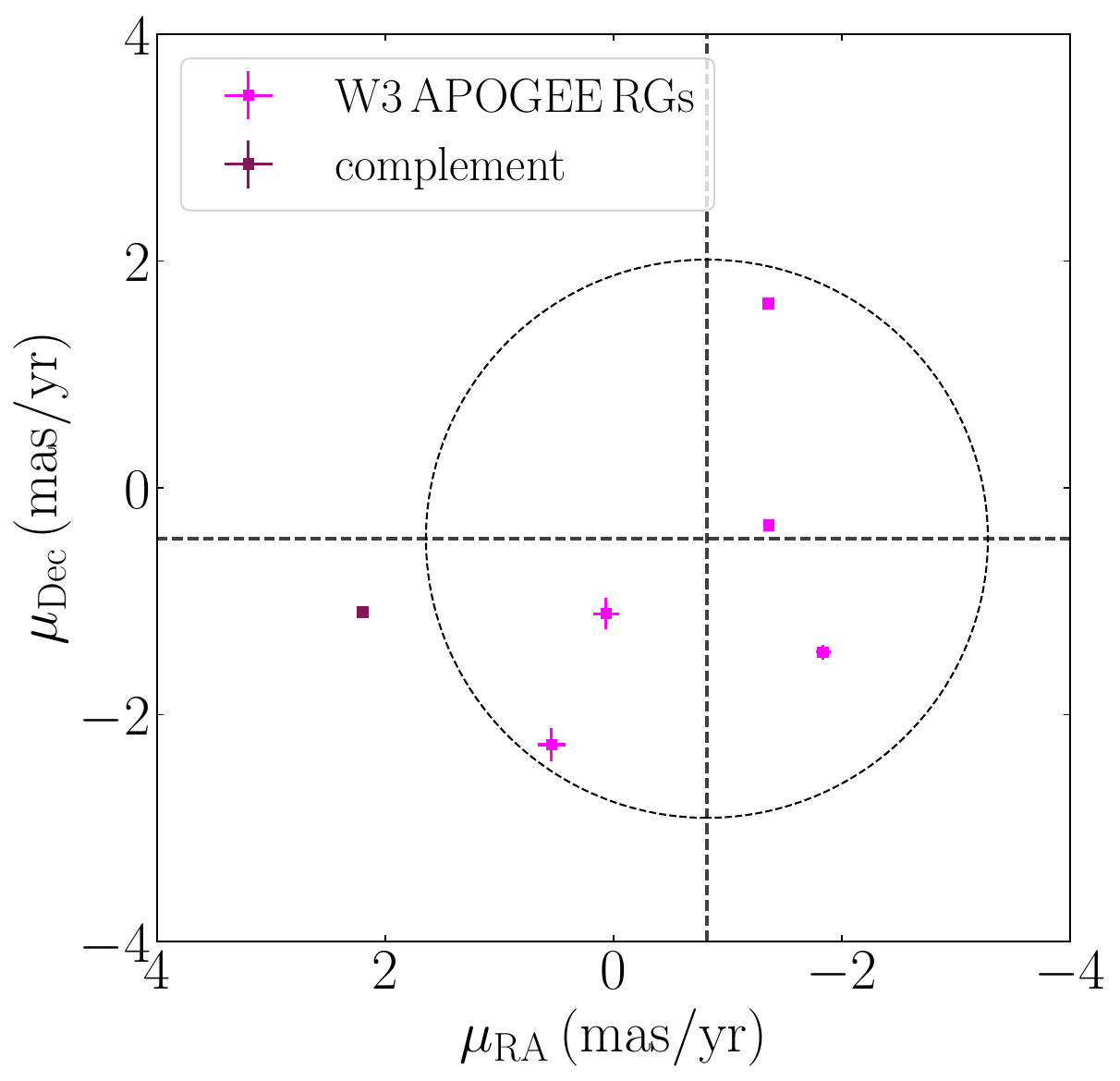}%\hspace*{-7pt}
\includegraphics[height=8.7cm]{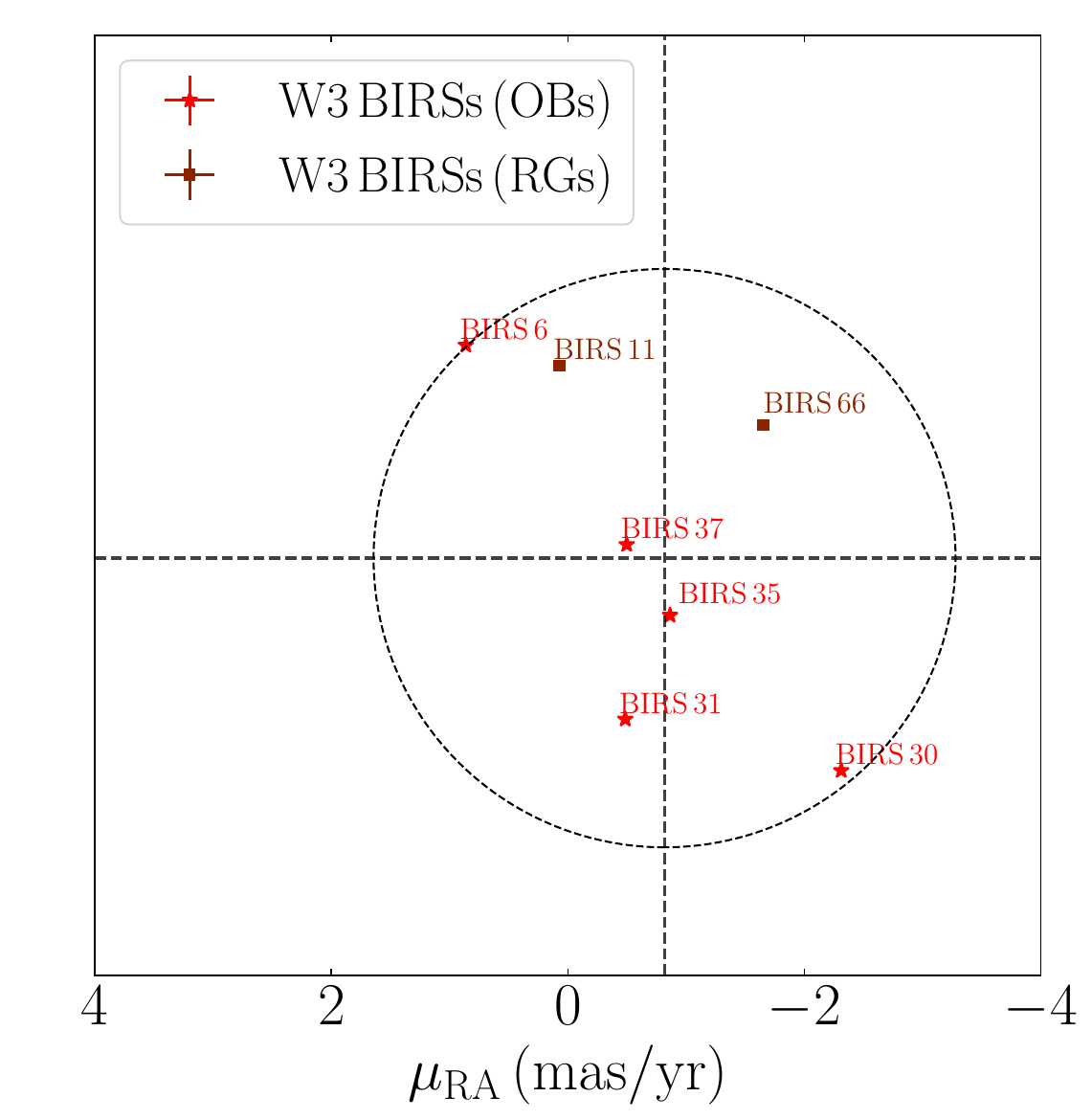}

\caption{Gaia proper motion in Declination ({\pmdec}) vs.\ Right Ascension ({\pmra}) for samples of stars seen projected on the W3 molecular cloud and with parallaxes consistent with the derived W3 parallax range (Section \ref{subsubsec:estimate}): OB stars (top left), IC 1795 cluster members (top right), APOGEE RGs (bottom left), and high-accuracy BIRS (bottom right). Note that {\pmra} increases to the left, similar to Equatorial coordinates. The common dashed vertical and horizontal lines indicate the weighted averages {\avgpmra} and {\avgpmdec}, respectively, of the spectroscopically-confirmed OB population. The common dashed circle represents the OB `proper motion centroid' 
as a $3\sigma$ radius around the average proper motion (see text), 
within which we define stars to be kinematically associated with W3. Stars found outside of this circle (the complement region) are shown using a darker color. In the top right, the plus sign and dotted circle indicate the weighted average proper motion and its $3\sigma$ radius for the IC 1795 members.}
\label{fig:pm_W3parallax}
\end{figure*}

%%%%%%%%%%%%%%%%%%%%%%%%
%                                                                       %
%                  PROPER MOTION                      %
%                                                                       %
%%%%%%%%%%%%%%%%%%%%%%%%

\subsection{Proper Motion and Kinematics-Based Cloud Membership}\label{subsec:propermotion}

\subsubsection{The proper motion centroid}
\label{subsubsec:centroid}

We used Gaia proper motions for the 43 high-precision spectroscopically-confirmed OB stars at the parallax of W3 to establish the fiducial kinematics of stars in W3.  These OB stars have Gaia $\langle{G}\rangle = 15.2 \pm 0.1\,\mags$ and $\langle\sigma_{\varpi}\rangle = 0.026 \pm 0.002\,\mas$. The IC 1795 cluster members tabulated in \citet{Roccatagliata2011} are embedded within protoplanetary disks and lightly obscured by dust, having significantly fainter $\langle{G}\rangle = 19.0 \pm 0.2\,\mags$ and consequently higher  $\langle\sigma_{\varpi}\rangle = 0.22 \pm 0.03\,\mas$, and they span a smaller area of sky. We therefore used these IC 1795 members only as a consistency check.

The upper row of Figure \ref{fig:pm_W3parallax} shows the proper motion in declination (\pmdec) vs.\ right ascension (\pmra) of the 43 spectroscopically-confirmed OB stars and the IC 1795 cluster members (top left and right panels, respectively). The common dashed vertical and horizontal lines indicate the weighted average of \pmra\ and \pmdec, respectively, of the OB members. We adopt these values, $\pmra = -0.82 \pm 0.1\,\masyr$ and $\pmdec = -0.5 \pm 0.1\,\masyr$, as representative of the proper motion of stars in the cloud. 

The common dashed circle centered on these and defined by radius $3\sigma$ ($2.5\,\masyr$), where $\sigma = \sqrt{ (\sigma_\pmra^2 + \sigma_\pmdec^2)/2 }$, represents the `proper motion centroid.' Stars found within the proper motion centroid are  taken to belong to W3 kinematically while those outside, in the complement region (`complement' stars), are less likely to belong.

% OB stars
The 43-star OB population is, by definition, well described by the OB proper motion centroid, concentrated within it. There are only two complement stars, \citetalias{Roman-Lopes2019} 27 (barely outside the circle to upper left) and \citetalias{Roman-Lopes2019} 20 (not shown for increased visibility), which we identify as a runaway pair in Section \ref{subsec:birs6}. The IC 1795 cluster members more concentrated within a $3 \sigma$ radius of $0.95\,\masyr$ around a slightly shifted average proper motion of $\pmra = -0.69 \pm 0.04\,\masyr$ and $\pmdec = -0.05 \pm 0.04\,\masyr$. All of these statistics are affected by the complement outliers because we have done no clipping.

\subsubsection{BIRS proper-motion-based cloud membership}
\label{subsubsec:pmmember}

% BIRS
We identified a small subsample of seven BIRS (BIRS 6, 11, 30, 31, 35, 37, and 66) with precise parallax measurements that are confidently at the W3 parallax by comparing them to the underlying parallax distribution of spectroscopically-confirmed OB stars that we obtained in Section \ref{subsec:parallax-systematics}. We show their proper motions in the bottom right panel of Figure \ref{fig:pm_W3parallax}, including those that we have identified as OB stars (red stars) and RGs (mahogany squares). They have proper motions consistent with the OB proper motion centroid but with little concentration. Neither of the RGs (BIRS 11 and 66) have cross matches to APOGEE RGs and were instead identified as RGs using various photometric-based diagrams.

We highlight BIRS 37 near the center of the OB proper motion, not in the upper left panel of Figure \ref{fig:pm_W3parallax} but a presumed OB star based on our photometric-based diagrams, that here is seen to share the proper motion (as well as the parallax) of W3. We additionally identify BIRS 15 as an OB star that similarly shares the W3 parallax and proper motion. BIRS 15 and 37 lack spectroscopic confirmation, making them new OB star candidates. As such, their IDs are boldfaced in Table \ref{table:birs-results}. We follow up on BIRS 37 in Section \ref{sec:newOBcandidates} where we search for new OB candidates in W3. 

As mentioned above, at the edge of the distribution is BIRS 6 (\citetalias{Roman-Lopes2019} 27), which we identify as part of an OB-runaway pair discussed in Section \ref{subsec:birs6}. 

% RG stars
Star formation in W3 is too recent for stars born in the cloud to have evolved to RGs. Nevertheless, there are some APOGEE stars whose parallaxes are consistent with the derived W3 parallax range (Section \ref{subsubsec:estimate}). These provide an alternative reference point for the large number of BIRS RGs including BIRS 11 and 66. This parallax-selected subset of the APOGEE stars has minimal spatial overlap with W3 and so we used a spatial cut to inspect the proper motions of the six that are projected on W3. As shown in the bottom left panel of Figure \ref{fig:pm_W3parallax}, five (magenta) have proper motions consistent with the W3 OB star centroid, but unlike the OB stars (upper left) they show no evidence of being concentrated in proper motion.

Our proper-motion-based possible cloud membership of the BIRS is encoded in the ninth column of Table \ref{table:birs-results} using the third digit. A star satisfying the W3 proper motion criterion is assigned the same value as the first digit and a zero if it does not (missing astrometry or a parallax uncertainty $> 0.2\,\mas$ is assigned a dash).

BIRS stars that are not imposters and satisfy both parallax and proper motion criteria for possible membership in the W3 cloud are marked with an asterisk in the ID column of Table \ref{table:birs-results}. Thirteen (BIRS 2, 6, 10, 11, 15, 23, 25, 28, 30, 31, 35, 37, 43) have a three digit code 111, two (BIRS 107, 125) have code 222, and five (BIRS 54, 60, 61, 66, 93) have code 333.  Only the seven with the most accurate parallaxes appear in the bottom right panel of Figure \ref{fig:pm_W3parallax}. 

%%%%%%%%%%%%%%%%%%%%%%%%%%%
%                                                                                %
%             POTENTIAL FOR MIGRATION                  %
%                                                                                %
%%%%%%%%%%%%%%%%%%%%%%%%%%%

\section{Potential for Migration and field stars}
\label{sec:migrate}

If the dispersion in the OB star proper motion diagram in Figure \ref{fig:pm_W3parallax} (upper left) represented the systematic peculiar motions, this would quantify the speed at which a star has migrated away from its formation site. With a proper motion dispersion of $\sim 0.8\,\masyr$ for the centroid distribution of stars, the corresponding migration speed is $13\arcmin/\Myr$ or $8\,\kms$, which is substantial (Section \ref{subsec:ucalc}). However, these stars come from a variety of formation sites and so this nuances this interpretation. Only the largest proper motion outliers have the potential to reveal runaway motion, which we investigate here.

%%%%%%%%%%%%%%%%%%%%
%                                                         %
%                       BIRS 6                       %
%                                                         %
%%%%%%%%%%%%%%%%%%%%

\subsection{Runaway OB Stars}
\label{subsec:birs6}

The O star exciting W3 North is \#7004 in \citet{Oey2005} ([OWK]7004), which we note is BIRS 6 and \#27 in \citet{Roman-Lopes2019} (RL 27). Based on X-ray observations by \citet{Feigelson2008}, W3 North appears not to have an associated cluster of less massive pre-MS stars like there is in W3 Main and W3(OH), and so although precise proper motion measurements were not yet available they suggested that this star was a possible runaway star that had migrated from its place of origin, either from the adjacent W4 cloud or W3 Main. We assess this suggestion using Gaia data. 

From Figure \ref{fig:pm_W3parallax} (lower right), the peculiar proper motion ($\delta\mu$) of BIRS 6 has a magnitude of $\sim 2.4\,\masyr$ from the direction of W3 Main. This makes it unlikely to have been ejected from the adjacent W4 cloud, as proposed \citep{Feigelson2008}. Using Equation \ref{eq:vtrans}, the corresponding transverse velocity ($23\,\kms$) is relatively small for a runaway star and would be interpreted as the terminal velocity of this star escaping the potential of W3 Main. To traverse the distance $0\fdg2$ from W3 Main to W3 North would take less than $0.3\,\Myr$ (Equation \ref{eq:ttrans}). This is less than the age of the OB stars in W3 Main \citep{Bik2012}, which appear to be older than $1\,\Myr$. As for BIRS 6 (RL 27) itself, it is a candidate spectroscopic binary with a spectral type of O6 V \citep{Oey2005} or O6.5-O7 IV-V \citep{Roman-Lopes2019}, indicating a short-lived star. It has perhaps started evolving off the MS and been ejected from the even denser regions of W3 Main revealed by Herschel \citep{RI2013, RI2015}.

The star with the highest peculiar proper motion among the W3 OB population LS I $+$61 266 \citep{Hardorp1959} (\#20 in \citealt{Roman-Lopes2019}; RL 20), an O7.5 IV-V star. We find that RL 20, with a peculiar proper motion magnitude of $\sim6.1\,\masyr$ in the opposite direction of RL 27, is proportionately further away spatially from W3 Main in the opposite direction to RL20. It seems plausible then that RL 20 and RL 27 were a pair ejected at the same time. RL 20 is slightly less massive than RL 27, which would help in understanding its kinematics (see \citealp{Gvaramadze2011} for a possible mechanism). 

We suggest that RL 20 is the star heating up the `Y' spatial feature away from the HDL to the west that can be seen in blue in the Herschel W3 press release image,\footnote{\url{http://www.herschel.fr/cea/hobys/en/Phocea/Vie_des_labos/Ast/ast_visu.php?id_ast=36}} in Herschel temperature and radiance (but not $\tau$) maps of thermal dust \citep{RI2013,Singh2022}, and in the WISE $12\,\microns$ map in figure 1 of \citet{Roman-Lopes2019}. There is a local peak of {\halpha} emission in figure 6 of \citet{Kiminki2015}, although that morphology has likely been affected by extinction. In the Canadian Galactic Plane Survey (CGPS; \citealp{Taylor2003}) 21 cm continuum image, there is also diffuse \HII\ emission as an extension of the HDL centered on this star and the adjacent Y feature. RL 20 is a young star apparently not near any dense material from which it might have been born (i.e., a runaway star as discussed) and possibly originated from the \HII\ region W3 H that contains IRS N2 \citep[see][]{Bik2012}. 

A puzzling case is BIRS 30 (IRS N3), which has a similar magnitude in peculiar proper motion as BIRS 6 and yet is fairly close to W3 Main. IRS N3, a B0 V --- B2 V star, is the brightest star in the diffuse \HII\ region (W3 J) but does not have enough ionizing radiation to power the radio emission \citep{Bik2012}. With our SED fitting, we find a B0.5 V spectral solution (Section \ref{subsec:refinedspec-BIRS}) with moderate extinction ($\AV = 1.6$\,mag). IRS N3 might be a young star that has not moved far, despite being an outlier in the proper motion distribution.

\begin{figure*}
\centering
\includegraphics[width=8.5cm]{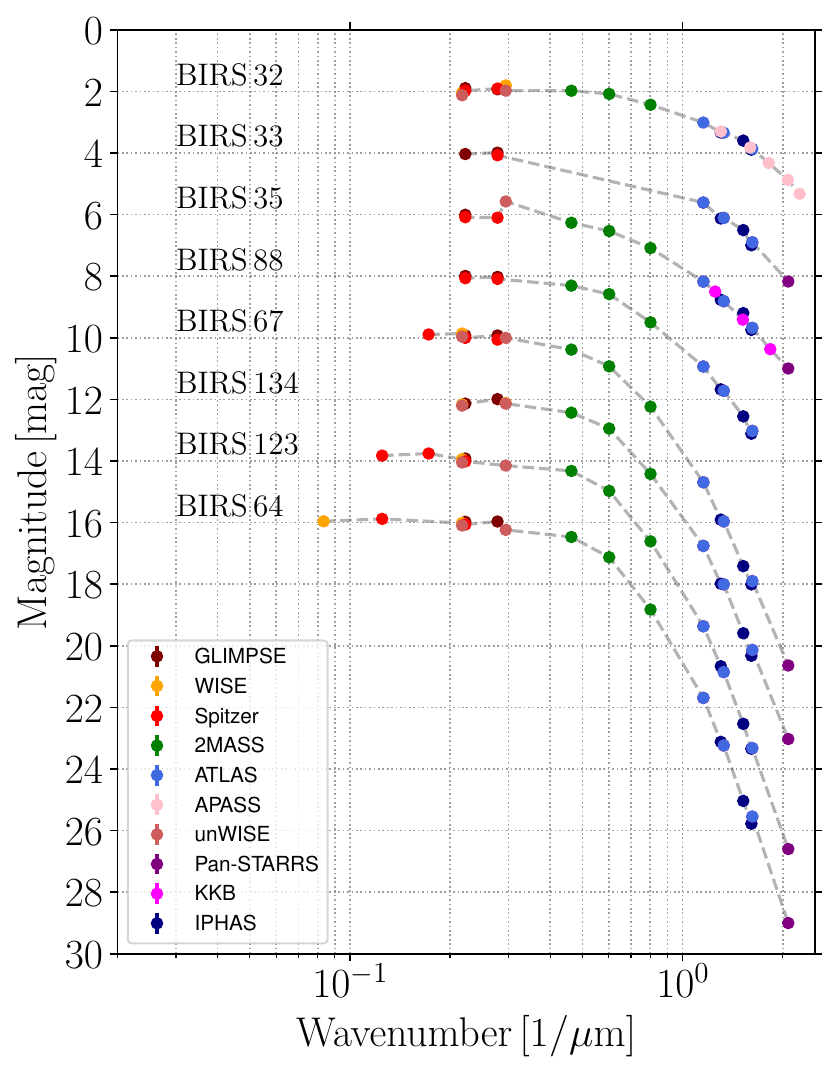}
\includegraphics[width=8.5cm]{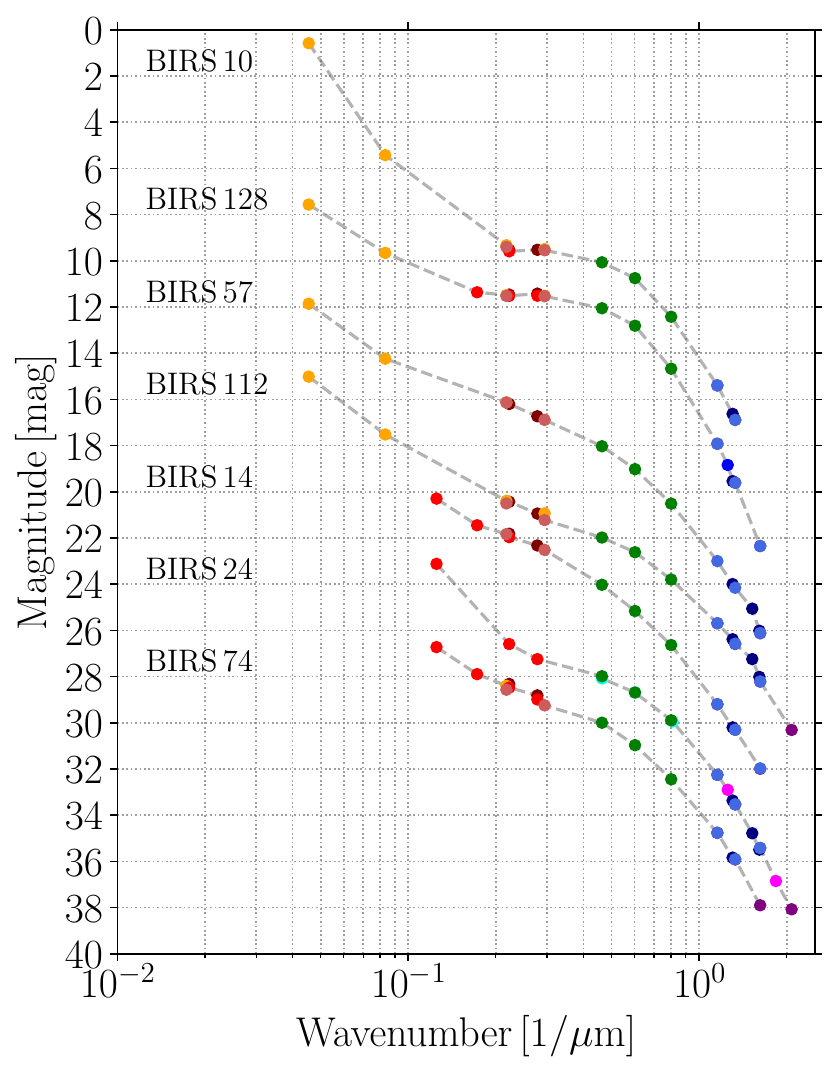}
\caption{Multi-band photometry in Vega magnitudes for representative examples of BIRS SEDs. The photometric data have been shifted vertically and connected linearly for readability. 
Left: SEDs that are well described by reddened stellar photospheres. 
Right: SEDs that have a \ire\ (BIRS 10 and 128) or are unusually steep (\uns).}
\label{fig:BIRS-SED-examples}
\end{figure*}

%%%%%%%%%%%%%%%%%%%%%
%                                                             %
%                      FIELD OBs                     %
%                                                             %
%%%%%%%%%%%%%%%%%%%%%

\subsection{Field OB Stars}
\label{subsec:fieldOBs}

It has been assumed that the OB stars in \citet{Kiminki2015} and \citet{Roman-Lopes2019} are in the W3 molecular cloud.  Note that Gaia astrometry was not used for the targeting of either survey. However, we found that the parallaxes of several B stars having parallax uncertainties $\lesssim 0.066\,\mas$ are inconsistent with the W3 parallax by more than $3\sigma$. These include some background B stars, \citetalias{Kiminki2015} 4, 47, 71, 72, and 83, and foreground A0 V star \citetalias{Kiminki2015} 74. For example, the placement of the background B star \citetalias{Kiminki2015} 72 on the $V$ vs.\ $(V-I)$ CMD in figure 3 (right) of \citet{Kiminki2015} is inconsistent with the displayed reddening tracks labelled by spectral type assuming the W3 distance: \citetalias{Kiminki2015} 72 is a B2 V star found at too faint an apparent magnitude, on the B7 V reddening track. Likewise, the foreground A0 V star is at too bright an apparent magnitude, between the B5 V and B7 V reddening tracks.

We fit their SEDs with extinction curves and verify their distance moduli in Section \ref{subsec:refinedspec-OB}.

%%%%%%%%%%%%%%%%%%%%%%%%%%%%
%                                                                                   %
%           PROPER MOTION  OF FIELD RGs               %
%                                                                                   %
%%%%%%%%%%%%%%%%%%%%%%%%%%%%

\subsection{Proper Motion of Field RGs}
\label{subsec:propermotionfield}

As another point of reference, we compared the proper motions of the BIRS in the field (i.e., not at the parallax of W3) to that of field RGs. In contrast to the OB stars, the field RGs are predominantly background stars that have proper motions more consistent with zero while the foreground RGs have higher proper motions. The BIRS imposters are amongst those with the largest proper motions and are primarily low-reddening foreground stars, whereas the bona fide BIRS distant field stars are significantly obscured by dust.

%%%%%%%%%%%%%%%%%%%%
%                                                         %
%                        SEDs                        %
%                                                         %
%%%%%%%%%%%%%%%%%%%%

\section{Multiband Photometric SEDs}\label{sec:SEDs}

We constructed multi-band SEDs of stars in the OB and RG fiducial samples and the BIRS. These SEDs range from the optical through the MIR and are in Vega magnitudes. We fit these SEDs in Section \ref{sec:SEDclass} to assess their spectral type and foreground interstellar extinction.

Most BIRS SEDs are well described by a stellar photosphere (Figure \ref{fig:BIRS-SED-examples}, left). However, at least 14\% are not-photospheric (Figure \ref{fig:BIRS-SED-examples}, right),  ten with significant MIR excesses (\ire s; BIRS 10, 25, 31, 82, 83, 93, 116, 128, 133, 137) and another nine with unusually steep SEDs (BIRS 14, 24, 26, 57, 74, 87, 89, 97, 112), noted as \ire\ and \uns, respectively, in the last column of Table \ref{table:birs-results}. These not-photospheric SEDs may be caused by some combination of nebular emission, circumstellar emission (e.g., Be stars), or crowding. The \uns\ stars are found below the reddened MS \band\ in our 2MASS and 2MASS-GLIMPSE CCDs and do not conform to the underlying assumptions of the RJCE technique for extinction.
%(see Appendix \ref{subsec:RJCEex}). 
The existence of a \ire\ or \uns\ appears to be uncorrelated with spectral type and cloud region. A few stars (BIRS 52, 96, 109) have unusual photometric anomalies that cannot be described by any of these attributes.

There is some overlap of the twelve BIRS that are LPVs (BIRS 52, 72, 73, 74, 83, 83, 84, 109, 119, 120, 130, 137) with the above lists.

In addition, eleven BIRS (9, 13, 16, 26, 33, 34, 40, 44, 45, 96, 97, 118) have \halpha\ emission according to IPHAS data, with BIRS 40 (A4 V) having the strongest. 

%%%%%%%%%%%%%%%%%%%%%%%
%                                                                    %
%                    FITTING SEDS                       %
%                                                                    % 
%%%%%%%%%%%%%%%%%%%%%%%

\section{Fitting SEDs to Refine Spectral Classifications}
\label{sec:SEDclass}

Three factors determine the observed SED of a star: the spectral type, distance modulus (\dm), and amount of foreground extinction. See Equation \ref{eq:mks} applied filter by filter, with the dm being in common. If the shape of an intrinsic SED can be reproduced within the uncertainties with a scaled extinction curve applied to an SED of earlier spectral type, then it would be impossible to determine the spectral type and amount of extinction separately. Knowing the parallax (whence dm), and the probable luminosity class from generic classification (Section \ref{sec:spectral-class}), helps to constrain the possible range in apparent magnitude.

Using SED fitting, we model spectroscopically-confirmed OB and APOGEE RG stars to provide the basis for a solution to this problem, which we then apply to the BIRS to refine their spectral classification.  The method can also be applied to other samples, such as the OB star candidates in Section \ref{subsec:obresults}.

%%%%%%%%%%%%%%%%%%%%%%%
%                                                                    %
%                    INTRINSIC SEDs                    %
%                                                                    % 
%%%%%%%%%%%%%%%%%%%%%%%

\subsection{Intrinsic SEDs}
\label{sec:SEDintr}

To construct intrinsic SEDs we used intrinsic colors in the 2MASS and SDSS filters (converted to the Vega system) computed by \citet{Covey2007} based on \citet{Pickles1998} spectra. 
The intrinsic colors were provided for solar-metallicity dwarfs (luminosity class V) and giants (luminosity class III) for the following colors: $(u-g)$, $(g-r)$, $(r-i)$, $(i-z)$, $(z-J)$, $(J-H)$, and $(H-\Ks)$.  Absolute magnitudes ($M_J$) were provided for each spectral class but some caution was registered against using the full range for quantitative science \citep{Covey2007}. We used these for giants of relevance, later than about spectral type G5, but for dwarfs we used absolute magnitudes ($M_J$) listed by \citet{Pecaut2013} in the revised electronic version of their table 5 (see footnote \ref{foot:EEM}). Given all-sky surveys in these filters, they provide a convenient basis for fitting observed SEDs even without interpolation.

The intrinsic SEDs of early OB V stars were poorly sampled in spectral type and so we supplemented them with an approximate Vega-magnitude SED created from the ratio of a blackbody relative to A0 V (Vega) according to effective temperatures (\Teff) in \citet{Pecaut2013} and scaled to $M_V$.\footnote{This approximation would not be valid in the UV (past the break at 4000 \AA) or for later spectral types for the RGs where the stellar photosphere contains strong spectral features producing photometric magnitudes that deviate from those for a simple blackbody shape.} We validated this approximation against the above SEDs based on \citet{Covey2007}; an example is shown in the top panel of Figure \ref{fig:SED_examples}. This allowed us to expand the grid of early OB types to include O3, O4, O5.5, O6, O6.5, O7, O7.5, O8, O8.5, O9.5, and B0.5 spectral types. 

%%%%%%%%%%%%%%%%%%%%%%%%%
%                                                                          %
%                EXTINCTION CURVES                    %
%                                                                          %
%%%%%%%%%%%%%%%%%%%%%%%%%

\subsection{Frequency-Dependent Dust Extinction, Distance Modulus, and Method of SED Fitting}
\label{sec:fitting}

We used the normalized extinction curves described in Section \ref{sec:extinction-curve} parameterized by \RV\ and \powerlaw.  These are scaled by a fitting parameter, \AKs. Where we had Gaia parallax measurements of sufficient quality, we used the parallax and its corresponding uncertainty to constrain the \dm\ to a small range (see below).

For each intrinsic stellar SED in the grids for the two populations, OB and RG, we simultaneously fit the \dm\ and scale factor of the dust extinction curve to model the observed optical ($griz$) and NIR ($JH\Ks$) photometric data. The NIR data are essential for constraining the combination of the absolute magnitude plus \dm\ (i.e., the vertical placement of the SED model). The optical data, in contrast, provide more leverage in assessing the extinction. MIR data often contained MIREs and small systematics in individual passbands that would cause substantial effects on the resulting fit. These, and optical data that were often available in other filters, were used only as a consistency check.

The optimal ranges in \dm\ for each of the stellar populations were quite different due to the considerable differences in the typical parallaxes and parallax uncertainties. Given the systematic errors in the intrinsic SEDs and extinction curves, we filtered on the quality of the photometric data included in the fit and carried out a simple unweighted fit that minimized the $\chi^2$ of the residuals between the observed SED and modeled spectrum. Where we had a priori information about the generic classification (Section \ref{sec:spectral-class}), we fit solutions only for the appropriate luminosity class. We assessed the relative goodness of fit using $\chi^2$.

%%%%%%%%%%%%%%%%%%%%%
%                                                             %
%                        OB STARS                   %
%                                                             %
%%%%%%%%%%%%%%%%%%%%%

\subsection{OB Stars}\label{subsec:refinedspec-OB}

We used the spectroscopically-confirmed OB stars for which \citet{Kiminki2015} provided the spectral classification and foreground extinction as the basis for honing and validating our SED fitting procedure. Without using parallax measurements to constrain the absolute magnitude of each star, we were unable to differentiate between dwarfs and giants just using the shape of the observed SED. For either luminosity class it was nearly always possible to find a \dm\ and scaling of foreground extinction that provided a close fit to the data and in fact the giant solution almost always provided a closer fit. Constraining \dm\ and restricting the luminosity class to dwarfs was essential for determining accurate spectral types.

To constrain \dm, we allowed a $\pm 3\sigma$ range in parallax. This along with our fiducial extinction curve (Section \ref{sec:extinction-curve}) parameterized by $\RV = 3.6$ (rather than 3.1), which is consistent with their photometric determination of \RV\ for these stars toward W3, 
enabled us to obtain spectral type solutions that were generally in agreement with \citet{Kiminki2015} within a few sub-types.

\begin{figure}[t!]
\centering
\includegraphics[width=8.5cm]{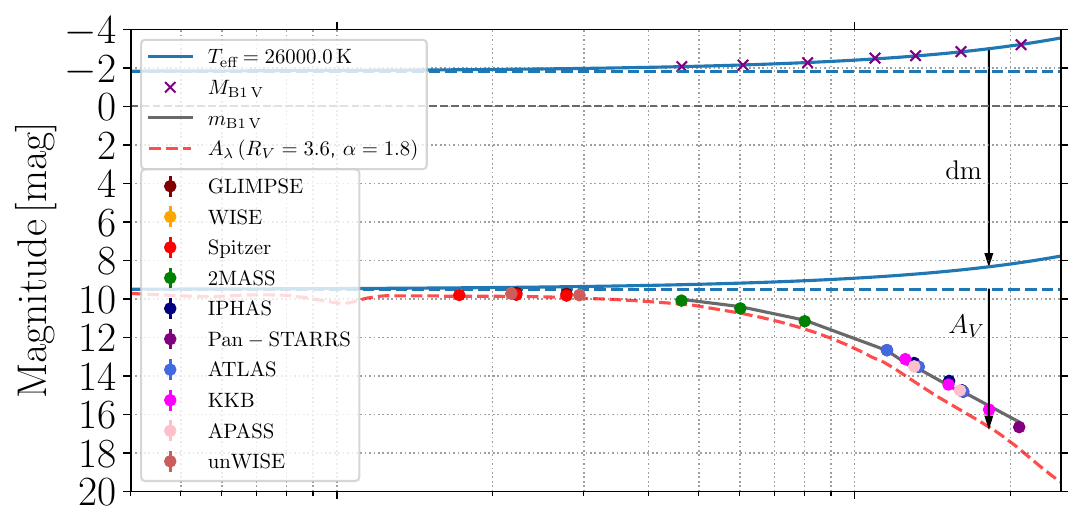}
\includegraphics[width=8.5cm]{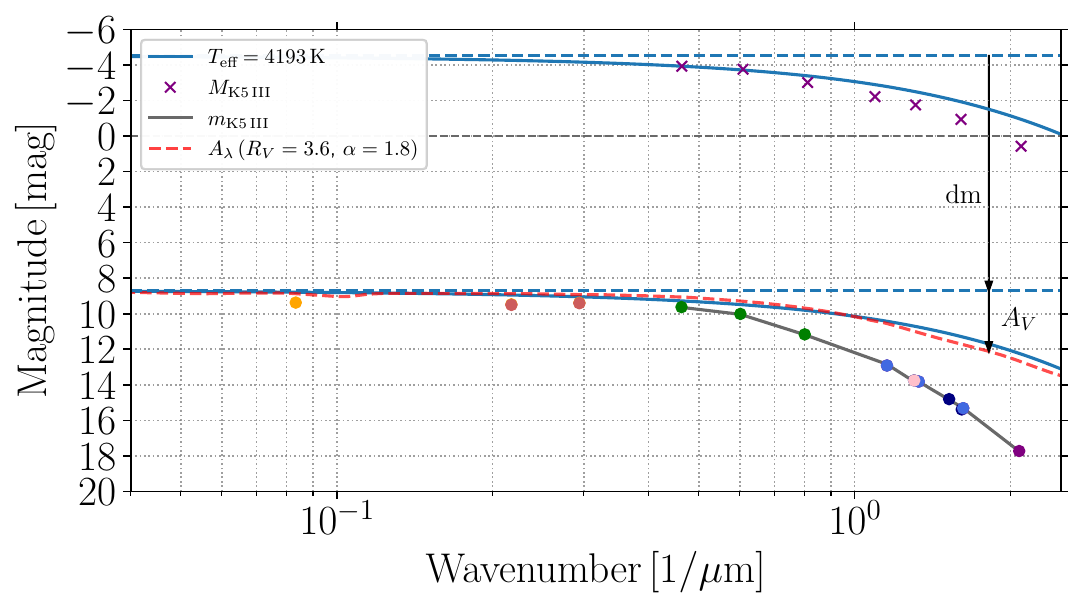}
\caption{Representative SEDs of spectroscopically-confirmed stars along with the component elements of the modeled solutions. 
Top: OB (\citetalias{Kiminki2015} 54). Bottom: APOGEE RG (2M02201295$+$6032102). 
Purple crosses: absolute magnitudes of best fit spectral type. Blue curve and dashed asymptote: blackbody-body based approximation (not used in fit). Red dashed: scaled extinction curve (relative to lower asymptote). Colored dots: apparent magnitudes (MIR values not used in fit). Grey curve: modeled fit. Black arrows: distance modulus, \dm, and visual extinction, \AV\ (see text).
}
\label{fig:SED_examples}
\end{figure}

Figure \ref{fig:SED_examples} (top) shows a representative example using \citetalias{Kiminki2015} 54, for which we find a solution in agreement with the known spectral type B1 V and extinction. The intrinsic SED used in the best fit is marked with purple crosses. Running through them, but not used in the fit, is the approximate theoretical blackbody-based spectrum shown with a blue solid curve (the blue dashed line is the asymptotic low-frequency limit in this approximation). The modeled (scaled) extinction curve is shown with a red dashed line. Together with \dm, these give the solid grey curve that reproduces the observed SED (the observed SED is above the modeled extinction curve because of the negative absolute magnitudes). Using the spectral type, \citet{Kiminki2015} found that this star has $\AV = 7.3 \pm 0.3\,\mags$. Our photometric-based SED fitting yields an extinction of $\AKs = 0.76\,\mags$, which corresponds to $\AV = 7.14\,\mags$, assuming $\AV/\AKs = 9.41$.

The residuals from fitting the SEDs of the \citet{Kiminki2015} OB population with good parallax measurements had a standard deviation of $0.09\,\mags$ and a median of $0.05\,\mags$. We verified that solutions for adjacent spectral-type implied different distance moduli but quite similar extinction scaling; i.e., the small constrained range in distance modulus allows for a small range in possible spectral types without significantly affecting the extinction. 

We compared our extinction estimates to the values tabulated by \citet{Kiminki2015} and found that our model reproduced {\AV} with a median offset of $-0.27\,\mags$ (ours were less) and a standard deviation of $0.6\,\mags$.
Fitted \AKs\ measurements are well correlated with values obtained using our modified RJCE technique.
%(Appendix \ref{subsec:RJCEex}).

We obtained predictions for six OB stars that were lacking specific spectral types: \citetalias{Kiminki2015} 12 (B5/B7), 13 (B5/B7), 20 (B5/B7), 22 (B3), 34 (B8), and 39 (B1). Furthermore, the spectroscopically-confirmed OB stars that we identified as field stars in Section \ref{subsec:fieldOBs}, based on their parallaxes, consistently have \dm\ solutions from SED fitting that indicate these stars are not in W3.

While our photometric model fitting works well on average, beyond uncertainties in the spectroscopic-based results that we attempt to reproduce,there are several sources of systematic error that affect individual solutions. Our method relies strongly on the Gaia parallax measurements for constraining the absolute magnitude and so any systematic errors in the parallaxes can affect the spectral type solution (e.g., see Section \ref{subsec:VES} for a discussion on VES 735 in KR 140). The absolute scale of the intrinsic SEDs, established through \MKs, is uncertain. Finally, the shape of the extinction curve is a parameterized approximation and might vary across the cloud.

%%%%%%%%%%%%%%%%%%%%%
%                                                             %
%                      BINARIES                      %
%                                                             %
%%%%%%%%%%%%%%%%%%%%%

\subsubsection{Unresolved Binaries}
\label{sec:unresolved}

Our method does not account for the possibility that some stars may be unresolved binaries. Not only can this affect our inferred spectral types but also it can have a significant effect on the predicted amount of ionization. To estimate the maximum effect on our results, we consider a binary with two stars of the same spectral type so that the absolute magnitude of the unresolved binary is $2.5 \log(2) = 0.75\,\mags$ brighter. We repeated our SED fitting method on the \citet{Kiminki2015} OBs after artificially dimming the apparent magnitudes by 0.75\,mag. This resulted in deduced spectral types that were typically only a couple of sub-types later and \AV\ values that were on average $\sim 0.5$\,\mags\ lower.

%%%%%%%%%%%%%%%%%%%%%%
%                                                                %
%                   RED GIANTS                       %
%                                                                %
%%%%%%%%%%%%%%%%%%%%%%

\subsection{APOGEE RGs}\label{subsec:refinedspec-RG}

With a viable procedure validated on the spectroscopically-confirmed OB stars, we extended our SED fitting to the APOGEE RGs that have been spectroscopically classified and for which we have high-quality parallax measurements for constraining the absolute magnitude. The parallax uncertainties of the APOGEE stars are typically quite small but larger than the OB stars. We allowed a $3\sigma$ range in parallax. 
We used the canonical RJCE technique, which assumes $\HG_0 = 0.08\,\mags$, to constrain the range in possible \AKs\ for the extinction curve scaling.  

We fit the scaling of the fiducial extinction curve (Section \ref{sec:extinction-curve})
using $griz$ plus 2MASS $J$, $H$, and {\Ks} photometry along with GLIMPSE CH1 and CH2 bands. 
Dropping the latter two bands makes only small differences.

Figure \ref{fig:SED_examples} (bottom) shows a representative example of the APOGEE star 2M02201295$+$6032102 ($\Teff = 4193$\,K), with best-fit spectral type K5 III. Because of their intrinsically red colors, the observed SEDs for these RGs are inevitably redder and below the modeled extinction curve.  

We fit 147 APOGEE SEDs, the vast majority of which (139; 95\%) we found to be of K and M spectral types. The residuals for the entire RG population fit had a standard deviation of $0.05\,\mags$ and a median $0.03\,\mags$, at least as good as for the OB stars. We found \Teff\ from APOGEE to be inversely correlated with our derived spectral type, with the exception of six erroneous stars of spectral type G0 and hotter, which we removed from further analysis. Fitted \AKs\ values for this stellar population were less correlated with the RJCE value compared to the OB stars.

%%%%%%%%%%%%%%%%%%%%%%%%%%
%                                                                             %
%             E(r-z) VS TEFF CORRELATION             %
%                                                                             %
%%%%%%%%%%%%%%%%%%%%%%%%%%

\subsubsection{Intrinsic Color of Red Giants as a Function of Effective Temperature} \label{sec:APOGEE_color_vs_Teff}

Intrinsic colors are fundamental physical properties of stars and are essential for quantifying color excesses and interstellar extinction. For cooler stars with strong absorption features, the intrinsic colors are challenging to measure and calibrate with respect to spectral type. For APOGEE stars of known temperature, the intrinsic colors change systematically with \Teff. This was anticipated for the RJCE technique \citep{Majewski2011} and verified using stars with zero reddening in the NIR \citep{Jian2017}. Based on our inspection of the OB SEDs, the fit of the extinction curve is extremely sensitive to the NIR channels. We explored this toward much shorter wavelengths where systematics are less prominent, also seeking to establish whether the extinction curve found using OB stars is consistent with that of RGs.\footnote{While {\Teff} is the primary stellar characteristic that determines the intrinsic color, secondary characteristics such as surface gravity (\logg) and metallicity (\FeH) can affect intrinsic colors as well. We used these stellar parameters to restrict our analysis to giants near the plane such that ${\logg}<3.5$ \citep{Worley2016} and ${\FeH}>-0.5$ \citep{Ramirez2005}. Most of our sample of APOGEE stars satisfy these criteria.}

\begin{figure}[t!]
\centering
\includegraphics[width=8.5cm]{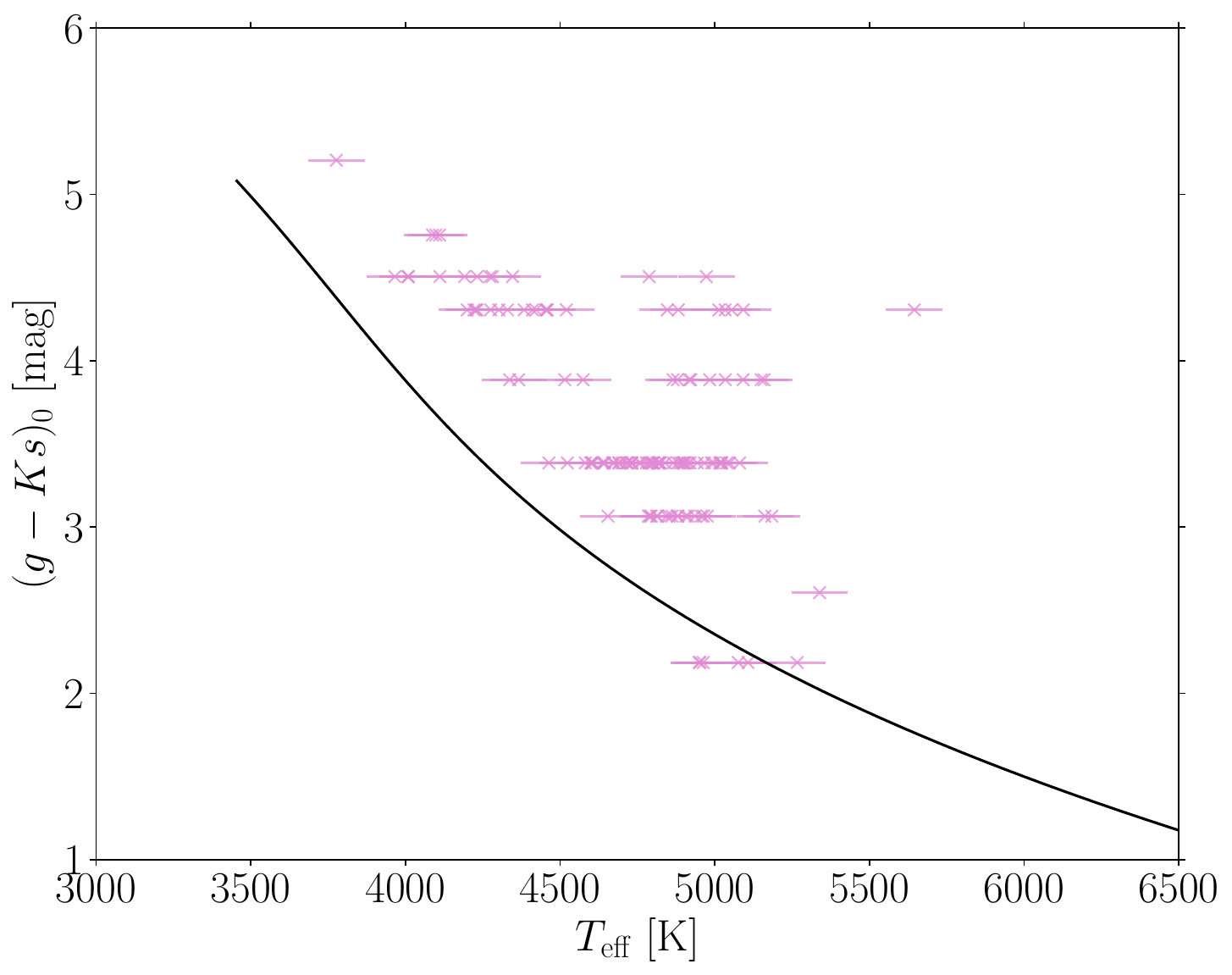}
\caption{$(g-\Ks)_0$ vs.\ \Teff\ for APOGEE giants toward W3 (magenta crosses) compared to the analytical description of Equation 10 in \citet{Casagrande2019} (solid black line).
}
\label{fig:(r-z)0_vs_Teff}
\end{figure} 

As one example, Figure \ref{fig:(r-z)0_vs_Teff} shows the intrinsic $(g-\Ks)_0$ color as a function of {\Teff} for the APOGEE RGs in our sample compared to the analytical description of Equation 10 in \citet{Casagrande2019} (solid black line).

%%%%%%%%%%%%%%%%%%%%%%%%%%%%%%%
%                                                                                             %
%      REFINING BIRS SPECTRAL CLASSIFICATION          %
%                                                                                             %
%%%%%%%%%%%%%%%%%%%%%%%%%%%%%%%

\subsection{BIRS}\label{subsec:refinedspec-BIRS}

We used our SED fitting procedure to refine the spectral typing of the BIRS. We only proceeded in instances where we had a priori information on the luminosity class (III or V) and high-quality parallax measurements to constrain the absolute magnitude. The parallax uncertainties for the BIRS are significantly larger than for the spectroscopically-confirmed OB and RG stars (see Figure \ref{fig:Splot}) and so we allowed only a $1\sigma$ range in parallax. We used our generic spectral classification from Section \ref{subsec:generic-BIRS-class} to determine which luminosity class was fit to each star and used the fiducial extinction curve from Section \ref{sec:extinction-curve}.
While many of the BIRS are RGs unlikely to be members of the W3 molecular cloud, they are predominantly behind the cloud and so experience extinction by dust within W3 (in addition to background and foreground extinction).

We were only able to obtain spectral classifications of 69 BIRS because many have poor or negative lower-limit parallax measurements while others are missing on various photometric-based diagrams leaving us unable to identify an a priori luminosity class.
Of the 69 BIRS (52\%) that have high-quality parallax measurements and were successfully fit with an SED model, 27 (39\%) were best fit with a dwarf solution and 42 (61\%) with a giant solution. This split is rather different than the percentages for the CCD-based spectral classifications because many of the a priori RGs have small enough parallaxes that their 1$\sigma$ ranges include negative parallaxes. The residuals have a standard deviation of $0.1\,\mags$ and a median $0.01\,\mags$. 

Our spectral classification results for the BIRS are summarized in columns 10 through 13 of Table \ref{table:birs-results}, including the dm, \AKs, \AV, and spectral classification. The spectral classifications of the BIRS imposters were found to be nearly all intermediate-to-low-mass dwarfs (five F V, two G V, three K V, and two M V), with the exception of BIRS 40 (A5 V) and BIRS 98 (F0 III).

\citet{Sprague2022} have determined $\log_{10}g$ (surface gravity) and $\log_{10}\Teff$ for about 650,000 stars using the Apogee Net II convolutional neural network pipeline, extending the coverage to massive stars with \Teff\ as high as 50,000\,K. There are cross matches to 20 B dwarfs from \citet{Kiminki2015}, 16 of which were among the stars used in the training of the network, and judging from the results for \Teff\ they cannot be used for precise spectral typing, all being systematically low, below $\log_{10}\Teff = 4.0$ (i.e., later than A0 V). Nevertheless, a high value of $\log_{10}g$ and (relatively) high $\log_{10}\Teff$ can be used to confirm that some BIRS are indeed early OB dwarfs. This is the case for the known OB stars BIRS 6, 28, 30, and 31 and by extension suggests that BIRS 25 and 47 are too (we had difficulty classifying BIRS 25 at all and BIRS 47 was ambiguous, in Table \ref{table:birs-results} inconsistently as K3 III). In contrast, the \citet{Sprague2022} stars with low $\log_{10}g$ should be RGs. This is the case for the known APOGEE RGs BIRS 85, 94, and 126, and by extension BIRS 84 should be too (we classified it as RG; LPV).

%%%%%%%%%%%%%%%%%%%%%%%%%%%%%%%%%%%%%%%%%%%%%%%%%%%%%%
%                                                                                          %
%            SEARCHING FOR NEW OB CANDIDATES         %
%                                                                                          %
%%%%%%%%%%%%%%%%%%%%%%%%%%%%%%%%%%%%%%%%%%%%%%%%%%%%%%

\begin{figure*}
\centering
\includegraphics[height=8cm]{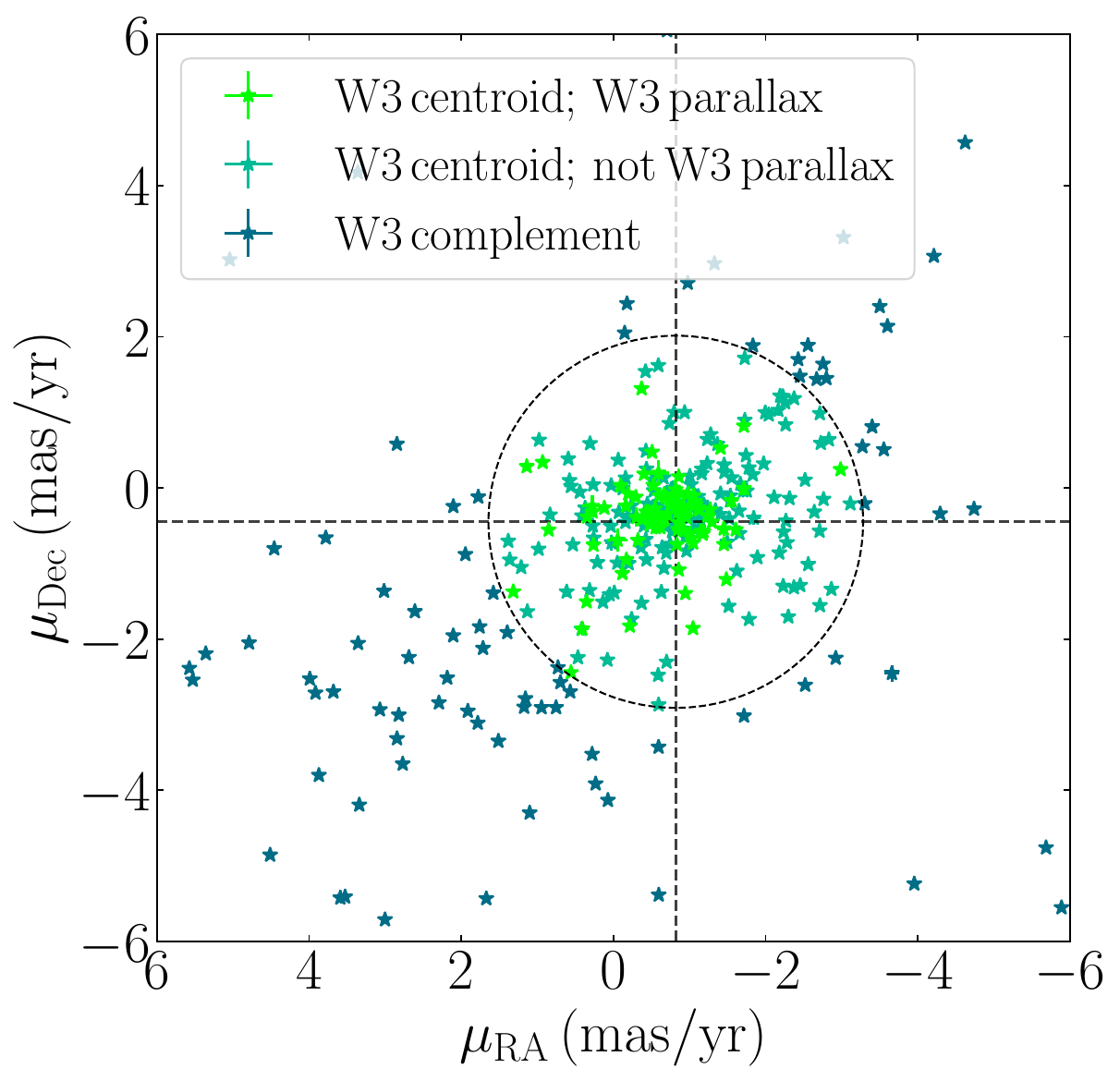}
\includegraphics[height=8cm]{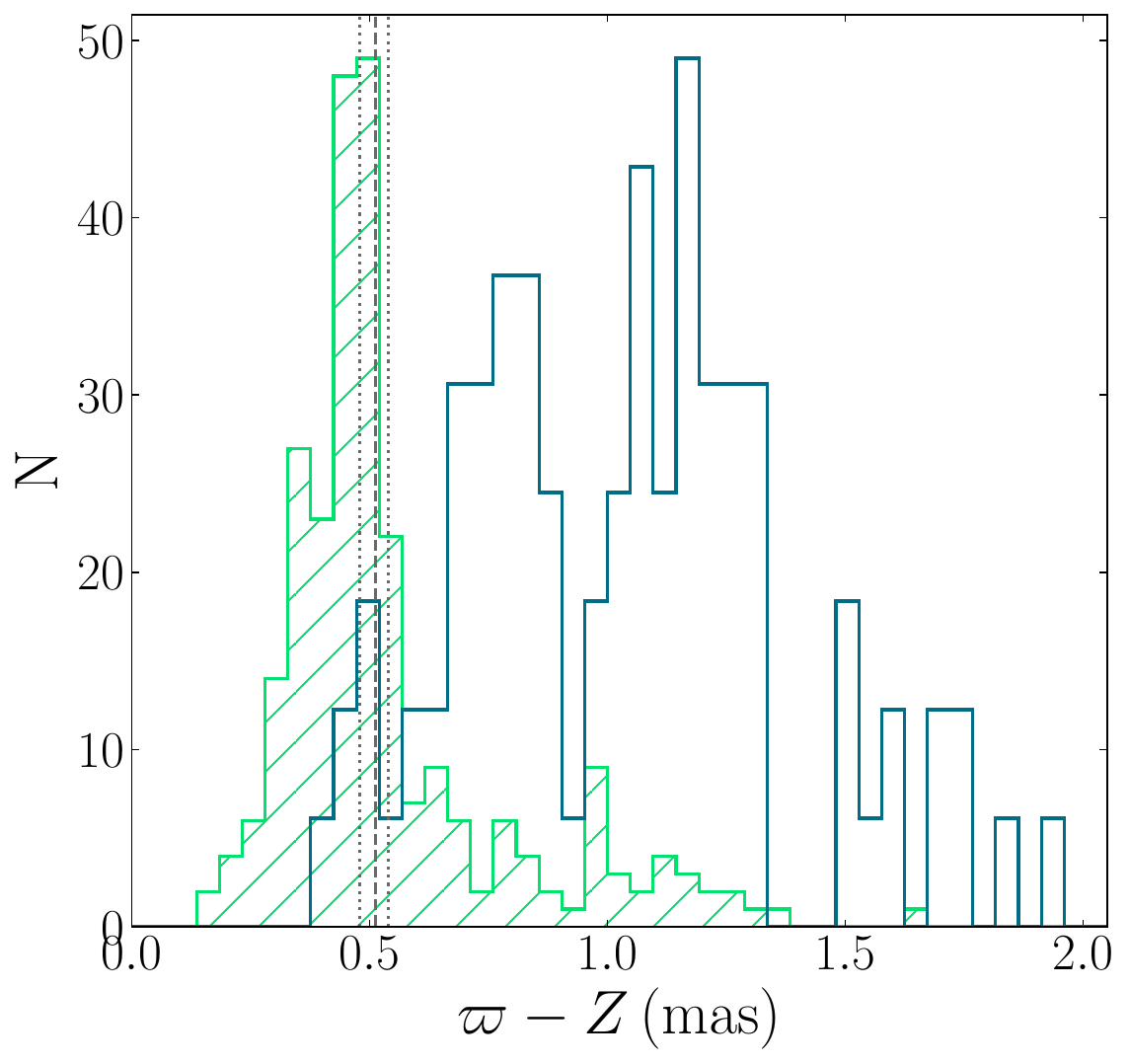}
\caption{Gaia astrometry of bright MS stars used in the identification of candidate OB stars in W3. Left panel: Proper motions. The dashed circle is the proper motion centroid of spectroscopically-confirmed OB stars from Figure \ref{fig:pm_W3parallax} (top left panel), within which we define stars to be kinematically associated with W3; lime indicates stars within the W3 parallax range while green indicates those that are outside of the W3 parallax range (see right panel). Stars found outside of this circle (i.e., the complement region) are shown using a dark green color. Right panel: Parallax distributions of stars shown in the left panel. Light hashed green indicates stars within the W3 proper motion centroid while dark green indicates those in the W3 proper motion complement region. The vertical dashed line denotes the W3(OH) maser parallax value and the dotted vertical lines denote the W3 parallax range used in this paper (see text).
}
\label{fig:OB-candidates-pm}
\end{figure*}

\section{Search for New OB Candidates in W3}\label{sec:newOBcandidates}

We used the tools that we developed here to search for new OB star candidates in W3. 

\subsection{Selection}
\label{subsec:obselection}

We started by isolating stars along the reddened MS \band\ of our 2MASS-GLIMPSE CCD (see Figure \ref{fig:RJCE_JK-HG2}). We invoked a spatial requirement to include only those stars projected on the cloud (Figure \ref{fig:tau-map}). Furthremore, we removed all stars cross matched to spectroscopically-confirmed OB stars. The 2MASS CCD remains a good alternative for increasing the number of candidates by removing the requirement of good quality MIR photometry. One could also use alternative RJCE colors on the x-axis including any NIR-MIR combination sampling the Rayleigh-Jeans portion of the stellar SED (but not $J$; \citealp{Majewski2011}). This yielded a preliminary list of 1567 candidates.

We then selected for potential bright OB candidates that would be brighter than a B9 V star ($\MKs=0.621\,\mags$) at the distance modulus of W3 ($\Ks=12\,\mags$), allowing for effects of extinction following the criterion $\Ks<0.9\,(H-[4.5])+12$, where 0.9 is approximately the slope of the reddening vector $\AKs/E\HG$ in the \Ks\ vs.\ \HG\ CMD. This reduced our preliminary list to 352 candidates.

Next, we identified those that are kinematically associated with W3 using their proper motions following the procedure in Section \ref{subsec:propermotion}. The left panel in Figure \ref{fig:OB-candidates-pm} shows their proper motions, where those inside the W3 proper motion centroid (dashed circle from Figure \ref{fig:pm_W3parallax}, top left panel) are considered to be kinematically associated with the cloud. As seen in the right panel of Figure \ref{fig:OB-candidates-pm}, those with proper motions consistent with W3 have a parallax distribution that peaks at the W3(OH) maser parallax while those that do not have a drastically different parallax distribution. This highlights the benefit of parallax in addition to proper motion requirements. 

Of the 260 preliminary candidates with proper motions consistent with W3, we selected 94 that are consistent with the parallax range in W3 ($0.48 < \varpi-Z < 0.54\,\mas$; see procedure outlined in Section \ref{subsubsec:estimate}).
Looking back to the left panel, the proper motions of these 94 (\lime) stars form a more compact distribution (standard deviation of 0.5\,\masyr\ from the centroid) compared to the remaining stars within the dashed circle (green, with standard deviation of 0.7\,\masyr\ from the centroid). Because all the lime points are inside the particular 3$\sigma$ proper motion centroid used, in effect we are selecting only on parallax. 

Lastly, we fit their SEDs using the refined sampling of intrinsic SEDs from Section \ref{sec:SEDintr} and the fiducial extinction curve (Section \ref{sec:extinction-curve}) to measure their foreground extinction and distance modulus while predicting their spectral type. Permitting a $1\,\mags$ range around the W3 distance modulus (11.46\,\mags) allowed us to identify eleven stars that were not at a suitable distance modulus, along with one that had an A0 spectral type, all of which discarded. 

Our final selection of 82 OB candidates is summarized in Table \ref{table:ob-candidates}, including the Gaia astrometric results (columns 2-8) and our SED fitting results (columns 9-12). These OB candidates have high-quality Gaia data, with a median (standard deviation) {\tt ruwe} of 1.01 (0.07) and a median (standard deviation) fidelity of 1.00 (0.007). We verified that all of our OB candidates were found in the OB region of the observational \hrd. As expected, there are no cross matches between our OB candidates and the aforementioned APOGEE stars \citep{Zasowski2013}. 

\input{tables/table2.tex}

\subsection{Discussion}
\label{subsec:obresults}

Among the 82 OB candidates we find the following MS spectral types via SED fitting: one O5.5, one O7.5, one O9.5, two B0.5, five B1, 17 B3, 15 B5/B7, 32 B8, and eight B9. It is possible that some of the earliest stars have evolved fairly vertically off the MS, though these would be considerably rarer compared to MS stars. If the star were an unresolved binary, it would appear brighter too.  Our classification method assumes single dwarfs and relies on brightness, and so a star that for either reason appeared brighter would be interpreted as having an earlier spectral type.  

The locations of our OB candidates are identified in Figure \ref{fig:tau-map} using open \lemon\ circles. The spatial coverage is considerably extended to the west. While the distribution is not uniform, it is not as concentrated as the distribution of YSOs in the same subregions found by \citet{RI2011}. 

We cross matched our OB candidates with the tables of YSOs and YSO candidates in \citet{RI2011} finding only five matches. This is expected: our color selection criteria were focused on finding MS stars and are very different than those used for YSOs. A search through SIMBAD for all 82 candidates also revealed these five stars. Three YSO candidates (Y*?; unclassified), \#s 2, 20, and 25 in our list, we classify as B8 V. 
Two YSOs (Y*O; classified B2.5 V and B1 V), \#s 81 and 82 in our list, we classify as B1 V, in good agreement. One further YSO (Y*O; unclassified by \citealp{Navarete2019}), \#\ 57 in our list, we classify as B9 V.

In our OB candidate list there are nine cross matches with the \citet{Sprague2022} list, \#s 28, 45, 47, 67, 72-4, 81, and 82, not surprisingly among our brighter OB MS candidates:\footnote{A SIMBAD search also turned up records for these nine stars.} all three that we classified as O stars, both B0.5, two of five B1, and just two of 17 B3. The \citet{Sprague2022} $\log_{10}\Teff$ is no higher than 4.15 and so again not a precise indicator of spectral type.  Nevertheless, the high value of $\log_{10}g$ and (relatively) high $\log_{10}\Teff$ confirm that these candidates are indeed early OB dwarfs as we have classified.

A more reliable way to classify the APOGEE-2 H-band spectra of O stars, validated by \citet{Roman-Lopes2018}, relies on measurements of the equivalent widths of H I Brackett lines and, where available for the hotter stars, equivalent widths of two lines of He II. This resulted in the confirmed O stars from \citet{Roman-Lopes2019} referred to as `RL' throughout this paper and not included in our candidates list.\footnote{Of the 15 toward W3, 14 have cross matches with \citet{Sprague2022} and again systematically low $\log_{10}\Teff$.}  A related classification method for B stars has been validated by \cite{Ramirez-Preciado2020} but results for W3 are pending.

\subsection{A few candidates of note: \#s 67, 47, 73, and 57}
\label{subsec:stars}

We found three intrinsically-luminous O-type candidates.
The O5.5 V candidate \#67 (IC 1795 150; \citealp{Ogura1976}) is seen projected on a narrow dust lane in the more obscured eastern part of the `Fishhead Nebula'\footnote{\url{https://apod.nasa.gov/apod/ap190731.html}.} and is quite heavily reddened with  $\AKs = 0.78$\,mag. Its SED strongly resembles that of BIRS 6, which we find to be an O5 V star with $\AKs = 0.83$\,mag. BIRS 6 (RL 27) is independently classified spectroscopically as O6.5-O7 IV-V \citep{Roman-Lopes2019}. 
About 10\arcmin\ to the WNW along the body of the fish is the bright `eye,' BD $+61$ 411 (IC 1795 89), classified as O6.5V((f))z by \citet{Maiz2016}, less heavily reddened, and therefore brighter in the optical.
BD $+61$ 411 is taken as the main ionizing star of IC 1795, but the geometrical placement and spectral classification of \#67 suggest that it is a luminous hot star contributing significantly too.
It will be important to confirm the classification of \#67 spectroscopically.

Candidate \#47, classified as O7.5 V with $\AKs=0.72\,\mags$, is on the western edge of W3 Main, and \#73, classified as O9.5 V with $\AKs=0.66\,\mags$, is on the southern end of the HDL.

At the other end of the spectral range of OB stars, candidate \#57 is classified as B9 V with $\AKs=0.47\,\mags$. This candidate lies below W3 Main and corresponds to BIRS 37 or G22 in the candidate list in table 4 in \citet{Navarete2019}. We previously reported BIRS 37 as a potential OB candidate in Section \ref{subsubsec:pmmember} 
(Figure \ref{fig:pm_W3parallax}, lower left panel).

%%%%%%%%%%%%%%%%%%%%%%%%%%%
%                                                                                %
%                 OTHER RECENT SEARCHES              %
%                                                                                %
%%%%%%%%%%%%%%%%%%%%%%%%%%%

\subsection{Other recent searches and targeting}
\label{subsec:othersearch}

While \citet{Kiminki2015} used an optical $V$ vs.\ $(V-I)$ CMD to identify reddened OB candidates, their spectroscopic follow-up yielded a low success rate of only $\sim$10\%. Similar to our optical $i$ vs.\ $(r-i)$ CMD in Figure \ref{fig:ivsri}, the reddened OB stars would overlap with the reddened RGs, causing contamination. Using an IR CCD such as Figure \ref{fig:RJCE_JK-HG2}, or even Figure \ref{fig:jhhk}, to separate the reddened MS and RG \bands\ would likely have increased their success rate substantially.

\citet{Navarete2019} (their table 4) identified 37 OB candidates in W3 based purely on Gaia astrometry. However, without the use of CCDs to isolate the reddened MS \band\ and CMDs to select the bright MS stars, their stellar sample is undoubtedly contaminated with non-OBs. We cross matched these 37 stars to 2MASS using a cross-match radius of $0{\farcs}5$, verifying the presence of nine reddened RG \band\ stars on the 2MASS CCD and an additional three in the ambiguous region between the reddened MS and RG bands, along with faint stars well below the A0 reddening vector on a {\Ks} vs.\ {\HK} CMD.\footnote{\citet{Navarete2019} used this contaminated stellar sample to establish the parallax of W3 (see Section \ref{subsubsec:expect}) but neither RGs nor non-OB MS stars would be persuasive distance indicators. A number of these stars show evidence of an IRE in the 2MASS CCD and might be low-mass Herbig Ae or Be stars. These stars might be useful distance indicators if the IREs are evidence that they are in W3. However, that point has not been discussed. Furthermore, their sample was restricted to a certain parallax range, which could introduce a bias. \label{foot:contam}}

Returning to the \citet{Sprague2022} list, there are 132 stars projected on the W3 molecular cloud of interest to us. Of these, 44 that are not among the nine matches with our OB candidate list or known B stars from \citet{Kiminki2015} have a sufficiently high $\log_{10}\Teff$ (though systematically low as discussed above) and $\log_{10}g$  to be potentially OB stars according to their results for \citet{Kiminki2015} stars. Compared to the 73 distinct candidates in our list, these 44 have different distributions in diagrams of interest: their {\JH} vs.\ {\HK} CCD is not as tight and extends to stars of lower reddening; the CMDs of relevance range to bright stars; the proper motions and parallaxes are much more scattered than we selected for (Figure \ref{fig:OB-candidates-pm}).  This simply reflects the different criteria used to select the APOGEE targets whose spectra  \citet{Sprague2022} analysed.

\citet{Zari2021} used 2MASS and Gaia $G$ band photometry in developing an all-sky target sample of OB stars for spectroscopy with SDSS V. In their `filtered target list,' we find only 21 of the 91 known \citet{Kiminki2015} OB stars, probably because of the \citet{Zari2021}  conservative selection criteria including parallax aimed at completeness for a brighter sample. Toward the W3 molecular cloud there are 149 stars, which reduces to 69 if the parallax range is restricted as for our candidate list and could be reduced more using proper motion. We also examined the placement of their candidates on the 2MASS-GLIMPSE CCD that we used in our OB star candidate selection; the distribution is not as tight (our criterion is somewhat more selective). Our candidates tend to be fainter and redder in the $G$ vs.\ $(G - \Ks)$ CMD and redder in the \JH\ vs.\ $(G - \Ks)$ CCD.

We cross matched their list with our OB candidate list, finding 24 matches, including the nine matches with \citet{Sprague2022}.  As mentioned, the nine comprise all three candidates that we classified as O stars, both B0.5, two of five B1, and just two of 17 B3.  The remaining 15 stars matched are two more of the five B1, and 13 more of the 17 B3.  The \citet{Zari2021} cross matches therefore comprise most of the stars earlier than B3 in our list, missing some stars that are fainter:
one B1 star with the highest extinction, one B3 star with the highest extinction (another B3 star with moderate extinction). They also miss all 55 of our candidates of later spectral type (15 B5/B7, 32 B8, eight B9), which are also fainter than their stars that are cross matches.

%%%%%%%%%%%%%%%%%%%%%%%%%%%%%%
%                                                                                           %
%          COMPARISON OF EXTINCTION TRACERS         %
%                                                                                          %
%%%%%%%%%%%%%%%%%%%%%%%%%%%%%%

\section{Individual distant Red Giants as probes of extinction}
\label{sec:distantprobes}

In this section we establish the potential of individual distant RGs as high spatial resolution probes accessing regions of high dust extinction.

\begin{figure*}[t!]
\centering
\includegraphics[width=8.5cm]{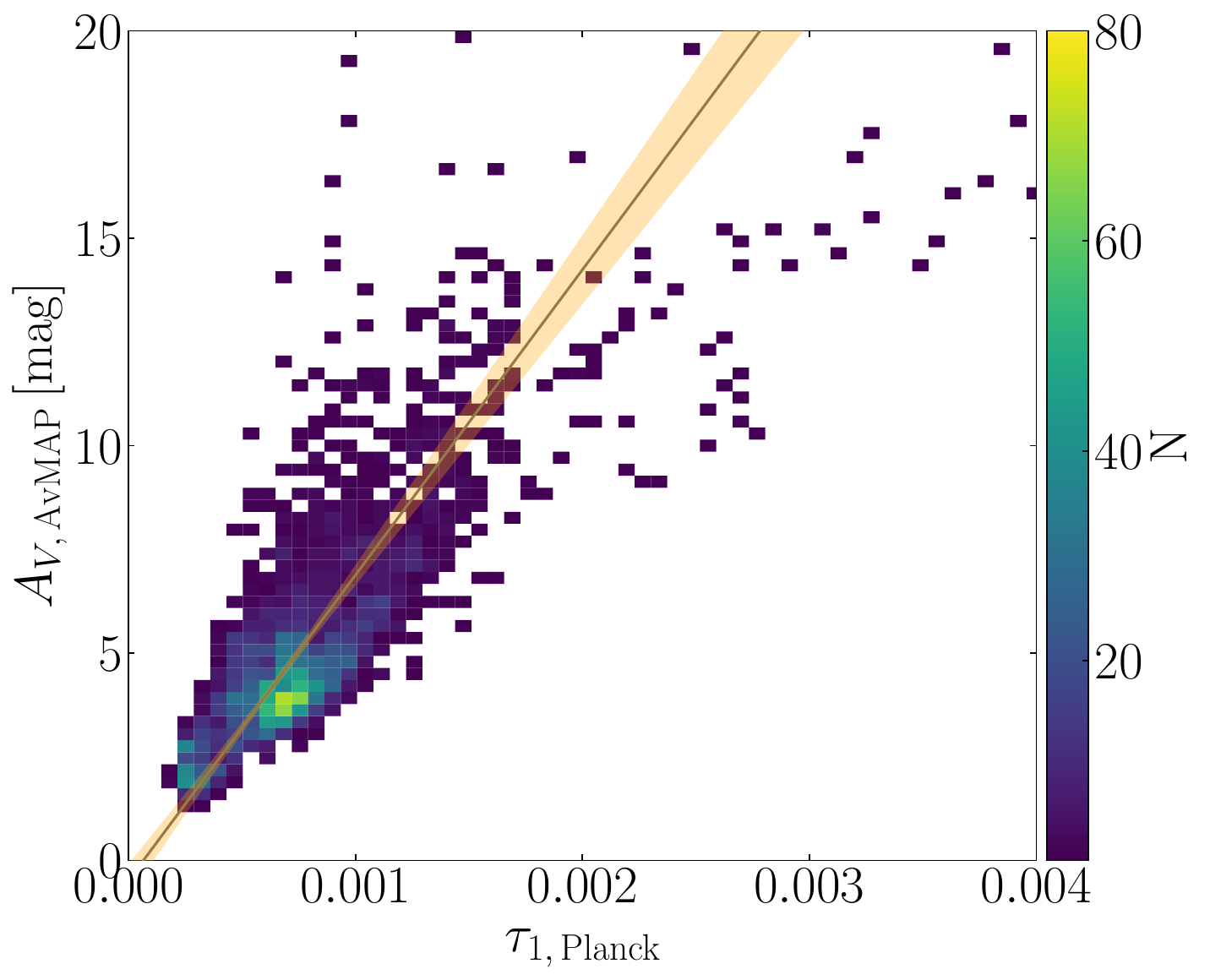}
\includegraphics[width=8.5cm]{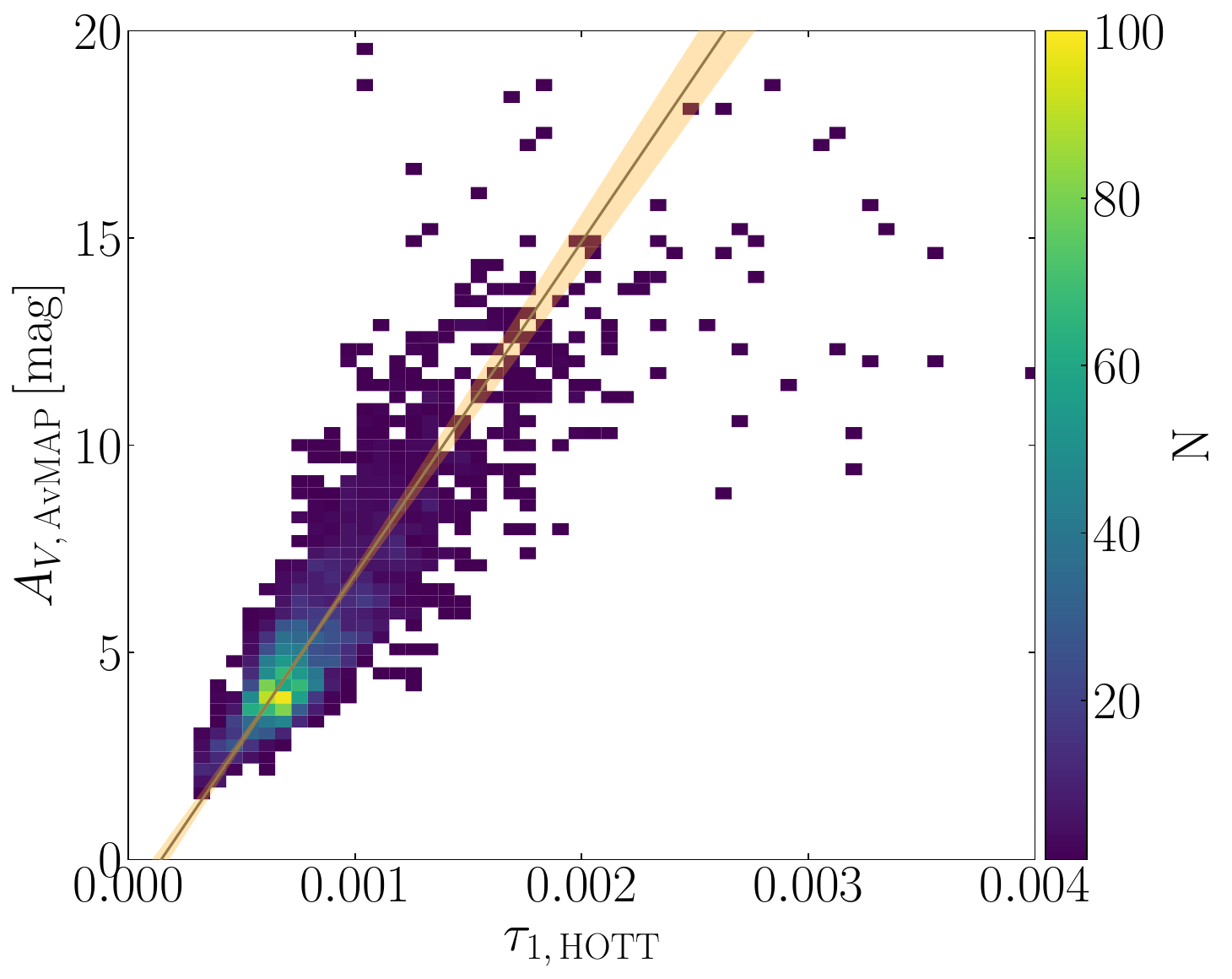}
\caption{2D histograms of $A_{V,\,\mathrm{AvMAP}}$ vs.\ $\tauu_\mathrm{,\,Planck}$ (left) and $A_{V,\,\mathrm{AvMAP}}$ vs.\ $\tauu_\mathrm{,\,HOTT}$ (right) toward W3. The optimal bisector fit for each is shown with a solid grey line with the fit uncertainty as solid yellow shading: $A_{V,\,\mathrm{AvMAP}} = (7.36 \pm 0.13) \times 10^3\, \tauu_\mathrm{,\,Planck} - (0.48 \pm 0.08)$ (left) and $A_{V,\,\mathrm{AvMAP}} = (8.04 \pm 0.14) \times 10^3 \, \tauu_\mathrm{,\,HOTT} - (1.2 \pm 0.1)$ (right).
}
\label{fig:AvMAP-vs-tau}
\end{figure*}

\subsection{Two lower resolution tracers of dust toward W3}
\label{subsec:lowerresdust}

\subsubsection{Reddening and extinction}

As a tracer of the total dust extinction, we adopted the `AvMAP' map {\tt avhk\_w3\_hso.fits} prepared by S.\ Bontemps that was used to plan the coverage of the observations of W3 with the Herschel Space Observatory \citep{Motte2010, RI2013}. The angular resolution is $\sim2{\farcm}1$ with $1{\farcm}5$ pixels. The `AvMAP' technique is based on the average reddening of background stars using 2MASS data \citep{Lombardi2001, Cambresy2002} in combination with predictions from the BGM to remove the effect of foreground stars \citep{Schneider2011}. Although it is a NIR measure, it is expressed as \AV. The AvMAP can be converted to a map $E(H-K)_{\mathrm{AvMAP}}$, a measure closer to the original data, using $\AV/E(H-K)=15.86$ \citep{Martin2012}.

For our fiducial extinction curve, $\AV/E(H-K)=15.33$, and so we adjusted the original map values in {\tt avhk\_w3\_hso.fits} to be consistent with that (a small systematic decrease by 0.9666).  We call this $A_{V,\,\mathrm{AvMAP}}$.

To produce a map of reddening $E\HG$ we used the above $E(H-K)_{\mathrm{AvMAP}}$ and multiplied it by $E\HG/E(H-K) = 1.86$ calculated from our fiducial extinction curve.  We call this $E\HG_{\mathrm{AvMAP}}$.

\subsubsection{Thermal dust emission optical depth}

The thermal dust emission optical depth ($\tau_\nu$) can be expressed as the product of the dust mass column density ($M_\mathrm{dust}$) and the dust emission (or absorption) cross section per unit mass ($\kappa_\nu$). Therefore, $\tau_\nu$ should be correlated with tracers of dust extinction, as has been observed for the high-Galactic latitude diffuse ISM \citep[e.g.,][]{Sturch1969, Knapp1974, Bohlin1978, Mirabel1979, Diplas1994, Rachford2009, Liszt2014, Green2018} as well as dense molecular clouds \citep[e.g.,][]{planck2011-7.13, Martin2012, Roy2013, Lombardi2014, Zari2016, Singh2022}.

The all-sky Planck dust map at 353 GHz ($\tau_{353}$) is widely used to estimate reddening \citep{planck2013-p06b} and can be extended to extinction by adopting an extinction curve. This has an angular resolution of 5\arcmin. We converted the $\tau_{353}$ map to $\tauu_\mathrm{,\,Planck}$ at a frequency of 1\,THz for consistency with $\tauu_\mathrm{,\,HOTT}$ for W3 \citep{Singh2022}, using their common assumption of a power law $\tau_\nu = \tau_0 (\nu/\nu_0)^\beta$ and the Planck $\beta$ map \citep{planck2013-p06b}. 

Note that $\tauu_\mathrm{,\,HOTT}$ has a significantly improved angular resolution of 36\farcs7. 

%%%%%%%%%%%%%%%%%%%%%%%%
%                                                                       %
%                  BONTEMPS VS TAU                   %
%                                                                       %
%%%%%%%%%%%%%%%%%%%%%%%%

\subsection{Comparison of Extinction and Thermal Emission Dust Maps toward W3}
\label{subsec:avtau}

To quantify the relationship of $A_{V,\,\mathrm{AvMAP}}$ and $\tauu$, we first brought the $A_{V,\,\mathrm{AvMAP}}$ map to the same resolution and projection as $\tauu_\mathrm{,\,Planck}$ and used a mask to restrict our comparison to be toward W3. Figure \ref{fig:AvMAP-vs-tau} (left) shows that $A_{V,\,\mathrm{AvMAP}}$ and $\tauu_\mathrm{,\,Planck}$ are strongly correlated. Systematic errors likely dominate over measurement errors and so we use a linear bisector fit \citep{Isobe1990, Akritas1996} rather than a simple linear least squares method. For the range $A_{V,\,\mathrm{AvMAP}}<20$\,\mags\ and $\tauu_\mathrm{,\,Planck}<0.002$ where $A_{V,\,\mathrm{AvMAP}}$ is less saturated, the best fit and its uncertainty are shown in the figure and caption. The dispersion around the fit is 0.99\,\mags. Beyond $A_{V,\,\mathrm{AvMAP}} \sim 10$\,\mags\, the fit should be used with caution. 

In the case of $\tauu_\mathrm{,\,HOTT}$, we smoothed its map to the $A_{V,\,\mathrm{AvMAP}}$  resolution. Figure \ref{fig:AvMAP-vs-tau} (right) shows $A_{V,\,\mathrm{AvMAP}}$ vs.\ $\tauu_\mathrm{,\,HOTT}$, including its best fit, which is not surprisingly quite similar. The dispersion around the fit, 0.15\,\mags, is smaller and the fits appears valid to higher extinctions of $A_{V,\,\mathrm{AvMAP}} \sim 15$\,mag.

%%%%%%%%%%%%%%%%%%%%%%
%                                                                %
%                  E(H-[4.5]) VS TAU                 %
%                                                                %
%%%%%%%%%%%%%%%%%%%%%%

\begin{figure*}[t!]
\centering
\includegraphics[width=8.5cm]{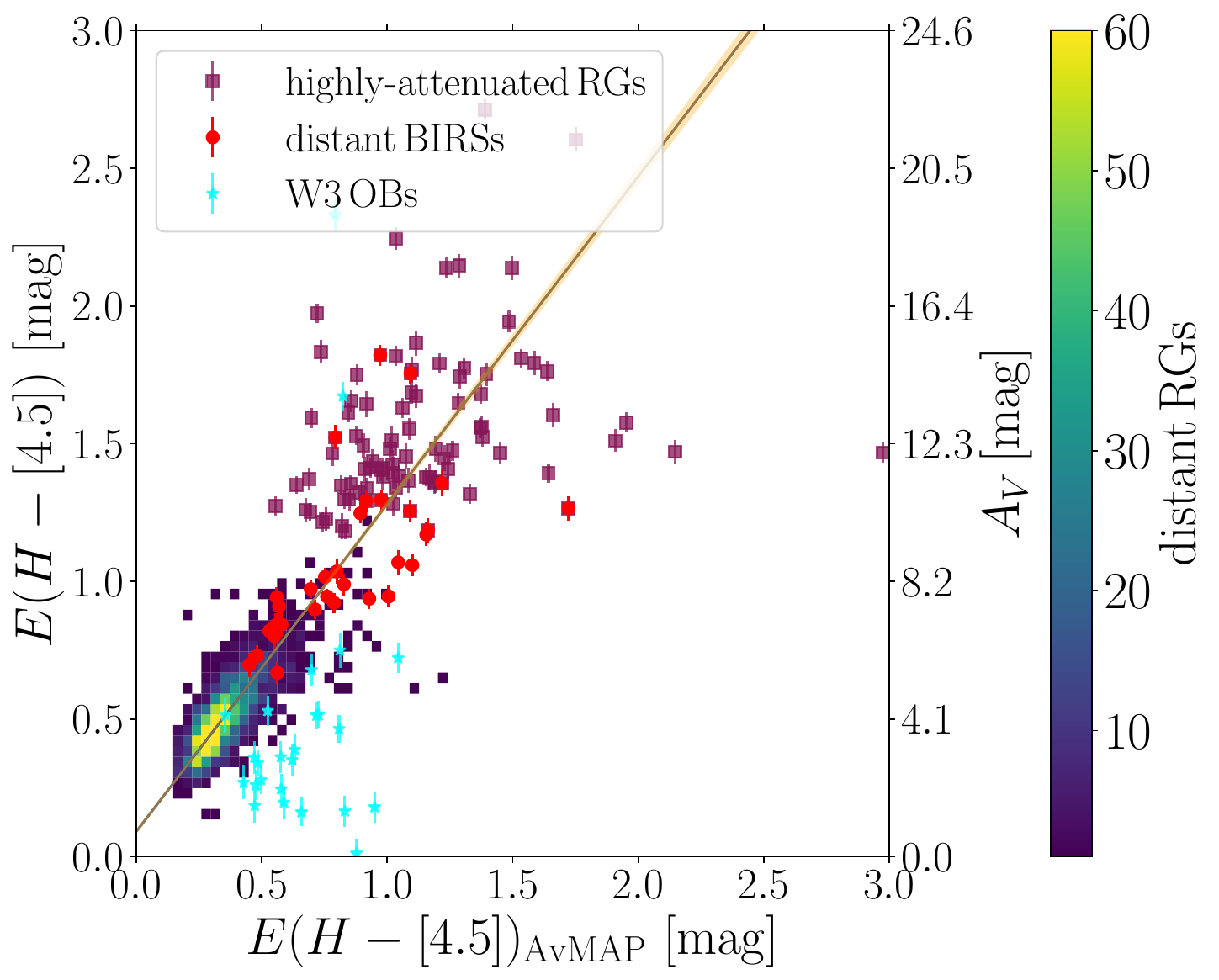}
\includegraphics[width=8.5cm]{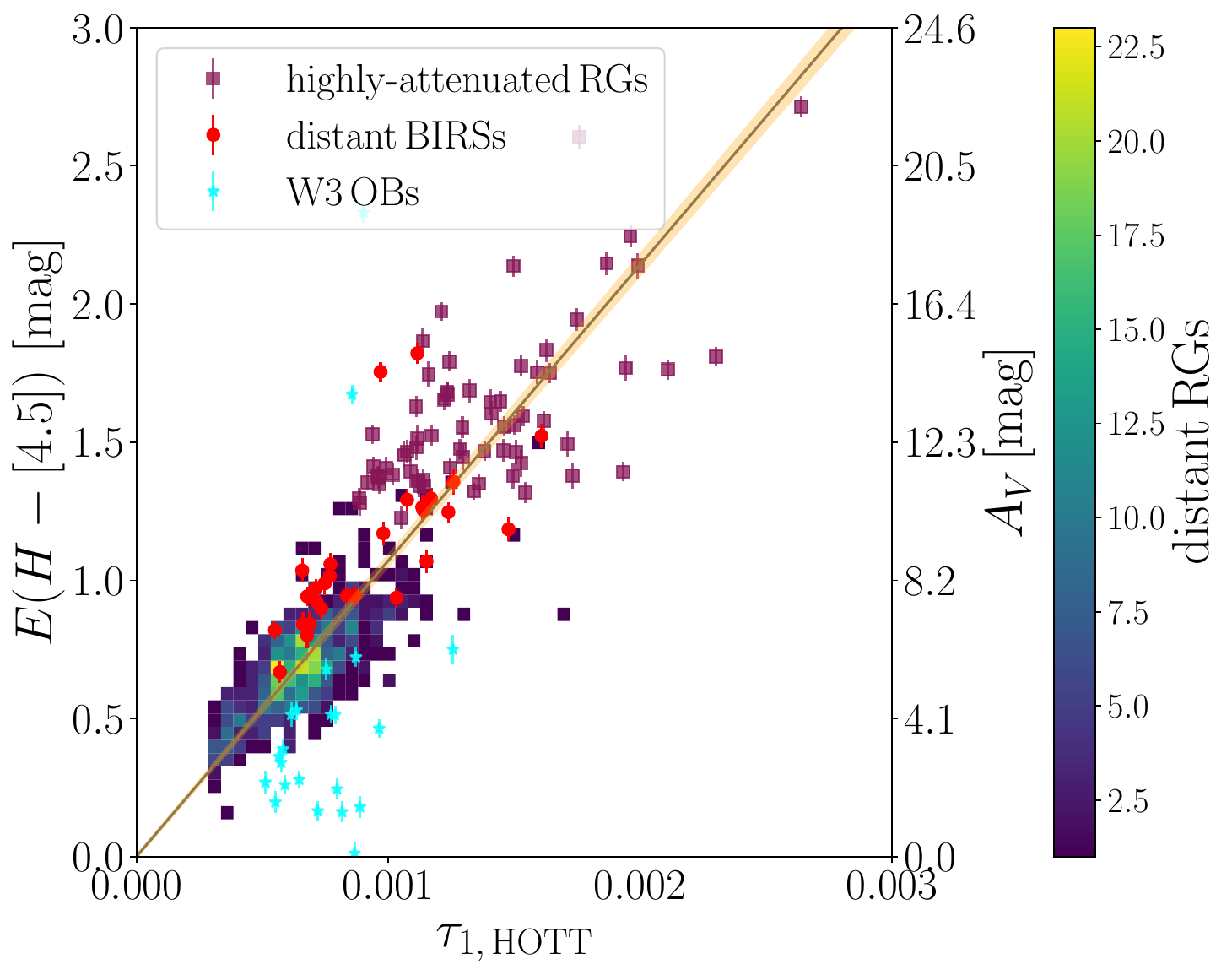}
\caption{(Left) RJCE-derived $E\HG$ for stars toward W3 vs.\ $E\HG_\mathrm{AvMAP}$.  Distant RGs (2D histogram) and highly-attenuated RGs lacking Gaia measurements (mahogany squares) are identified (see text for selection). Distant BIRS behind the cloud (red circles) and spectroscopically-confirmed OB stars at the distance of W3 (cyan stars) are shown (see text). The y-axis on the right shows \AV\ assuming our fiducial extinction curve 
(Section \ref{sec:extinction-curve}).
(Right) RJCE-derived $E\HG$ vs.\ HOTT \tauu.}
\label{fig:EHG2_stars_vs_AvMAP}
\end{figure*}

\subsection{Comparison of Individual Stellar Extinction Measurements and Dust Maps}
\label{subsec:stellarprobes}

The above dust maps quantify the total extinction along a given line of sight, whereas $E\HG$ traces the dust foreground to an individual star. Therefore, we concentrated on intrinsically bright RGs that according to the Besan\c{c}on Galactic model are predominantly at large distances (see rectangle in Figure \ref{fig:hrd-besancon} in Section \ref{subsubsec:HRD-BGM}), so that these two independent measurements sample consistent dust columns. We selected distant RGs using our observational \hrd\ via $-8 \leq \MKs \leq -1\,\mags$ and $0.35 \leq \JK_0 \leq 1.15\,\mags$ (see rectangle in the Figure \ref{fig:hrd-obs}) together with a parallax criterion of $\varpi < 0.3\,\mas$ ($d > 3.33\,\kpc$) and a parallax uncertainty of $<0.1\,\mas$. The selected distant RGs and BIRSs have sufficiently high extinction such that $E\HG$ is not particularly sensitive to the choice of intrinsic color; we adopted $\HG_0 = 0.08$. For the OB stars used for comparison, we adopted our modified RJCE technique.

Given the Gaia optical sensitivity \citep{Gaia2021a}, 
the parallax requirements limited the extinction range of our sample.  
Highly-attenuated stars, even if lacking Gaia measurements, were identified using the reddened RG band in the 2MASS-GLIMPSE CCD (Figure \ref{fig:RJCE_JK-HG2}), along with the color criteria $\JK>2.5\,\mags$ and $\HG>1.25\,\mags$. 

\subsubsection{Stellar vs.\ AvMAP Color Excess}

We assess how well our RJCE-derived $E\HG$ stellar measurements compare to those from AvMAP in Figure \ref{fig:EHG2_stars_vs_AvMAP} (left), including distant RGs (2d histogram) and highly-attenuated RGs lacking Gaia measurements (mahogany squares). There is a clear linear correlation that is extended by the highly-attenuated RGs as expected, albeit with a larger scatter. We measure a best-fit slope of $1.19 \pm 0.05$ and a small y-intercept of $0.09 \pm 0.02$. The slope should be close to unity, given that our individual RGs could be among those used to produce the AvMAP. The deviation from unity might arise from systematics in the AvMAP technique compared to individual probes or in the conversion of AvMAP to $E\HG_\mathrm{AvMAP}$.
The highly-attenuated RGs are slightly more concentrated above the line of best fit, possibly indicating that AvMAP is becoming saturated, missing the high peaks that can be probed by these selected point sources.

In our analysis of BIRS parallax-based cloud membership (see Section \ref{subsec:parallax-membership}), we found that 36 BIRSs are behind the cloud. These distant BIRSs (red circles) follow the empirical correlation because they adequately trace the entire line-of-sight dust column. By contrast, the spectroscopically-confirmed OB stars at the distance of W3 (cyan stars) do not trace the full dust column even within the cloud and so fall well below the correlation.\footnote{The lone OB star \citetalias{Kiminki2015} 65 that is well above the empirical correlation amongst the highly-attenuated RGs has a MIRE that biases our stellar color excess measurement to a greater value.}

\subsubsection{Stellar Color Excess vs.\ HOTT}

Having established a strong correlation between our stellar $E\HG$ measurements and \AV, we calibrate the relationship between stellar color excess and dust optical depth using the latest HOTT dust map. Figure \ref{fig:EHG2_stars_vs_AvMAP} (right)
shows $E\HG$ vs.\ $\tauu_\mathrm{,\,HOTT}$ using the same populations of stars as those in Figure \ref{fig:EHG2_stars_vs_AvMAP}, measuring a best fit of $E\HG = (1.07 \pm 0.04) \times 10^3\, \tauu_\mathrm{,\,HOTT} + (0.00 \pm 0.02)\,\mags$. These results are not strongly dependent on the parallax cut: while lowering the upper limit in parallax increases the number of stars that are below the best-fit line, the slope and y-intercept are consistent within a $3\sigma$ uncertainty range. 

\subsection{Consideration of Systematic Errors}

The best-fit y-intercepts of the $\AV_\mathrm{AvMAP}$ vs.\ $\tauu_\mathrm{,\,HOTT}$ and $E\HG$ vs.\ $E(H-[4.5])_{\rm AvMAP}$ correlations imply an underestimated \AV\ measurement that may be due to zero-point levels in AvMAP. This possibility is consistent with the zero y-intercept measured for $E\HG$ vs.\ $\tauu_\mathrm{,\,HOTT}$ in Figure \ref{fig:EHG2_stars_vs_AvMAP}, which does not include AvMAP. 
 
Using both the Planck and HOTT dust maps, we found evidence that the slope of $\AV_\mathrm{AvMAP}/\tauu$ and $E\HG/\tauu$ varied by $\sim 10\%$ over the W3 field. In particular, the slope is higher in the northeast corner of W3 compared to the southwest corner. This spatial dependence in the slope may be related to changes in dust $T$ or $\beta$. 

While we target distant RGs to probe consistent dust columns between $E\HG$ and $\tauu_\mathrm{,\,HOTT}$, there remain systematic differences between these measurements. While $E\HG$ is a pencil-beam measurement, $\tauu_\mathrm{,\,HOTT}$ is interpolated over coarse $14\arcsec$ pixels. Furthermore, the ISM is a turbulent medium with small-scale density fluctuations that likely cause spatial variations in $\tauu_\mathrm{,\,HOTT}$ on a scale significantly smaller than the $36\farcs7$ angular resolution of the HOTT continuum map. See also section 11 and 12 in \citet{Singh2022} for a discussion of systematics in deriving $\tauu_\mathrm{,\,HOTT}$. 

\subsection{Environmental changes}

There is persuasive evidence that the ratio of thermal dust optical depth to extinction changes with the environment. For instance, in the Taurus molecular cloud, the dust optical depth is well correlated with NIR extinction but this ratio (i.e., the slope of the correlation) is about two times higher in the molecular phase compared to the atomic phase \citep{planck2011-7.13}. Similarly, the submillimeter ratios toward 14 small subregions surrounding the Vela molecular cloud were found to be $2-4$ times higher compared to the local high-Galactic latitude atomic ISM \citep{Martin2012}. Additional evidence for variation is provided using the ratios reported in \cite{Lombardi2014}, \cite{Zari2016}, and \cite{Singh2022}.
There is also evidence that the optical depth vs.\ extinction relations may be nonlinear at high column densities \citep{Roy2013}, although this might be complicated by the common assumption of a single dust temperature along the line of sight \citep{Roy2014}. 
Such collective evidence indicates that dust is evolving in these high-density regions, as expected from dust grain evolution \citep[e.g.,][]{Ossenkopf1994, Stognienko1995} and that a single dust model will not necessarily apply to all clouds or environments.

While the submillimeter $\tau_\nu$ is often converted to gas column density, the conversion factor is poorly understood \citep{Singh2022} and furthermore such conversion factors (e.g., hydrogen column density per unit $\tau_\nu$, color excess, or extinction) might change at the highest densities of a molecular cloud \citep{Roy2013}. To avoid any further assumptions, we have explored only empirical dust-to-dust relationships. 

%%%%%%%%%%%%%%%%%%%%%%%
%                                                                   %
%                     CONCLUSIONS                    %
%                                                                   %
%%%%%%%%%%%%%%%%%%%%%%%

\section{Conclusions and Summary}\label{sec:conclusions}

In this work, we explore reddened stellar populations and dust extinction toward the W3 giant molecular cloud using multi-band photometry and Gaia DR3 astrometry. In particular, we examine the spectroscopic classification of Bright Infrared Stars (BIRS; \citealp{Elmegreen1980}) and use the tools developed here to search for new OB star candidates in W3. The following summarizes our results.

% BIRS
\textit{Generic BIRS properties:} We found that many of the BIRS positions were significantly affected by an inaccurate plate solution and provided revised Gaia DR3 coordinates. We identified 17 BIRS `imposters' that are neither particularly bright nor red based on modern ATLAS photometry, contradicting their original `BIRS' designation. Comparing the placement of the BIRS to reddened bands of OB stars (OBs) and red giants (RGs) on various photometric-based diagrams allowed us to classify 98 BIRS as 68 RGs, 12 OBs, and 18 low-mass main sequence (MS) stars.

% parallax + field stars
\textit{Parallax and field OBs:} We examined the Gaia DR3 parallax distribution of spectroscopically confirmed OB stars, carefully assessing systematic errors in the parallax zero point ($Z$), finding a central W3 parallax of $0.511\,\mas$ that is consistent with the W3(OH) maser parallax \citep{Xu2006}. We identified seven B stars behind W3, conflicting with previous assumptions that all spectroscopic OB stars reside within the cloud \citep{Kiminki2015, Roman-Lopes2019}. 

% proper motion
\textit{Proper motion and runaway OBs:} Using high-precision OB stars at the W3 parallax, we measured a fiducial cloud proper motion of $\pmra = -0.82 \pm 0.1\,\masyr$ and $\pmdec = -0.5 \pm 0.1\,\masyr$ with a $3\sigma$ radius of $2.5\,\masyr$. We identified BIRS 6 (O6.5-O7 IV-V; RL 27 in \citealp{Roman-Lopes2019}) as a potential runaway star with a magnitude in the peculiar proper motion of $\sim 2.4\,\masyr$ from the direction of W3 Main. We also noted that RL 20 (O7.5 IV-V) as having a peculiar proper motion of $\sim 6.1\,\masyr$ opposite to RL 27. Given their opposite displacements relative to W3 Main toward the \HII\ region W3 H, BIRS 6 (RL 27) and RL 20 may comprise an OB runaway pair.

% SED fitting
\textit{SED fitting:} We fit extinction curves to the optical-to-NIR portion of SEDs using the `Astrodust' model \citep{Hensley2020} for $\lambda \geq$ 2\,\micron\ and the CCM model \citep{Cardelli1989} for $\lambda <$ 2\,\micron, adopting $\RV=3.6$ and $\powerlaw=1.8$. This enabled us to simultaneously measure distance moduli, foreground extinctions, and photometric-based spectral classifications. We reproduced the spectral types of \citet{Kiminki2015} OBs within a few sub-types, refining the spectral type of six OBs and 69 BIRS. In particular, we predicted the spectral classification of the high proper motion star BIRS 20 \citep{Elmegreen1980} to be an M6 V dwarf.

% OB candidates
\textit{Search for new OB star candidates in W3:} We used the tools that we developed here to search for new OB star candidates in W3, we identified three O-type and 79 B-type candidates with significant foreground extinction. We highlighted the importance of using CCDs to isolate the reddened MS \band\ and CMDs to select those that are the brightest. Our results yield a sample that is significantly less contaminated than other recent searches that (i) neglected to use CCDs to separate the reddened MS and RG bands \citep{Kiminki2015} and (ii) are based purely on Gaia astrometry \citep{Navarete2019}.  The 24 candidates in our list earlier the B3 V are also in the candidate list developed by \citet{Zari2021} using different criteria, and so should be priority targets.   

% E(H-[4.5]) vs tau
\textit{Empirical dust-to-dust relationships:} We established several dust-to-dust empirical relationships between distinct tracers of the dust columm:
\begin{enumerate}
\item Dust optical depth maps at 1\,THz: Planck ($\tauu_\mathrm{,\,Planck}$; \citealp{planck2013-p06b}) and HOTT ($\tauu_\mathrm{,\,HOTT}$ \citealp{Singh2022}).
\item Map products from AvMAP \citep{Schneider2011}): $A_{V,\,\mathrm{AvMAP}}$ and $E\HG_{\,\mathrm{AvMAP}}$.
\item Our color excess measurements for individual stars: $E\HG$.
\end{enumerate}

We found that $A_{V,\,\mathrm{AvMAP}}$ is well correlated with both $\tauu_\mathrm{,\,Planck}$ and $\tauu_\mathrm{,\,HOTT}$, with best-fit linear solutions of $A_{V,\,\mathrm{AvMAP}} = (7.36 \pm 0.13) \times 10^3\, \tauu_\mathrm{,\,Planck} - (0.48 \pm 0.08)$ and $A_{V,\,\mathrm{AvMAP}} = (8.04 \pm 0.14) \times 10^3 \, \tauu_\mathrm{,\,HOTT} - (1.2 \pm 0.1)$. 

We established that our individual-star $E\HG$ is well correlated with the map product $E\HG_{\,\mathrm{AvMAP}}$ (slope of $1.19 \pm 0.05$ and y-intercept of $0.09 \pm 0.02$). 

We measured a best-fit relationship directly between $E\HG$ (individual stars) and $\tauu_\mathrm{,\,HOTT}$ (highest resolution dust map) as $E\HG = (1.07 \pm 0.04) \times 10^3\, \tauu_\mathrm{,\,HOTT} + (0.00 \pm 0.02)\,\mags$, consistent with the relationships above. 

We discussed the possible connection between the y-intercepts and AvMAP, spatial dependences, and systematic errors.

\textit{Future work:} There are several avenues for future work, three of which we highlight here.

% KR 140
VES 735 is the young O8.5 V(e) exciting star of KR 140 \citep{Kerton1999, Ballantyne2000, Kerton2001, Kerton2008} that we found to have an inconsistent parallax with W3. 
If KR 140 is indeed part of the W3 molecular cloud complex, then perhaps the parallax of VES 735 is incorrect?
Searching for evidence of a young star cluster that formed around the same time as VES 735 and establishing the parallax and proper motion of said cluster would help to shed light on whether KR 140 shares the same distance and 3D motion as W3. 

% OB candidates
Following up with our OB star candidates with targeted spectroscopy is likely to significantly expand the number of OB stars in W3, particularly toward the west end of the cloud. Once confirmed, these young stars could be used to help establish a more thorough analysis of the cloud distance and dynamics. Furthermore, our selection criteria can be used to identify new OB star candidates in other (giant) molecular clouds.

% temperature correlation of submillimeter ratio
Lastly, we found evidence that the ratio between $E\HG$ and \tauu, including both Planck and HOTT dust maps, may have a spatial dependence over the cloud. A detailed analysis of the possible spatial dependence and whether this correlates with properties like dust temperature, $\beta$, or column density would be informative for dust evolution models.

%%%%%%%%%%%%%%%%%%%%%%%%%
%                                                                          %
%                 ACKNOWLEDGMENTS                   %
%                                                                          %
%%%%%%%%%%%%%%%%%%%%%%%%%

\section{acknowledgments}

We thank the anonymous referee for their constructive feedback on our paper. This research was supported by the Natural Sciences and Engineering Research Council (NSERC) of Canada. JLC acknowledges support from the Ontario Graduate Student Scholarship. Some of this work was carried out by co-authors while summer interns or senior undergraduate researchers at CITA, University of Toronto. For their contributions re data products as mentioned in the text, we thank S.\ Bontemps, J.J.\ Kavelaars, D.\ Lang, N.\ Law, E.\ Magnier, A.\ Rivera-Ingraham, C.\ Sasaki, S.\ Sivanandam, and H.K.C.\ Yee. This work made use of the NASA Astrophysics Data System. 

% software
\software{Astropy \citep{Astropy2013,Astropy2018}, TOPCAT \citep{Taylor2005}, matplotlib \citep{Hunter2007}, numpy \citep{Harris2020}, scipy \citep{Virtanen2020}.}

%%%%%%%%%%%%%%%%%%%%%
%                                                              %
%                    APPENDIX                        %
%                                                              %
%%%%%%%%%%%%%%%%%%%%%

\appendix
\counterwithin{table}{section}
\counterwithin{figure}{section}

%%%%%%%%%%%%%%%%%%%%%%%%
%                                                                       %
%                   BIRS ASTROMETRY                  %
%                                                                       %
%%%%%%%%%%%%%%%%%%%%%%%%

\section{Modern Astrometry and Photometry of the BIRS}\label{app:astrometry}

To obtain updated astrometric positions of the BIRS, we identified the original BIRS positions by comparing digitized versions of the annotated photographic images (see Section \ref{subsec:astrometry}) to 2MASS $J$ images. The original BIRS positions were often significantly offset from the nearest 2MASS objects, so we used the digital WCS and nearby asterisms to confirm the identifications of corresponding 2MASS stars. We verified that the stars were visibly red using IPHAS $r$ and $i$ images and that the positional offsets were consistent in a given region of the digitized photographic image.

We used the General Catalog Query Engine as part of the NASA/IPAC Infrared Science Archiv\footnote{\url{http://irsa.ipac.caltech.edu/cgi-bin/Gator/nph-scan?submit=Select\&projshort=2MASS}} to identify the target BIRS and obtain its 2MASS coordinates. These coordinates were matched to Pan-STARRS to inspect their $griz$ optical images using the Pan-STARRS image cutout service\footnote{\url{http://ps1images.stsci.edu/cgi-bin/ps1cutouts}} to verify colors and coordinates. Once we were confident that we had identified the correct stars in 2MASS and Pan-STARRS, we matched their coordinates to Gaia DR3 for their final positions.

Table~\ref{table:birs-positions} lists the Gaia J2000 positions of the BIRS and their respective offsets. 
Also tabulated are the original $R$ and $I$ photographic magnitudes \citep{Elmegreen1980} and the modern ATLAS $r$ and $i$ AB photometric magnitudes
(Section \ref{subsec:RI-photometry}).

\input{tables/tableA1.tex}

%%%%%%%%%%%%%%%%%%%%%%%%%
%                                                                          %
%                quality filtering of photometry            %
%                                                                          %
%%%%%%%%%%%%%%%%%%%%%%%%%

\section{Quality control filtering of photometric data}
\label{app:qualityfilter}

Because of the usual naming of infrared and submillimeter passbands, all passbands will be designated here using \micron\ rather than disparately nm or \AA.

\subsection{Optical}

ATLAS is an all-sky astrometric and photometric reference catalog (Refcat2) that combines data from Gaia, Pan-STARRS, SkyMapper, the ATLAS Pathfinder survey, and reprocessed APASS images for optical-IR photometry that have been transformed to the Pan-STARRS passbands \citep{Tonry2018}. The Refcat2 magnitudes are weighted means of the contributing catalogs and strongly resemble Pan-STARRS photometry north of declination $\delta=-30\deg$ and fainter than magnitude $m=14\,\mags$ (AB) \citep{Tonry2018}. The catalogs comprising the ATLAS data required numerous conditions for quality photometry \citep{Tonry2018}, so we used a simple requirement of ${\SNR}>20$ for any given passband, finding consistency with the Pan-STARRS PSF photometry.

Pan-STARRS is an optical-NIR survey of the sky north of declination $\delta=-30\deg$ in the passbands 
$g_{\rm P1}$ (0.481\,\micron), $r_{\rm P1}$ (0.617\,\micron), $i_{\rm P1}$ (0.752\,\micron), $z_{\rm P1}$ (0.866\,\micron), and $y_{\rm P1}$ (0.962\,\micron) using a 1.8 m telescope in Hawaii \citep{Chambers2016}. The data have 5$\sigma$ detection levels of 23.3, 23.2, 23.1, and 22.2 {\mags} (AB) with saturation levels of 14.5, 15, 15, and 14 {\mags} (AB) in the $g_{\rm P1}$, $r_{\rm P1}$, $i_{\rm P1}$, $z_{\rm P1}$, and $y_{\rm P1}$ bands, respectively. We used the PSF photometry and required that ${\tt qualityFlag}{<}63$, ${\tt XFlags}{<}115000$, where {\tt X} designates the $grizy$ passbands, and ${\SNR}>20$ for any given band.

IPHAS is an optical survey of the northern Galactic plane in narrow-band {\halpha} emission 
(0.6568\,\micron) and broadband optical photometry in the  $r'$ (0.6240\,\micron) and $i'$ (0.7743\,\micron) passbands using the 2.5\,m INT \citep{Drew2005}. This survey has a higher saturation level than that of Pan-STARRS and is therefore sensitive to brighter sources. We required that ${\tt fieldGrade}={\tt A}$\texttt{++} or ${\tt A}$\texttt{+}, ${\tt XSaturated}={\tt False}$, and ${\tt Xdeblend}={\tt False}$, where ${\tt X}$ designates the $r'$ or $i'$ passbands.

APASS is an all-sky optical-NIR survey in Johnson $B$ 
(0.4448\,\micron), $V$ (0.5505\,\micron), and Sloan $u'$ (0.3557\,\micron), $g'$ (0.4825\,\micron), $r'$ (0.6261\,\micron), $i'$ (0.7672\,\micron), and $z'$ (0.9097\,\micron) passbands using twin 20\,cm astrographs in New Mexico and Chile that provide homogenous all-sky optical photometry to bridge the gap between existing optical surveys \citep{Henden2014, Henden2016}. We required that $\SNR>20$ in a given band.

We made use of unpublished exploratory imaging data taken for us in the KR 140 region by H.K.C.\ Yee with the Canada France Hawaii Telescope (CFHT) 12K camera in December 2001 and January 2002 in the B, V, R (Mould), and I filters. These included some short exposures to mitigate saturation for brighter stars.  We merged original reductions of individual images, courtesy of J.J.\ Kavelaars and E.\ Magnier using a standard CFHT pipeline.

We reduced Sloan $r$, $i$, and $z$ imaging data from the experimental robotic Dunlap Institute Telescope (DIT) located in New Mexico taken for 10 pointings in December 2011 and January 2012 by N.\ Law and S.\ Sivinandam.  We wrote a standard pipeline for reduction. While this DIT photometry included short exposures to mitigate saturation, it was ultimately superseded by ATLAS.

Gaia DR3 \citep{Gaia2021a} provides optical photometry at 
$G$ (0.330--1.050\,\micron), $G_{BP}$ (0.330--0.680\,\micron), and $G_{RP}$ (0.640--1.050\,\micron) passbands, obtained using a 1.45 m space telescope \citep{Gaia2016}. The Gaia $(G_{BP}-G_{RP})$ colors are measured using spectra dispersed over a sky area of $2{\times}3\arcsec$ \citep{Riello2021} and are likely unreliable for crowded sources \citep{Rybizki2022}. The white-light $G$-band has a broad bandpass and red objects like the BIRS can cause an increase in the effective wavelength, causing a bias to bluer $(G_{BP} - G_{RP})$ colors \citep{Riello2021}. We therefore make use of Gaia astrometry (see Section \ref{subsec:astrometry}) but not photometry.  Gaia photometry is used by \citet{Zari2021} for their OB star target selection for SDSS V (Section \ref{subsec:othersearch}).

\subsection{Near infrared (NIR)}

2MASS is an all-sky NIR survey of the $J$ 
(1.235\,\micron), $H$ (1.662\,\micron), and {\Ks} (2.159\,\micron) passband using two 1.3 m telescopes in the northern (Arizona) and southern (Chile) hemispheres \citep{Skrutskie2006}. The data have 10$\sigma$ detection levels of 15.8, 15.1, and 14.3 {\mags} (Vega) with saturation levels of 9, 8.5, and 8.0 {\mags} (Vega) in the $J$, $H$, and {\Ks} bands, respectively. We required that ${\tt cc\_flg}=0$, ${\tt bl\_flg}<2$, ${\tt ph\_qual}=A$ or $B$, and $\SNR>20$ in any given band.

\subsection{Mid infrared (MIR)}

The GLIMPSE-360 survey \citep{Whitney2008} extended the original GLIMPSE survey \citep{Benjamin2003} from the inner Galactic plane ($295\deg<{\ell}<65\deg$) to the outer Galaxy ($65\deg<{\ell}<265\deg$) in the two short-wavelength IRAC bands of 
$G1$ (3.6\,\micron) and $G2$ (4.5\,\micron) \citep{Fazio2004} as part of the Spitzer Warm Mission Exploration Science program.

We also used the four-band Spitzer/IRAC MIR photometry from PI survey imaging reduced by A.\ Rivera-Ingraham for her study of YSOs in W3 \citep{RI2011}. These agree well with the GLIMPSE-360 data that are available only in the two shorter wavelength bands. IRAC has better spatial resolution than the corresponding WISE data, making it useful in crowded regions.

WISE is an all-sky MIR survey that obtained data using a 40 cm aperture space telescope in the 
$W1$ (3.4\,\micron), $W2$ (4.6\,\micron), $W3$ (12\,\micron), and $W4$ (22,\micron) passbands with an angular resolution of 6{\farcs}1, 6{\farcs}4, 6{\farcs}5, and 12{\farcs}0, respectively \citep{Wright2010}. The spacecraft continued to collect data in the two short-wavelength bands ($W1$ and $W2$) after the depletion of cryogenic coolant, renamed the Near-Earth Object Wide-field Infrared Survey Explorer (NEOWISE; \citealp{Mainzer2011, Mainzer2014}). The allWISE mission combines data from the cryogenic WISE and post-cryogenic NEOWISE phases \citep{Cutri2014}. We required that ${\tt ph\_qual}=A$ or $B$ and ${\tt wXsat}=0$, where {\tt X} designates the W$1{-}4$ bands, and $\SNR>20$ in a given band.

unWISE is a catalog of five years worth of reprocessed WISE photometry in the two short-wavelength bands ($W1$, $W2$) with deeper photometry at higher resolution and improved modeling in crowded fields \citep{Lang2014, Schlafly2019}. 

%%%%%%%%%%%%%%%%%%%%%%%%%%%
%                                                                                %
%                HIGH PROPER MOTION STARS           %
%                                                                                %
%%%%%%%%%%%%%%%%%%%%%%%%%%%

\section{High Proper Motion Field Star BIRS 20}\label{subsec:BIRS20}

BIRS 20 has the greatest proper motion amongst the foreground stars. Using data from a nearly 60-year baseline, including nine epochs of the multi-color DSS, our recalibration of \citet{Elmegreen1980}, 2MASS, IPHAS, Spitzer/IRAC (GLIMPSE360), and Pan-STARRS, we measured both BIRS 20 and a single comparison star in the nearby asterism to derive the relative proper motion ($\mu_{\rm RA}$, $\mu_{\rm Dec}$) of ($261.4, -222.9$)\,{\masyr}~, with an uncertainty of about 1\,{\masyr}.  There is a small correction, of order the uncertainty, from the proper motion of the comparison star, ($-1.977 \pm 0.069, -0.629 \pm 0.105$), to be added to the above, yielding ($259.4, -223.5$)\,{\masyr} for the corrected BIRS 20 proper motion. This agrees with the subsequent much more accurate Gaia DR3 values of ($258.13 \pm 0.04, -225.47 \pm 0.06$)\,\masyr\ (Table \ref{table:birs-results}).
The Gaia DR3 J2000 coordinates are cited in Table~\ref{table:birs-positions}. 

The Gaia DR3 zero point corrected  $\varpi - Z = 48.36\,{\pm}\,0.05\,\mas$ (Table \ref{table:birs-results}) corresponds to a parallactic distance of $\approx20.7\,\pc$ (geometric distance $20.60\,{\pm}\,0.04\,\pc$ \citep{Bailer-Jones2018}).  The corresponding tangential motion is about 34~km~s$^{-1}$ (Equation \ref{eq:vtrans}).

In Section \ref{subsec:2MASS-CCD} we found that the 2MASS colors of BIRS 20 were consistent with M6 V.  Here, we refine our spectral classification using the \citet{Pecaut2013} version 2022.04.16 full suite of intrinsic colors and absolute magnitudes of M dwarfs,  along with the Gaia parallax, to predict the apparent magnitudes for comparison to the full SED defined by our observed photometric magnitudes.  Given the proximity of the star, we assumed that the extinction, which would impact the optical data most, is negligible.

\begin{figure}[t!]
\centering
\includegraphics[width=8.5cm,trim = 0 0 0 -1.5cm]{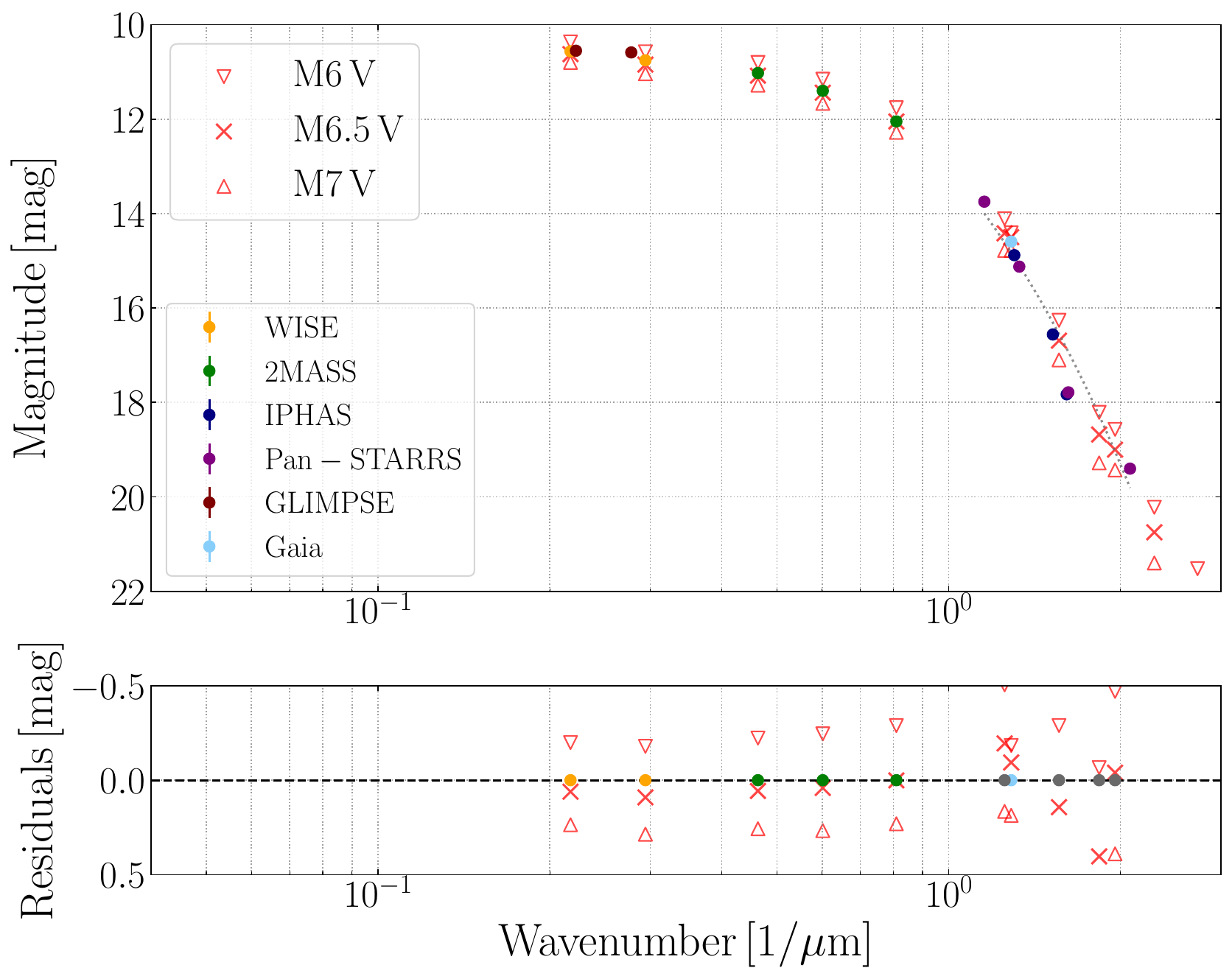}
\caption{Top: BIRS 20 SED and predicted apparent magnitudes, given the parallax of BIRS 20, for three late M spectral types (crosses and triangles -- see legend and text). Bottom:  residuals between the observed magnitudes (circles and interpolated dotted line in the optical) and predicted magnitudes (crosses and triangles).
}
\label{fig:BIRS20_SED}
\end{figure}

Figure \ref{fig:BIRS20_SED} shows the data (top panel) and the residuals (bottom). Because the optical filters differ, we measured the residuals in the optical relative to the interpolated dotted line passing near the data, which ignores a possible kink in the SED at $r$ \citep{Covey2007}. The optimal solution for spectral classification is M6.5 V. 

%%%%%%%%%%%%%%%%%%%%%%%%%%%
%                                                                                %
%                           MODIFIED RJCE                        %
%                                                                                %
%%%%%%%%%%%%%%%%%%%%%%%%%%%

\section{Modified RJCE Extinction}
\label{subsec:RJCEex}

In constructing the \hrd\ in Section \ref{subsec:HRD}, and in other applications, we need to evaluate the extinction and color excesses of all stellar populations. To achieve this we developed a modification of the RJCE technique.

The RJCE technique begins with the baseline {\HG} color excess

\begin{equation}
\label{eq:rjce}
    E\HG = \HG - \HG_0 \, ,
\end{equation}

{\noindent}where $\HG_0$ is the intrinsic color of the stellar photosphere. This color combination lies in the Rayleigh-Jeans portion of most stellar spectra. Some stars are incompatible with this basic assumption of the RJCE technique, having an SED that does not look like a simple blackbody photosphere in this spectral range (`not-photospheric' stars), e.g., with an IR color excess or with a roughly power-law SED (in Vega magnitudes) declining monotonically from the MIR through the optical. These stars need to be flagged and removed from the analysis.

Note that $E\HG$ is a property only of the normalized extinction curve adopted, within the common scale factor adjusting for the actual dust column (Section \ref{sec:extinction-curve}). Therefore, all other measurements of extinction, at a single passband or a color excess, are proportional to $E\HG$. For the \hrd\ we need to evaluate $\AKs$ and $E\JK$, which can be written

\begin{equation}
\label{eq:AKs}
    \AKs=F \, E\HG
\end{equation}

{\noindent}and

\begin{equation}
\label{eq:EJK}
    \JK - \JK_0 \equiv E\JK = f \, E\HG \, .
\end{equation}

For the adopted extinction curves with $\powerlaw=1.8$ we find $F$ and $f$ to be $0.874$ and $1.488$, respectively, which are reasonably close to the values of $0.918$ and $1.377$ adopted in original applications of the RJCE technique to RGs \citep{Indebetouw2005,Zasowski2013}. There is certainly evidence that the shape of the NIR extinction curve in this range varies with position in the Galaxy \citep{Zasowski2009}.

To use Equations \ref{eq:AKs} and \ref{eq:EJK} directly we would need to know $E\HG$, and hence $\HG_0$, in Equation \ref{eq:rjce}. The original RJCE method assumes an intrinsic color of $\HG_0 = \hga$ appropriate for the bright RG stars to which it has been applied \citep{Majewski2011,Zasowski2013}. However, for field stars $\HG_0$ is unknown, because $\HG_0$ changes with the spectral type (which is unknown), which can affect the inferred extinction. For example, for RGs with lower effective temperatures than RCs, $\HG_0$ increases to at least 0.17\,mag \citep{Jian2017}, resulting in a lower estimate of the dust extinction as measured by this color excess (alternatively, if $\HG_0$ is kept constant at $\hga$, then the extinction is overestimated). There is a similar effect for cool MS stars, for which $\HG_0$ increases to at least 0.19\,mag \citep{Jian2017}.  For an A0 V star like Vega, $\HG_0$ is about 0 and for OB stars, $\HG_0$ becomes negative \citep{Pecaut2013}. For these hotter stars, this results in a larger estimate of the extinction than keeping $\HG_0$ constant.

If there were a known functional relationship between $\JK_0$ and $\HG_0$, then Equation \ref{eq:EJK} could be solved simultaneously with this constraint.  A simple example that is sufficient for our purposes is to assume this relationship is linear, passing through the $[\JK_0, \HG_0]$ pairs [\jka,\hga] and [$\jkb,\hgb$], approximating RCs and OBs, respectively, and thus having slope $\sjg = $ \sjgeval\ for the adopted pairs. For any star, this allows us to express $\JK_0$ as

\begin{equation}
\label{eq:JK0}
    \JK_0 = \frac{ \JK - f(\HG -\hga) - f \, \jka/\sjg}{1-f/\sjg} \, .
\end{equation}

{\noindent}This clearly reduces to Equation \ref{eq:EJK} when $\HG_0$ is constant (i.e., $\sjg = \infty$). Also, if the second pair were [0,0], the numerator would simply be $\JK - f\HG$.

Once we have $[\JK_0$ from Equation \ref{eq:JK0}, we can obtain $E\HG$ from Equation \ref{eq:EJK} and then \AKs\ from Equation \ref{eq:AKs}, which is the factor scaling the normalized extinction curve. Given the assumed normalized extinction curve, we can obtain any other extinction or color excess as well.

%%%%%%%%%%%%%%%%%%%%%%%
%                                                                   %
%                    APPLICATIONS                     %
%                                                                   %
%%%%%%%%%%%%%%%%%%%%%%%

\subsection{Applications}
\label{sec:rjceap}

An obvious application is that the data in the 2MASS-GLIMPSE CCD (Figure \ref{fig:RJCE_JK-HG2}) collapse to the assumed line relating $\JK_0$ and $\HG_0$. As long as the pairs are chosen reasonably, a sensible result is obtained. More interestingly, the 2MASS CCD collapses to a sensible line too.

\begin{figure}
\centering
\includegraphics[width=7cm]{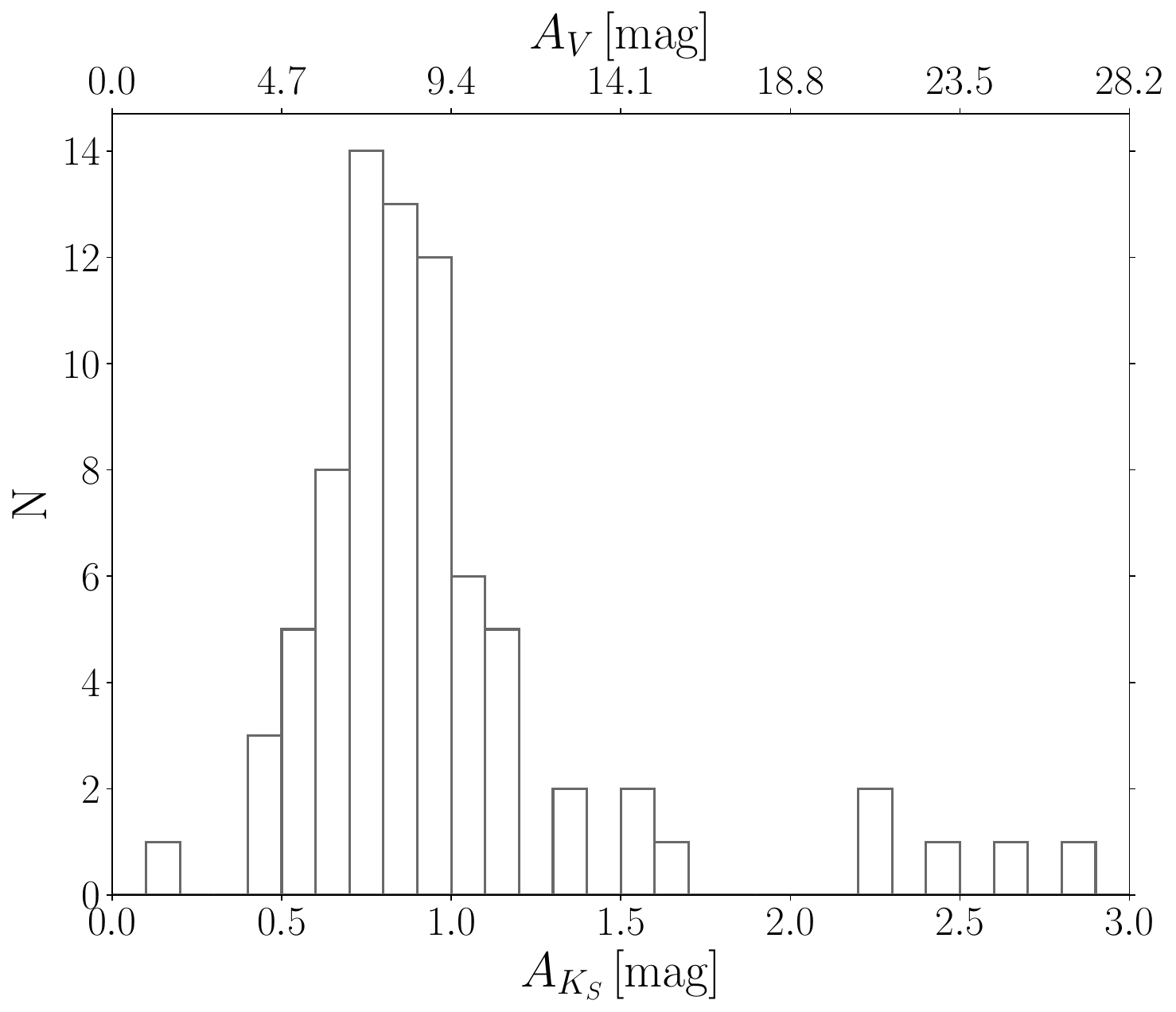}
\caption{Distribution of \AKs\ (lower x-axis) for the BIRS using our modified RJCE method and the corresponding values of {\AV} (upper x-axis) assuming $\AV/\AKs=9.41$ using our adopted extinction curve (see text).}
\label{fig:RJCE_AKhist}
\end{figure}

As a second application, in Figure \ref{fig:RJCE_AKhist} we show the distribution of extinction values for the BIRS derived using our modified RJCE technique. The lower x-axis is \AKs while the upper x-axis is \AV, assuming $\AV/\AKs = 9.41$ for the extinction curve with \RV\ = 3.6 and \powerlaw\ = 1.8. The BIRS have a range in \AKs\ of $0.15\,\mags$ to $3.39\,\mags$, with a median \AKs\ of $0.87\,\mags$ and a standard deviation of $0.54\,\mags$.

%%%%%%%%%%%%%%%%%%%%%%%%
%                                                                      %
%              VES 735 + companion                    %
%                                                                      %
%%%%%%%%%%%%%%%%%%%%%%%%

\section{VES 735 and its Projected Companion in KR 140}\label{subsec:VES}

KR 140 is a young \HII\ region believed to be ionized by the spontaneously-formed O8.5 V(e) exciting star, VES 735 \citep{Kerton1999, Ballantyne2000, Kerton2001, Kerton2008}. While this region is isolated to the west of the HDL, the radial velocities of VES 735 and the nebular \halpha\ and CO emission all match to within $2\,\kms$ and are all likely to be components of the same Perseus arm molecular complex \citep{Kerton1999}. KR 140 lacks precise distance measurements, although extinction measurements have placed KR 140 toward the near side of the cloud \citep{Ballantyne2000}.

% VES and Gaia
The parallax and proper motion of VES 735 suggest that the KR 140 region may have a different distance and bulk proper motion than the HDL. The Gaia parallax is $0.69 \pm 0.03\,\mas$, which is larger than the OB and maser parallax by $5\sigma$. The Gaia data for this star have a high-quality fidelity (1.0) but a larger ruwe (1.94) than the suggested cutoff value ($\lesssim 1.4$). At face value, the parallax places VES 735 more than 400 pc closer than the HDL, which conflicts with the significant amount of foreground extinction and the coherent CO velocities between the HDL and KR 140. Compared to classic runaway stars, VES 735 has a low peculiar motion ($\pmra = -0.75 \pm 0.03\,\masyr$ and $\pmdec = -1.10 \pm 0.03\,\masyr$) and it is still within the \HII\ region that it is creating.

% SED fits
We fit the observed SED of VES 735 with a standard extinction curve parameterized by $\RV=3.6$ and $\alpha=1.8$, allowing a small range in distance modulus based on the parallax. The ATLAS photometric data were inconsistent with those of APASS and so were disregarded in the fit; the results are consistent with fitting the 2MASS data alone, implying systematic effects in the ATLAS photometry. We found a spectral type of O9 V that is consistent with the literature within half a sub-type and an extinction of $\AKs=0.65\,\mags$ ($\AV=6.16\,\mags$) and a distance modulus of $10.75\,\mags$. 

We repeated the VES 735 fit by constraining the distance modulus instead to a $0.1\,\mags$ range around that of W3, yielding a distance modulus of $11.45\,\mags$ and a negligible change in the amount of foreground extinction ($\AKs=0.66\,\mags$, $\AV=6.21\,\mags$) for an O6.5 V spectral-type solution. While this is an earlier-than-expected spectral type, the absolute magnitude is only $\sim 0.6\,\mags$ brighter than expected for an O8.5 V star and may be indicative of a more luminous pre-MS star or one that has started to evolve off the MS. An increased luminosity and/or earlier spectral type would also boost the number of ionizing photons; the \HII region appears to be unbounded on the near side \citep{Ballantyne2000} and so this could in principle be accommodated.

% VESc 
VES 735 has a projected companion (angular offset $\sim 4{\farcs}5$), which we refer to as simply VESc 735, about which very little appears to be known. The measured parallax is $0.45 \pm 0.05\,\mas$, which is consistent with the OB and maser parallaxes of W3. With a proper motion of $\pmra = -0.60 \pm 0.04\,\masyr$ and $\pmdec = -0.44 \pm 0.05\,\masyr$, the peculiar proper motion ($0.22\,\masyr$) is small compared to the dispersion in the proper motion of OB stars. This star has a high-quality ruwe (0.47) but a somewhat low fidelity (0.55) that nevertheless meets our quality requirements. Fitting the observed SED with an extinction curve with the same parameterization as VES 735 yields a B5/B7 spectral type with an extinction of $\AKs = 0.64\,\mags$ ($\AV = 6.04\,\mags$) and distance modulus of $11.48\,\mags$. This extinction is very similar to that of VES 735, which supports the possibility that both stars are at the same distance as W3.

%%%%%%%%%%%%%%%%%%%%%%%
%                                                                   %
%                   BIBLIOGRAPHY                     %
%                                                                   %
%%%%%%%%%%%%%%%%%%%%%%%

\newpage
\bibliographystyle{aasjournal}
\bibliography{bibliography.bib,Planck_bib}

\end{document}

%% file: commands.tex
\newcommand{\Ks}{\ifmmode{K_s} \else{$K_s$}\fi}              % K_s magnitude
\newcommand{\JH}{\ifmmode{(J-H)} \else{$(J-H)$}\fi}          % (J-H) colour
\newcommand{\JK}{\ifmmode{(J-K_s)} \else{$(J-K_s)$}\fi}      % (J-K) colour
\newcommand{\HG}{\ifmmode{(H-[4.5])} \else{$(H-[4.5])$}\fi}  % (H-4.5) colour
\newcommand{\HK}{\ifmmode{(H-K_s)} \else{$(H-K_s)$}\fi}      % (H-K) colour magnitude
\newcommand{\ri}{\ifmmode{(r-i)} \else{$(r-i)$}\fi}          % (r-i) colour
\newcommand{\rz}{\ifmmode{(r-z)} \else{$(r-z)$}\fi}          % (r-z) colour
\newcommand{\band}{band}
\newcommand{\bands}{bands} % in CCDs
\newcommand{\hrd}{HR diagram}

\newcommand{\hga}{0.08} % end point for RC stars from Zasowski 2013
\newcommand{\jka}{0.6} 
\newcommand{\hgb}{-0.04} % end point for B3 V stars from EEM
\newcommand{\jkb}{-0.12} 
\newcommand{\sjgeval}{[\jka $- (\jkb)$]/[\hga $- (\hgb)$]}
\newcommand{\sjg}{s_{JG}}

\newcommand{\ire}{MIRE} % infrared excess 
\newcommand{\mire}{MIRE} % infrared excess but not \uns
\newcommand{\uns}{UnS} % unusually steep SED

\newcommand{\pmra}{\ifmmode{\mu_\mathrm{RA}} \else{$\mu_\mathrm{RA}$}\fi}  % proper motion in R.A.
\newcommand{\pmdec}{\ifmmode{\mu_\mathrm{Dec}} \else{$\mu_\mathrm{Dec}$}\fi} % proper motion in Dec.
\newcommand{\avgpmra}{\ifmmode{\langle\mu_\mathrm{RA}\rangle} \else{$\langle\mu_\mathrm{RA}\rangle$}\fi}  % average proper motion in R.A.
\newcommand{\avgpmdec}{\ifmmode{\langle\mu_\mathrm{Dec}\rangle} \else{$\langle\mu_\mathrm{Dec}\rangle$}\fi} % average proper motion in Dec.
\newcommand{\dm}{\ifmmode{\mathrm{dm}} \else{dm}\fi}  % distance modulus

\newcommand{\Alambda}{\ifmmode{A_\lambda} \else{$A_\lambda$}\fi}       % extinction at lambda wavelength
\newcommand{\RV}{\ifmmode{R_\mathrm{V}} \else{$R_\mathrm{V}$}\fi}      % R_V parameter
\newcommand{\AV}{\ifmmode{A_\mathrm{V}} \else{$A_\mathrm{V}$}\fi}      % extinction in V-band
\newcommand{\AJ}{\ifmmode{A_\mathrm{J}} \else{$A_\mathrm{J}$}\fi}      % extinction in J-band
\newcommand{\AH}{\ifmmode{A_\mathrm{H}} \else{$A_\mathrm{H}$}\fi}      % extinction in H-band
\newcommand{\AKs}{\ifmmode{A_\mathrm{K_s}} \else{$A_\mathrm{K_s}$}\fi} % extinction in Ks-band
\newcommand{\mlow}{\ifmmode{m_l} \else{$m_l$}\fi}                      % low-frequency Vega apparent magnitude
\newcommand{\Mlow}{\ifmmode{M_l} \else{$M_l$}\fi}                      % low-frequency Vega absolute magnitude
\newcommand{\powerlaw}{\ifmmode{\alpha} \else{$\alpha$}\fi}            % power-law exponent of NIR extinction curve
\newcommand{\NHI}{\ifmmode {N_{\rm HI}}\else $N_{\rm HI}$\fi}          % atomic hydrogen column density
\newcommand{\NH}{\ifmmode {N_{\rm H}}\else $N_{\rm H}$\fi}             % hydrogen column density
\newcommand{\tauu}{\ifmmode {\tau_{1}}\else $\tau_{1}$\fi}             % Herschel dust optical depth
\newcommand{\Kintrinsic}{\ifmmode{{K_{s,0}}} \else{${K_{s,0}}$}\fi} % extinction in Ks-band
\newcommand{\MKs}{\ifmmode{M_\mathrm{K_s}} \else{$M_\mathrm{K_s}$}\fi}     % absolute K_s
\newcommand{\mKs}{\ifmmode{m_\mathrm{K_s}} \else{$m_\mathrm{K_s}$}\fi}     % apparent K_s

\newcommand{\lambdaeff}{\ifmmode {\lambda_{\rm eff}}\else $\lambda_{\rm eff}$\fi}         % effective wavelength
\newcommand{\halpha}{\ifmmode {{\rm H}\alpha}\else H$\alpha$\fi}         % H-alpha
\newcommand{\HI}{\ifmmode \mathrm{\ion{H}{1}} \else \ion{H}{1} \fi}

\newcommand{\HII}{\ifmmode \mathrm{\ion{H}{2}} \else \ion{H}{2} \fi}

\newcommand{\mags}{\ifmmode{\rm mag} \else{mag}\fi}                               % mag
\newcommand{\kms}{\ifmmode{\mathrm{km\,s^{-1}}} \else {$\mathrm{km\,s^{-1}}$}\fi} % km/s
\newcommand{\microns}{\ifmmode{{\mu}\mathrm{m}} \else {${\mu}$m}\fi}              % microns
\newcommand{\nm}{\ifmmode{\mathrm{nm}} \else {nm}\fi}                             % nanometers
\newcommand{\cmsq}{\ifmmode{\mathrm{cm\,^{-2}}} \else {cm$^{-2}$}\fi}             % cm^-2
\newcommand{\mas}{\ifmmode{\rm mas} \else{mas}\fi}                                % mas
\newcommand{\masyr}{\ifmmode{\rm mas\,yr^{-1}} \else{mas\,yr$^{-1}$}\fi}          % mas/yr
\newcommand{\kpc}{\ifmmode{\rm kpc} \else{kpc}\fi}                                % kpc
\newcommand{\pc}{\ifmmode{\rm pc} \else{pc}\fi}                                   % pc
\renewcommand{\deg}{\ifmmode {^\circ}\else $^\circ$\fi}                           % degrees
\newcommand{\Msun}{\ifmmode {M_\odot}\else $M_\odot$\fi}                          % degrees
\newcommand{\uas}{\ifmmode{\mu{\rm as}} \else{{$\mu$as}}\fi}                      % mas
\newcommand{\Myr}{\ifmmode{\mathrm{Myr}} \else {$\mathrm{Myr}$}\fi}               % Myr

\newcommand{\SNR}{\ifmmode{\rm S/N} \else{S/N}\fi}                          % signal to noise ratio
\newcommand{\Teff}{\ifmmode{T_\mathrm{eff}} \else{$T_\mathrm{eff}$}\fi}     % effective temperature
\newcommand{\logg}{\ifmmode{\mathrm{log}\,g} \else{log\,$g$}\fi}            % logarithm of surface gravity
\newcommand{\FeH}{\ifmmode{\mathrm{[Fe/H]}} \else{[Fe/H]}\fi}               % [Fe/H] metallicity
\newcommand{\ruwe}{\ifmmode{\rm \tt ruwe} \else{\tt ruwe}\fi}               % signal to noise ratio
\newcommand{\rest}{\ifmmode{r_\mathrm{est}} \else{$r_\mathrm{est}$}\fi}     % geometric distance

\newcommand{\lime}{lime} % color of discarded points within the dashed circle in figure 17 left

\newcommand{\lemon}{lime} % color of location circles in figure 4

%% file: tables/table1.tex
%\begin{turnpage}
\begin{deluxetable*}{lccccccccccccc}
%{lC{1.6cm}C{1.8cm}C{2.0cm}C{2.0cm}C{0.5cm}C{0.5cm}C{0.7cm}C{1cm}C{0.001cm}C{0.6cm}C{0.9cm}C{0.8cm}C{3cm}}
\tablecaption{Summary of Gaia and SED Fitting Results for the BIRS \label{table:birs-results}}
\tablehead{
 & & \multicolumn{7}{c}{Gaia DR3 Results} & & \multicolumn{4}{c}{SED Fitting Results}  \\
 \cline{3-9} \cline{11-14}
\colhead{BIRS} &
\colhead{Alt.} & 
\colhead{$\varpi-Z$\tablenotemark{\footnotesize{c}}} & 
\colhead{\pmra}      & 
\colhead{\pmdec}     &
\colhead{ruwe}       & 
\colhead{fv2}        &
\colhead{$Z$}        &
\colhead{Cloud}      &
\colhead{}           &
\colhead{dm}       &
\colhead{{\AKs}}     &
\colhead{{\AV}}      &
\colhead{Object\tablenotemark{\footnotesize{d}}}     \\
\colhead{ID\tablenotemark{\footnotesize{a}}} &
\colhead{ID\tablenotemark{\footnotesize{b}}}           & 
\colhead{[mas]}      & 
\colhead{[\masyr]}   & 
\colhead{[\masyr]}   &
\colhead{}           & 
\colhead{}           &
\colhead{[mas]}           &
\colhead{Class.}     &
\colhead{}           &
\colhead{[mag]}      &
\colhead{[mag]}      &
\colhead{[mag]}      &
\colhead{Class.} \\
\colhead{(1)}        &
\colhead{(2)}        &
\colhead{(3)}        & 
\colhead{(4)}        & 
\colhead{(5)}        &
\colhead{(6)}        & 
\colhead{(7)}        &
\colhead{(8)}        &
\colhead{(9)}        & 
\colhead{}           & 
\colhead{(10)}       & 
\colhead{(11)}       & 
\colhead{(12)}       & 
\colhead{(13)}
}
\startdata
1                     & \nodata               & 0.39$\pm$0.07  & -1.26$\pm$0.06  & -0.13$\pm$0.07   & 1.10    & 0.51    & -0.05   & 100     & & 12.52 & 0.51 & 4.75 & K3 III\textsuperscript{1,4}\\[-0ex]
2\textsuperscript{*}  & \nodata               & 0.52$\pm$0.06  & -1.53$\pm$0.05  & -0.51$\pm$0.06   & 1.00    & 1.00    & -0.06   & 111     & & 11.35 & 0.64 & 5.98 & K3 III\textsuperscript{1,2,3,4}\\[-0ex]
3                     & \nodata               & \nodata        & \nodata         & \nodata          & 1.32    & 0.02    & \nodata & 1-{}-   & & \nodata & \nodata & \nodata & \nodata\\[-0ex]
4                     & \nodata               & \nodata        & \nodata         & \nodata          & \nodata & \nodata & \nodata & \nodata & & \nodata & \nodata & \nodata & diffuse\textsuperscript{e}\\[-0ex]
5                     & \nodata               & \nodata        & \nodata         & \nodata          & \nodata & \nodata & \nodata & \nodata & & \nodata & \nodata & \nodata & diffuse\textsuperscript{e}\\[-0ex]
6\textsuperscript{*}  & RL27                  & 0.49$\pm$0.02  & 0.86$\pm$0.02   & 1.36$\pm$0.02    & 1.08    & 1.00    & -0.03   & 111     & & 11.52 & 0.83 & 7.77 & O5 V\textsuperscript{1,2,3,4}\\[-0ex]
\enddata
\tablenotetext{}{\textbf{Notes.} The full version of this table is available electronically in machine-readable format.} \vspace{-1.5ex}
\tablenotetext{a}{Asterisks (*) specify the bona fide (not imposter) BIRS based on ATLAS photometry that are associated with W3 via Gaia astrometry (see Section \ref{subsec:propermotion}). Boldface indicates OB candidates in W3 meeting strict IR photometry and Gaia astrometry requirements (see Section \ref{sec:newOBcandidates}).} \vspace{-1.5ex}
\tablenotetext{b}{Obtained from the literature, including IRS for \citetalias{Bik2012} infrared sources, KKB for \citetalias{Kiminki2015} OB stars, RL for \citetalias{Roman-Lopes2019} OB stars, A for APOGEE RGs \citep{Zasowski2013}, and G for Gaia LPV sources \citep{Gaia2019}.} \vspace{-1.5ex}
\tablenotetext{c}{The measured parallax zero point ($Z$) is included in the listed parallax.} \vspace{-1.5ex}
\tablenotetext{d}{Derived using (1) the SED model fitting results, and placement of star in the following diagrams: (2) observational HRD, (3) 2MASS-GLIMPSE CCD, (4) 2MASS CCD, and (5) ATLAS CMD. For those identified as LPVs, we note their confidence levels in parentheses.} \vspace{-1.5ex}
\tablenotetext{e}{Using Pan-STARRS and 2MASS images.} \vspace{-1.5ex}
\tablenotetext{f}{Derived using intrinsic colors of M-dwarfs \citep{Pecaut2013}.} 
\end{deluxetable*}
%\end{turnpage}

%% file: tables/table2.tex
\begin{deluxetable*}{lcccccccccccc}
\tablecaption{OB Star Candidates in W3 \label{table:ob-candidates}}
\tablehead{
 & \multicolumn{7}{c}{Gaia DR3 Results} & & \multicolumn{4}{c}{SED Fitting Results} \\
 \cline{2-8} \cline{10-13}
\colhead{ID}           &
\colhead{R.A.}         & 
\colhead{Dec.}         & 
\colhead{$\varpi-Z$\textsuperscript{a}}     & 
\colhead{\pmra}        &
\colhead{\pmdec}       & 
\colhead{ruwe}         &
\colhead{fv2}          &
\colhead{}             &
\colhead{dm}         &
\colhead{\AKs}         & 
\colhead{\AV}          & 
\colhead{Spec.}        \\
\colhead{}              &
\colhead{[J2000]}       &
\colhead{[J2000]}       & 
\colhead{[mas]}         & 
\colhead{[\masyr]}      &
\colhead{[\masyr]}      & 
\colhead{}              &
\colhead{}              &
\colhead{}              &
\colhead{[mag]}         & 
\colhead{[mag]}         &
\colhead{[mag]}         &
\colhead{Type}\\
\colhead{(1)}              &
\colhead{(2)}              &
\colhead{(3)}              & 
\colhead{(4)}              & 
\colhead{(5)}              &
\colhead{(6)}              & 
\colhead{(7)}              &
\colhead{(8)}              &
\colhead{}              &
\colhead{(9)}             & 
\colhead{(10)}             &
\colhead{(11)}             &
\colhead{(12)}
}
\startdata
 1 & 02:16:51.740 & +61:38:45.455 & 0.52$\pm$0.02 & -0.83$\pm$0.02 & -0.75$\pm$0.02 & 1.00 & 1.00 & & 11.34 & 0.41 & 3.40 & B8 V \\
 2 & 02:17:51.832 & +61:34:41.747 & 0.54$\pm$0.02 & -0.19$\pm$0.01 & -0.08$\pm$0.02 & 1.00 & 1.00 & & 11.18 & 0.32 & 2.63 & B8 V \\
 3 & 02:17:53.927 & +61:04:53.533 & 0.52$\pm$0.02 & -0.84$\pm$0.02 & -0.24$\pm$0.02 & 1.01 & 1.00 & & 11.32 & 0.40 & 3.32 & B8 V \\
 4 & 02:17:55.450 & +61:07:09.153 & 0.52$\pm$0.02 & -0.85$\pm$0.01 & -0.32$\pm$0.02 & 1.12 & 1.00 & & 11.24 & 0.36 & 3.02 & B3 V \\
 5 & 02:18:11.044 & +61:09:10.913 & 0.52$\pm$0.01 & -0.83$\pm$0.01 & -0.19$\pm$0.02 & 1.05 & 1.00 & & 11.32 & 0.46 & 3.81 & B3 V \\
 6 & 02:18:12.642 & +61:57:38.490 & 0.50$\pm$0.03 & -1.45$\pm$0.02 & -0.75$\pm$0.03 & 1.16 & 1.00 & & 11.30 & 0.62 & 5.17 & B3 V \\
\enddata
\tablenotetext{}{\textbf{Notes.} The full version of this table is available electronically in machine-readable format.} \vspace{-1.5ex}
\tablenotetext{a}{The measured parallax zero point ($Z$) is included in the listed parallax.} \vspace{-1.5ex}
\tablenotetext{b}{We previously identified candidate \# 57 as a possible OB candidate based purely on its astrometry (see Section \ref{subsubsec:pmmember}).} \vspace{-1.5ex}
\end{deluxetable*}

%% file: tables/tableA1.tex
%\newpage
%\begin{turnpage}
\begin{deluxetable*}{lcccccccc}
\tablecaption{Modern Positions and IR Photometry of the BIRS \label{table:birs-positions}}
\tablehead{
\colhead{BIRS}             &
\colhead{R.A.}             & 
\colhead{Dec.}             & 
\colhead{$\Delta$R.A.}     & 
\colhead{$\Delta$Dec.}     &
\colhead{$R$}              & 
\colhead{$I$}              &
\colhead{$r$}              &
\colhead{$i$}             \\
\colhead{ID}               &
\colhead{[J2000]}          & 
\colhead{[J2000]}          & 
\colhead{[\arcsec]}        & 
\colhead{[\arcsec]}        &
\colhead{[mag]}            & 
\colhead{[mag]}            &
\colhead{[mag]}            &
\colhead{[mag]}           \\
\colhead{(1)}              &
\colhead{(2)}              &
\colhead{(3)}              & 
\colhead{(4)}              & 
\colhead{(5)}              &
\colhead{(6)}              & 
\colhead{(7)}              &
\colhead{(8)}              &
\colhead{(9)}
}
\startdata
1  & 2:27:17.4 & +62:14:00.3   & 6.32  & 0.15    & 16.8       & 13.4 & 17.57$\pm$0.01 & 16.19$\pm$0.01 \\
2  & 2:27:15.1 & +62:13:48.8   & 7.31  & -0.11   & 17.3       & 13.3 & 17.47$\pm$0.01 & 15.85$\pm$0.01 \\
3  & 2:26:56.4 & +62:15:43.2   & 7.78  & 0.41    & \nodata    & 14.2 & \nodata & \nodata \\
4  & \nodata & \nodata & \nodata & \nodata & \nodata & \nodata & \nodata & \nodata \\
5  & \nodata & \nodata & \nodata & \nodata & \nodata & \nodata & \nodata & \nodata \\
6  & 2:26:49.6 & +62:15:35.1   & 6.63   & 0.15   & \nodata    & 11.8 & 13.50$\pm$0.07 & 12.21$\pm$0.04  \\
\enddata
\tablenotetext{}{\textbf{Notes.} The full version of this table is available electronically in machine-readable format. 
%\peter{Does not need to be rotated. Does not need so many rows.}
}
\end{deluxetable*}
%\end{turnpage}
%\clearpage